\documentclass[sigconf,screen]{acmart}

\setcopyright{rightsretained}
\acmPrice{}
\acmDOI{10.1145/3597926.3598038}
\acmYear{2023}
\copyrightyear{2023}
\acmSubmissionID{issta23main-p36-p}
\acmISBN{979-8-4007-0221-1/23/07}
\acmConference[ISSTA '23]{Proceedings of the 32nd ACM SIGSOFT International Symposium on Software Testing and Analysis}{July 17--21, 2023}{Seattle, WA, USA}
\acmBooktitle{Proceedings of the 32nd ACM SIGSOFT International Symposium on Software Testing and Analysis (ISSTA '23), July 17--21, 2023, Seattle, WA, USA}
\received{2023-02-16}
\received[accepted]{2023-05-03}

\usepackage{xspace}
\usepackage{xcolor}
\usepackage{graphicx}
\usepackage[normalem]{ulem} %
\usepackage{amsmath}
\usepackage{enumitem}
\usepackage{semantic} %
\usepackage{siunitx}
\usepackage{scalefnt}
\usepackage{microtype}

\usepackage{hyperref}
\usepackage{breakurl}

\usepackage{booktabs}
\usepackage{multirow}
\usepackage{makecell}
\usepackage{ragged2e}

\usepackage{caption}
\usepackage{subcaption}
\usepackage{wrapfig}

\usepackage{pifont}
\usepackage{listings}

\usepackage{algorithm}
\usepackage[noend]{algpseudocode} %

\usepackage{tikz}
\usepackage{forest} %
\usetikzlibrary{arrows.meta}
\usetikzlibrary{backgrounds}
\usetikzlibrary{calc}
\usetikzlibrary{decorations.pathreplacing}
\usetikzlibrary{fit}
\usetikzlibrary{positioning}
\usetikzlibrary{shapes.geometric}

\usepackage{balance} %
\usepackage{datetime}
\usepackage{fmtcount} %

\newcommand{\XSpace}[1]{}
\newcommand{\XComment}[1]{}
\newcommand{\EditAdd}[1]{\textcolor{green}{[#1]}}
\newcommand{\EditRm}[1]{\textcolor{red}{[\sout{#1}]}}
\newcommand{\EditMod}[2]{\textcolor{red}{[\sout{#1}]}\textcolor{green}{[#2]}}
\newcommand{\DefMacro}[2]{\expandafter\newcommand\csname rmk-#1\endcsname{#2}}
\newcommand{\UseMacro}[1]{\csname rmk-#1\endcsname}

\newcommand{\MyPara}[1]{\vspace{2pt}\noindent\textbf{#1}.}
\newcommand{\MyParaOnly}[1]{\noindent\textbf{#1}}
\newcommand{\reducedstrut}{\vrule width 0pt height .9\ht\strutbox depth .9\dp\strutbox\relax}
\newcommand{\InputWithSpace}[1]{\bgroup\def\arraystretch{1.1}\input{#1}\egroup}
\newcommand{\Code}[1]{{\ifmmode{\mathtt{#1}}\else$\mathtt{#1}$\fi}}
\newcommand{\CodeIn}[1]{\texttt{\small #1}}
\newcommand{\ColorBack}[1]{%
\begingroup
\setlength{\fboxsep}{0pt}%
\colorbox{purple!20}{\reducedstrut#1\/}%
\endgroup
}
\newcommand{\BigNum}[1]{\num[group-separator = {,}, group-four-digits]{#1}} %
\newcommand{\Capitalize}[1]{\expandafter\MakeUppercase {#1}}
\newcommand{\Num}[1]{#1}    %
\newcommand{\Hrule}[1][gray]{%
\par\addvspace{1pt}%
\begingroup\color{#1}%
\hrule
\endgroup
\addvspace{1pt}%
}

\newcommand{\specialcell}[2][c]{%
\begin{tabular}[#1]{@{}c@{}}#2\end{tabular}}
\newcolumntype{R}[1]{>{\RaggedLeft\arraybackslash}p{#1}}
\newcolumntype{L}[1]{>{\RaggedRight\arraybackslash}p{#1}}

\newcommand{\ltrue}{\top} %
\newcommand{\lfalse}{\bot} %

\newcommand{\limply}{\rightarrow} %
\newcommand{\liff}{\leftrightarrow} %
\newcommand{\lcofactor}{\downarrow} %

\newcommand{\llimply}{\Rightarrow} %
\newcommand{\lliff}{\Leftrightarrow} %
\newcommand{\lsat}{\vDash} %
\newcommand{\lnsat}{\nvDash} %

\newcommand{\lqall}[1]{\forall #1\ldotp~} %
\newcommand{\ScaleForall}[1]{\vcenter{\hbox{\scalefont{#1}$\forall$}}}
\newcommand{\lqAll}[2]{%
\vphantom\sum
\mathchoice{\ScaleForall{2}}{\ScaleForall{1.4}}{\ScaleForall{1}}{\ScaleForall{0.75}}_{\substack{#1}} #2\ldotp~} %
\newcommand{\lqexist}[1]{\exists #1\ldotp~} %
\newcommand{\ScaleExists}[1]{\vcenter{\hbox{\scalefont{#1}$\exists$}}}
\newcommand{\lqExist}[2]{%
\vphantom\sum
\mathchoice{\ScaleExists{2}}{\ScaleExists{1.4}}{\ScaleExists{1}}{\ScaleExists{0.75}}_{\substack{#1}} #2\ldotp~} %
\newcommand{\lqallOnly}[1]{\forall #1 } %
\newcommand{\lqexistOnly}[1]{\exists #1 } %

\newcommand{\lAnd}[1]{\bigwedge_{\substack{#1}}} %
\newcommand{\lOr}[1]{\bigvee_{\substack{#1}}} %

\newcommand{\set}[1]{\{#1\}} %
\newcommand{\Set}[1]{\left\{#1\right\}} %
\newcommand{\tuple}[1]{\langle#1\rangle} %
\newcommand{\Tuple}[1]{\left\langle#1\right\rangle} %

\newcommand{\sunion}{\cup} %
\newcommand{\sinter}{\cap} %

\definecolor{gray}{RGB}{211,211,211}
\newcommand{\jbasicstyle}{\small\sffamily} %
\newcommand{\textcode}[1]{{#1}}
\newcommand{\jnumberstyle}{\scriptsize}
\newcommand{\Hilight}{\makebox[0pt][l]{\color{gray}\rule[-3pt]{0.80\linewidth}{9pt}}}
\newcommand{\lstbg}[3][0pt]{{\fboxsep#1\colorbox{#2}{\strut #3}}}

\lstdefinelanguage{pseudo}
{
morekeywords={},
keywordstyle=\bfseries,
lineskip=-0.1em,
numbers=left, %
numberstyle=\jnumberstyle,
numbersep=4pt,
basicstyle=\jbasicstyle,
breaklines=true,
breakautoindent=true,
tabsize=2,
columns=fullflexible,
morecomment=*[l][\textsl]{//},
mathescape=true,
xleftmargin=10pt,
}

\lstdefinelanguage{todo-comment}
{
morekeywords={},
keywordstyle=\bfseries,
lineskip=-0.1em,
numbers=none,
basicstyle=\jbasicstyle,
breaklines=true,
breakautoindent=true,
tabsize=2,
columns=fullflexible,
morecomment=*[l][\textsl]{//},
mathescape=true,
xleftmargin=-10pt,
}

\lstdefinelanguage{bug}
{
language=java,
numbers=left,
basicstyle=\scriptsize\ttfamily,
numberstyle=\scriptsize,
breaklines=true,
columns=fullflexible,
xleftmargin=16pt,
showstringspaces=false,
escapeinside={(*@}{@*)}
}

\lstdefinelanguage{jog}
{
language=java-pretty,
emph={@Pattern,@Constant,before,after}, emphstyle=\color{blue}
}

\lstdefinelanguage{jog-in-tikz}
{
language=jog,
numbers=none,
basicstyle=\scriptsize\ttfamily,
xleftmargin=0pt,
}

\lstdefinelanguage{sketchy-display}
{
language=sketchy,
numbers=none,
}

\lstdefinelanguage{sketchy-appendix}
{
language=java,
numbers=left,
basicstyle=\scriptsize\ttfamily,
numberstyle=\scriptsize,
breaklines=true,
columns=fullflexible,
xleftmargin=16pt,
showstringspaces=false,
emph={@Entry}, emphstyle=\color{blue},
emph={intId,intVal,arithmetic,relation,logic,@Entry,SUB,ADD,MUL,DIV,LE,GE,LT,GT,EQ,NE,AND,OR}, emphstyle=\color{blue},
emph={[2]eval}, emphstyle={[2]\color{purple}\bfseries},
escapeinside={(*@}{@*)}
}

\lstdefinelanguage{syntax}
{
numbers=none,
basicstyle=\footnotesize\ttfamily,
breaklines=true,
columns=fullflexible,
xleftmargin=16pt,
showstringspaces=false,
}

\lstdefinelanguage{semantics}
{
morekeywords={if,else,where}, keywordstyle=\bfseries,
numbers=none,
basicstyle=\footnotesize\ttfamily,
breaklines=true,
columns=fullflexible,
xleftmargin=16pt,
showstringspaces=false,
}

\lstdefinelanguage{cpp-pretty-diff}
{
language=cpp-pretty,
morecomment=[f][\lstbg{red!20}]{-},
morecomment=[f][\lstbg{blue!20}]+,
keepspaces=true
}

\lstdefinelanguage{cpp-pretty}
{
language=C++,
numbers=left,
basicstyle=\footnotesize\ttfamily,
numberstyle=\footnotesize,
breaklines=true,
columns=fullflexible,
xleftmargin=16pt,
showstringspaces=false,
}

\lstdefinelanguage{java-pretty}
{
language=Java,
numbers=left,
basicstyle=\footnotesize\ttfamily,
numberstyle=\footnotesize,
breaklines=true,
columns=fullflexible,
xleftmargin=16pt,
showstringspaces=false,
}

\lstdefinelanguage{java-display}
{
language=java-pretty,
numbers=none,
}

\makeatletter
\let\OldStatex\Statex
\renewcommand{\Statex}[1][3]{%
\setlength\@tempdima{\algorithmicindent}%
\OldStatex\hskip\dimexpr#1\@tempdima\relax}
\makeatother

\algnewcommand\OR{\textbf{or} }

\algnewcommand\AND{\textbf{and} }

\algnewcommand\True{\textbf{true}}

\algnewcommand\False{\textbf{false}}

\algnewcommand\Break{\textbf{break}}%

\algdef{SE}[DOWHILE]{Do}{EndDoWhile}{\algorithmicdo}[1]{\algorithmicwhile\ #1}

\algdef{SL}[INPUT]{Input}{0}{\textbf{Input: }}
\algdef{SL}[OUTPUT]{Output}{0}{\textbf{Output: }}

\makeatletter
\algrenewcommand\alglinenumber[1]{\footnotesize #1:} %
\algrenewcommand\ALG@beginalgorithmic{\footnotesize}
\algrenewcommand\algorithmiccomment[2][\footnotesize]{{#1\hfill\(\triangleright\) #2}} %
\makeatother

\newcommand{\Title}{Pattern-Based Peephole Optimizations with Java JIT Tests}
\newcommand{\Tool}{\textsc{JOG}\xspace}

\newcommand{\Alive}{Alive\xspace}
\newcommand{\AliveTwo}{Alive2\xspace}
\newcommand{\Renaissance}{\textit{Renaissance}\xspace}

\newcommand{\eAST}{eAST\xspace}
\newcommand{\eASTs}{eASTs\xspace}
\newcommand{\API}{API\xspace}
\newcommand{\APIs}{APIs\xspace}
\newcommand{\JIT}{JIT\xspace}
\newcommand{\COne}{C1\xspace}
\newcommand{\CTwo}{C2\xspace}
\newcommand{\HotSpot}{HotSpot\xspace}
\newcommand{\JVM}{JVM\xspace}
\newcommand{\JVMs}{JVMs\xspace}
\newcommand{\OpenJDK}{OpenJDK\xspace}
\newcommand{\LLVM}{LLVM\xspace}
\newcommand{\DSL}{DSL\xspace}
\newcommand{\DSLs}{DSLs\xspace}
\newcommand{\optimize}{optimize\xspace}
\newcommand{\Optimize}{Optimize\xspace}
\newcommand{\optimizes}{optimizes\xspace}
\newcommand{\Optimizes}{Optimizes\xspace}
\newcommand{\optimized}{optimized\xspace}
\newcommand{\Optimized}{Optimized\xspace}
\newcommand{\optimization}{optimization\xspace}
\newcommand{\Optimization}{Optimization\xspace}
\newcommand{\optimizations}{optimizations\xspace}
\newcommand{\Optimizations}{Optimizations\xspace}
\newcommand{\ptn}{pattern\xspace}
\newcommand{\Ptn}{Pattern\xspace}
\newcommand{\ptns}{patterns\xspace}
\newcommand{\Ptns}{Patterns\xspace}
\newcommand{\shadow}{shadow\xspace}
\newcommand{\shadows}{shadows\xspace}
\newcommand{\Shadow}{Shadow\xspace}
\newcommand{\Shadows}{Shadows\xspace}
\newcommand{\shadowing}{shadowing\xspace}
\newcommand{\Shadowing}{Shadowing\xspace}
\newcommand{\shadowed}{shadowed\xspace}
\newcommand{\Shadowed}{Shadowed\xspace}
\newcommand{\Composite}{Composite\xspace}
\newcommand{\Compositing}{Compositing\xspace}
\newcommand{\Composited}{Composited\xspace}

\newcommand{\aTool}{jattack}
\newcommand{\aAttack}{Boom}
\newcommand{\aHole}[1]{\ensuremath{[\![}#1\ensuremath{]\!]}}
\newcommand{\aConfig}[1]{\ensuremath{\langle} #1 \ensuremath{\rangle}}

\let\oldding\ding%
\renewcommand{\ding}[2][1]{\scalebox{#1}{\oldding{#2}}}%
\newcommand{\cicrlenumber}[1]{\ding[1.1]{\number\numexpr181 + #1}}

\newcommand{\annotation}[2]{\cicrlenumber{#1}\label{#2}}
\newcommand{\highlightApi}[1]{{\color{blue}#1}}
\newcommand{\highlightEval}[1]{{\color{purple}\textbf{#1}}}
\colorlet{colorhole1}{red!20}
\colorlet{colorhole2}{green!20}
\colorlet{colorhole3}{yellow!20}
\colorlet{colorhole4}{teal!20}
\colorlet{colorhole5}{brown!20}

\DefMacro{test-num-classes}{8}
\DefMacro{test-num-methods}{45}

\DefMacro{ptn-pSub1}{SUB1}
\DefMacro{ptn-pSub2}{SUB2}
\DefMacro{ptn-pSub3}{SUB3}
\DefMacro{ptn-pSubXMinus_XPlusY_}{SUB4}
\DefMacro{ptn-pSub_XMinusY_MinusX}{SUB5}
\DefMacro{ptn-pSubXMinus_YPlusX_}{SUB6}
\DefMacro{ptn-pSub7}{SUB7}
\DefMacro{ptn-pSub8}{SUB8}
\DefMacro{ptn-pSub9}{SUB9}
\DefMacro{ptn-pSub10}{SUB10}
\DefMacro{ptn-pSub11}{SUB11}
\DefMacro{ptn-pSub12}{SUB12}
\DefMacro{ptn-pSub13}{SUB13}
\DefMacro{ptn-pSubAssociative1}{SUB14}
\DefMacro{ptn-pSubAssociative2}{SUB15}
\DefMacro{ptn-pSubAssociative3}{SUB16}
\DefMacro{ptn-pSubAssociative4}{SUB17}
\DefMacro{ptn-pSubNegRShiftToURShift}{SUB18}
\DefMacro{ptn-pNewSubCMinus_C2MinuxX_}{SUB19}
\DefMacro{ptn-pNewSubAddSub1539}{SUB20}
\DefMacro{ptn-pNewSubAddSub1560}{SUB21}
\DefMacro{ptn-pNewSubNotXMinusNotY}{SUB22}
\DefMacro{ptn-pNewSubAddSub1564}{SUB23}
\DefMacro{ptn-pNewSub_XOrY_Minus_XXorY_}{SUB24}
\DefMacro{ptn-pNewSub_AOrB_Minus_AAndB_}{SUB25}
\DefMacro{ptn-pNewSub_APlusB_Minus_AOrB_}{SUB26}
\DefMacro{ptn-pNewSub_APlusB_Minus_AXorB_}{SUB27}
\DefMacro{ptn-pNewSub_AAndB_Minus_AOrB_}{SUB28}
\DefMacro{ptn-pNewSub_AXorB_Minus_AOrB_}{SUB29}
\DefMacro{ptn-pNewSubAddSub1574}{SUB30}
\DefMacro{ptn-pNewSub_XOrY_MinusX}{SUB31}
\DefMacro{ptn-pNewSub_Op1And_NegX__MinusOp1}{SUB32}
\DefMacro{ptn-pNewSub_Op1AndC_MinusOp1}{SUB33}
\DefMacro{ptn-pNewSubXMinus_XAndY_}{SUB34}
\DefMacro{ptn-pNewNotXMinus_NotXMinY_}{SUB35}
\DefMacro{ptn-pNewNotXMinus_NotXMinY_Sym}{SUB36}
\DefMacro{ptn-pNew_NotXMinY_MinusNotX}{SUB37}
\DefMacro{ptn-pNew_NotXMinY_MinusNotXSym}{SUB38}
\DefMacro{ptn-pNewNotXMinus_NotXMaxY_}{SUB39}
\DefMacro{ptn-pNewNotXMinus_NotXMaxY_Sym}{SUB40}
\DefMacro{ptn-pNew_NotXMaxY_MinusNotX}{SUB41}
\DefMacro{ptn-pNew_NotXMaxY_MinusNotXSym}{SUB42}
\DefMacro{ptn-pAdd1}{ADD1}
\DefMacro{ptn-pAdd2}{ADD2}
\DefMacro{ptn-pAdd3}{ADD3}
\DefMacro{ptn-pAdd3Sym}{ADD4}
\DefMacro{ptn-pAdd4}{ADD5}
\DefMacro{ptn-pAdd4Sym}{ADD6}
\DefMacro{ptn-pAdd5}{ADD7}
\DefMacro{ptn-pAdd6}{ADD8}
\DefMacro{ptn-pAdd7}{ADD9}
\DefMacro{ptn-pAdd8}{ADD10}
\DefMacro{ptn-pAdd9}{ADD11}
\DefMacro{ptn-pAddAssociative1}{ADD12}
\DefMacro{ptn-pAddAssociative2}{ADD13}
\DefMacro{ptn-pAddAssociative3}{ADD14}
\DefMacro{ptn-pAddAssociative4}{ADD15}
\DefMacro{ptn-pAddURShiftThenLShiftToRRotation}{ADD16}
\DefMacro{ptn-pAddURShiftThenLShiftToRRotationSym}{ADD17}
\DefMacro{ptn-pAddNotXPlusOne}{ADD18}
\DefMacro{ptn-pAddMoveConstantRight}{ADD19}
\DefMacro{ptn-pAddConstantAssociative}{ADD20}
\DefMacro{ptn-pAddPushConstantDown}{ADD21}
\DefMacro{ptn-pAddPushConstantDownSym}{ADD22}
\DefMacro{ptn-p_AMaxB_Plus_AMinB_}{ADD23}
\DefMacro{ptn-p_AMaxB_Plus_AMinB_Sym}{ADD24}
\DefMacro{ptn-p_XLShiftC1_Or_XURShiftC2_}{OR1}
\DefMacro{ptn-p_XURShiftC1_Or_XLShiftC2_}{OR2}
\DefMacro{ptn-p_XLShiftS_Or_XURShift_ConMinusS__}{OR3}
\DefMacro{ptn-p_XURShiftS_Or_XLShift_ConMinusS__}{OR4}
\DefMacro{ptn-pMinAssociative}{MIN1}
\DefMacro{ptn-pAddDistributiveOverMin}{MIN2}
\DefMacro{ptn-pAssociativeThenAddDistributiveOverMin}{MIN3}
\DefMacro{ptn-pNewAddNotXPlusOneToNegX1}{ADD25}
\DefMacro{ptn-pNewAddNotXPlusOneToNegX2}{ADD26}
\DefMacro{ptn-pNewAddNotXPlusOneToNegX3}{ADD27}
\DefMacro{ptn-pNewAddNotXPlusOneToNegX4}{ADD28}
\DefMacro{ptn-pNewAddAddSub1165}{ADD29}
\DefMacro{ptn-pNewAddAddSub1156}{ADD30}
\DefMacro{ptn-pNewAddAddSub1202}{ADD31}
\DefMacro{ptn-pNewAddAddSub1295}{ADD32}
\DefMacro{ptn-pNewAddAddSub1295Sym}{ADD33}
\DefMacro{ptn-pNewAddAddSub1309}{ADD34}
\DefMacro{ptn-pNewAddAddSub1309Sym}{ADD35}
\DefMacro{ptn-pNewAdd_APlusC1_Plus_C2MinusB_}{ADD36}
\DefMacro{ptn-pNewAddXModC0PlusXDivC0ModC1MulC0}{ADD37}
\DefMacro{ptn-pNewAddAddSub1040}{ADD38}
\DefMacro{ptn-pNewAddAddSub1043}{ADD39}
\DefMacro{ptn-pNew_XOrC2_PlusC}{ADD40}
\DefMacro{ptn-pNewXPlus_ConMinusY_}{ADD41}
\DefMacro{ptn-pNewXPlus_ConMinusY_Sym}{ADD42}
\DefMacro{ptn-pNewDeMorganLawOrToAnd}{OR5}
\DefMacro{ptn-pNewDeMorganWithReassociationOrToAnd}{OR6}
\DefMacro{ptn-pNewDeMorganWithReassociationOrToAndSym}{OR7}
\DefMacro{ptn-pNew_AAndB_OrNot_AOrB_}{OR8}
\DefMacro{ptn-pNew_AAndB_OrNot_AOrB_Sym}{OR9}
\DefMacro{ptn-pNew_AAndB_OrNot_BOrA_}{OR10}
\DefMacro{ptn-pNew_AAndB_OrNot_BOrA_Sym}{OR11}
\DefMacro{ptn-pNew_AXorB_OrNot_AOrB_}{OR12}
\DefMacro{ptn-pNew_AXorB_OrNot_AOrB_Sym}{OR13}
\DefMacro{ptn-pNew_AXorB_OrNot_BOrA_}{OR14}
\DefMacro{ptn-pNew_AXorB_OrNot_BOrA_Sym}{OR15}
\DefMacro{ptn-pNew_AAndNotB_Or_NotAAndB_1}{OR16}
\DefMacro{ptn-pNew_AAndNotB_Or_NotAAndB_2}{OR17}
\DefMacro{ptn-pNew_AAndNotB_Or_NotAAndB_3}{OR18}
\DefMacro{ptn-pNew_AAndNotB_Or_NotAAndB_4}{OR19}
\DefMacro{ptn-pNew_Not_AOrB_AndC_Or_Not_AOrC_AndB_}{OR20}
\DefMacro{ptn-pNew_Not_AOrB_AndC_Or_Not_BOrC_AndA_}{OR21}
\DefMacro{ptn-pNew_Not_AOrB_AndC_OrNot_AOrC_}{OR22}
\DefMacro{ptn-pNew_Not_AOrB_AndC_OrNot_AOrC_Sym}{OR23}
\DefMacro{ptn-pNew_Not_AOrB_AndC_OrNot_BOrC_}{OR24}
\DefMacro{ptn-pNew_Not_AOrB_AndC_OrNot_BOrC_Sym}{OR25}
\DefMacro{ptn-pNew_Not_AOrB_AndC_OrNot_COr_AXorB__}{OR26}
\DefMacro{ptn-pNew_Not_AOrB_AndC_OrNot_COr_AXorB__Sym}{OR27}
\DefMacro{ptn-pNew__NotAAndB_AndC_OrNot__AOrB_OrC_}{OR28}
\DefMacro{ptn-pNew__NotAAndB_AndC_OrNot__AOrB_OrC_Sym}{OR29}
\DefMacro{ptn-pNew_NotAAndBAndC_Or_Not_AOrB_}{OR30}
\DefMacro{ptn-pNew_NotAAndBAndC_Or_Not_AOrB_Sym}{OR31}
\DefMacro{ptn-pNew_NotAAndBAndC_Or_Not_AOrC_}{OR32}
\DefMacro{ptn-pNew_NotAAndBAndC_Or_Not_AOrC_Sym}{OR33}
\DefMacro{ptn-pMul1}{MUL1}
\DefMacro{ptn-pMul2}{MUL2}
\DefMacro{ptn-pMul3}{MUL3}
\DefMacro{ptn-pMul4}{MUL4}
\DefMacro{ptn-pMul5}{MUL5}
\DefMacro{ptn-p_NegX_AndMinus1}{AND1}
\DefMacro{ptn-p_XPlusCon1_LShiftCon0}{LSHIFT1}
\DefMacro{ptn-p_XRShiftC0_LShiftC0}{LSHIFT2}
\DefMacro{ptn-p_XURShiftC0_LShiftC0}{LSHIFT3}
\DefMacro{ptn-p__XRShiftC0_AndY_LShiftC0}{LSHIFT4}
\DefMacro{ptn-p__XURShiftC0_AndY_LShiftC0}{LSHIFT5}
\DefMacro{ptn-p_XAndRightNBits_LShiftC0}{LSHIFT6}
\DefMacro{ptn-p_XAndC0_RShiftC1}{RSHIFT1}
\DefMacro{ptn-p_XURShiftA_URShiftB}{URSHIFT1}
\DefMacro{ptn-p__XLShiftZ_PlusY_URShiftZ}{URSHIFT2}
\DefMacro{ptn-p_XAndMask_URShiftZ}{URSHIFT3}
\DefMacro{ptn-p_XLShiftZ_URShiftZ}{URSHIFT4}
\DefMacro{ptn-p_XRShiftN_URShift31}{URSHIFT5}
\DefMacro{ptn-pXRotateLeftC}{ROTATELEFT1}
\DefMacro{ptn-pNewDeMorganLawAndToOr}{AND2}
\DefMacro{ptn-pNewDeMorganWithReassociationAndToOr}{AND3}
\DefMacro{ptn-pNewDeMorganWithReassociationAndToOrSym}{AND4}
\DefMacro{ptn-pNew_AOrB_And_NotAAndB_}{AND5}
\DefMacro{ptn-pNew_AOrB_And_NotAAndB_Sym}{AND6}
\DefMacro{ptn-pNew_AOrB_And_NotBAndA_}{AND7}
\DefMacro{ptn-pNew_AOrB_And_NotBAndA_Sym}{AND8}
\DefMacro{ptn-pNew_AOrNotB_And_NotAOrB_1}{AND9}
\DefMacro{ptn-pNew_AOrNotB_And_NotAOrB_2}{AND10}
\DefMacro{ptn-pNew_AOrNotB_And_NotAOrB_3}{AND11}
\DefMacro{ptn-pNew_AOrNotB_And_NotAOrB_4}{AND12}
\DefMacro{ptn-pNew_Not_AAndB_OrC_And_Not_AAndC_OrB_}{AND13}
\DefMacro{ptn-pNew_Not_AAndB_OrC_And_Not_BAndC_OrA_}{AND14}
\DefMacro{ptn-pNew_Not_AAndB_OrC_AndNot_AAndC_}{AND15}
\DefMacro{ptn-pNew_Not_AAndB_OrC_AndNot_AAndC_Sym}{AND16}
\DefMacro{ptn-pNew_Not_AAndB_OrC_AndNot_BAndC_}{AND17}
\DefMacro{ptn-pNew_Not_AAndB_OrC_AndNot_BAndC_Sym}{AND18}
\DefMacro{ptn-pNew__NotAOrB_OrC_AndNot__AAndB_AndC_}{AND19}
\DefMacro{ptn-pNew__NotAOrB_OrC_AndNot__AAndB_AndC_Sym}{AND20}
\DefMacro{ptn-pNew_NotAOrBOrC_And_Not_AAndB_}{AND21}
\DefMacro{ptn-pNew_NotAOrBOrC_And_Not_AAndB_Sym}{AND22}
\DefMacro{ptn-pNew_NotAOrBOrC_And_Not_AAndC}{AND23}
\DefMacro{ptn-pNew_NotAOrBOrC_And_Not_AAndCSym}{AND24}

\DefMacro{table-pattern-stats-summary-row0-num-patterns}{162}
\DefMacro{table-pattern-stats-summary-row0-num-OpenJDK}{68}
\DefMacro{table-pattern-stats-summary-row0-num-LLVM}{92}
\DefMacro{table-pattern-stats-summary-row0-num-OWN}{2}
\DefMacro{table-pattern-stats-summary-row0-num-prs}{7}

\DefMacro{table-pattern-stats-summary-float-env}{table}
\DefMacro{table-pattern-stats-summary-caption}{Summary of \ptns that we wrote in \Tool.\label{tab:pattern-stats-summary}\vspace{-10pt}}
\DefMacro{table-pattern-stats-summary-fontsize}{small}
\DefMacro{table-pattern-stats-summary-col-num-patterns}{\#\Ptns}
\DefMacro{table-pattern-stats-summary-col-num-OpenJDK}{\#\OpenJDK}
\DefMacro{table-pattern-stats-summary-col-num-LLVM}{\#\LLVM}
\DefMacro{table-pattern-stats-summary-col-num-OWN}{\#Original}
\DefMacro{table-pattern-stats-summary-col-num-prs}{\#PRs}

\DefMacro{table-pattern-stats-row0-name}{SUB1}
\DefMacro{table-pattern-stats-row0-origin}{OpenJDK}
\DefMacro{table-pattern-stats-row0-type}{SubINode}
\DefMacro{table-pattern-stats-row0-shadows}{}
\DefMacro{table-pattern-stats-row0-composites}{}
\DefMacro{table-pattern-stats-row0-pr}{}
\DefMacro{table-pattern-stats-row1-name}{SUB2}
\DefMacro{table-pattern-stats-row1-origin}{OpenJDK}
\DefMacro{table-pattern-stats-row1-type}{SubINode}
\DefMacro{table-pattern-stats-row1-shadows}{}
\DefMacro{table-pattern-stats-row1-composites}{}
\DefMacro{table-pattern-stats-row1-pr}{}
\DefMacro{table-pattern-stats-row2-name}{SUB3}
\DefMacro{table-pattern-stats-row2-origin}{OpenJDK}
\DefMacro{table-pattern-stats-row2-type}{SubINode}
\DefMacro{table-pattern-stats-row2-shadows}{pNewSubAddSub1574}
\DefMacro{table-pattern-stats-row2-composites}{}
\DefMacro{table-pattern-stats-row2-pr}{}
\DefMacro{table-pattern-stats-row3-name}{SUB4}
\DefMacro{table-pattern-stats-row3-origin}{OpenJDK}
\DefMacro{table-pattern-stats-row3-type}{SubINode}
\DefMacro{table-pattern-stats-row3-shadows}{}
\DefMacro{table-pattern-stats-row3-composites}{}
\DefMacro{table-pattern-stats-row3-pr}{}
\DefMacro{table-pattern-stats-row4-name}{SUB5}
\DefMacro{table-pattern-stats-row4-origin}{OpenJDK}
\DefMacro{table-pattern-stats-row4-type}{SubINode}
\DefMacro{table-pattern-stats-row4-shadows}{}
\DefMacro{table-pattern-stats-row4-composites}{}
\DefMacro{table-pattern-stats-row4-pr}{}
\DefMacro{table-pattern-stats-row5-name}{SUB6}
\DefMacro{table-pattern-stats-row5-origin}{OpenJDK}
\DefMacro{table-pattern-stats-row5-type}{SubINode}
\DefMacro{table-pattern-stats-row5-shadows}{}
\DefMacro{table-pattern-stats-row5-composites}{}
\DefMacro{table-pattern-stats-row5-pr}{}
\DefMacro{table-pattern-stats-row6-name}{SUB7}
\DefMacro{table-pattern-stats-row6-origin}{OpenJDK}
\DefMacro{table-pattern-stats-row6-type}{SubINode}
\DefMacro{table-pattern-stats-row6-shadows}{}
\DefMacro{table-pattern-stats-row6-composites}{}
\DefMacro{table-pattern-stats-row6-pr}{}
\DefMacro{table-pattern-stats-row7-name}{SUB8}
\DefMacro{table-pattern-stats-row7-origin}{OpenJDK}
\DefMacro{table-pattern-stats-row7-type}{SubINode}
\DefMacro{table-pattern-stats-row7-shadows}{}
\DefMacro{table-pattern-stats-row7-composites}{}
\DefMacro{table-pattern-stats-row7-pr}{}
\DefMacro{table-pattern-stats-row8-name}{SUB9}
\DefMacro{table-pattern-stats-row8-origin}{OpenJDK}
\DefMacro{table-pattern-stats-row8-type}{SubINode}
\DefMacro{table-pattern-stats-row8-shadows}{}
\DefMacro{table-pattern-stats-row8-composites}{}
\DefMacro{table-pattern-stats-row8-pr}{}
\DefMacro{table-pattern-stats-row9-name}{SUB10}
\DefMacro{table-pattern-stats-row9-origin}{OpenJDK}
\DefMacro{table-pattern-stats-row9-type}{SubINode}
\DefMacro{table-pattern-stats-row9-shadows}{}
\DefMacro{table-pattern-stats-row9-composites}{}
\DefMacro{table-pattern-stats-row9-pr}{}
\DefMacro{table-pattern-stats-row10-name}{SUB11}
\DefMacro{table-pattern-stats-row10-origin}{OpenJDK}
\DefMacro{table-pattern-stats-row10-type}{SubINode}
\DefMacro{table-pattern-stats-row10-shadows}{}
\DefMacro{table-pattern-stats-row10-composites}{}
\DefMacro{table-pattern-stats-row10-pr}{}
\DefMacro{table-pattern-stats-row11-name}{SUB12}
\DefMacro{table-pattern-stats-row11-origin}{OpenJDK}
\DefMacro{table-pattern-stats-row11-type}{SubINode}
\DefMacro{table-pattern-stats-row11-shadows}{}
\DefMacro{table-pattern-stats-row11-composites}{}
\DefMacro{table-pattern-stats-row11-pr}{}
\DefMacro{table-pattern-stats-row12-name}{SUB13}
\DefMacro{table-pattern-stats-row12-origin}{OpenJDK}
\DefMacro{table-pattern-stats-row12-type}{SubINode}
\DefMacro{table-pattern-stats-row12-shadows}{}
\DefMacro{table-pattern-stats-row12-composites}{}
\DefMacro{table-pattern-stats-row12-pr}{}
\DefMacro{table-pattern-stats-row13-name}{SUB14}
\DefMacro{table-pattern-stats-row13-origin}{OpenJDK}
\DefMacro{table-pattern-stats-row13-type}{SubINode}
\DefMacro{table-pattern-stats-row13-shadows}{}
\DefMacro{table-pattern-stats-row13-composites}{}
\DefMacro{table-pattern-stats-row13-pr}{}
\DefMacro{table-pattern-stats-row14-name}{SUB15}
\DefMacro{table-pattern-stats-row14-origin}{OpenJDK}
\DefMacro{table-pattern-stats-row14-type}{SubINode}
\DefMacro{table-pattern-stats-row14-shadows}{}
\DefMacro{table-pattern-stats-row14-composites}{}
\DefMacro{table-pattern-stats-row14-pr}{}
\DefMacro{table-pattern-stats-row15-name}{SUB16}
\DefMacro{table-pattern-stats-row15-origin}{OpenJDK}
\DefMacro{table-pattern-stats-row15-type}{SubINode}
\DefMacro{table-pattern-stats-row15-shadows}{}
\DefMacro{table-pattern-stats-row15-composites}{}
\DefMacro{table-pattern-stats-row15-pr}{}
\DefMacro{table-pattern-stats-row16-name}{SUB17}
\DefMacro{table-pattern-stats-row16-origin}{OpenJDK}
\DefMacro{table-pattern-stats-row16-type}{SubINode}
\DefMacro{table-pattern-stats-row16-shadows}{}
\DefMacro{table-pattern-stats-row16-composites}{}
\DefMacro{table-pattern-stats-row16-pr}{}
\DefMacro{table-pattern-stats-row17-name}{SUB18}
\DefMacro{table-pattern-stats-row17-origin}{OpenJDK}
\DefMacro{table-pattern-stats-row17-type}{SubINode}
\DefMacro{table-pattern-stats-row17-shadows}{}
\DefMacro{table-pattern-stats-row17-composites}{}
\DefMacro{table-pattern-stats-row17-pr}{}
\DefMacro{table-pattern-stats-row18-name}{SUB19}
\DefMacro{table-pattern-stats-row18-origin}{LLVM}
\DefMacro{table-pattern-stats-row18-type}{SubINode}
\DefMacro{table-pattern-stats-row18-shadows}{}
\DefMacro{table-pattern-stats-row18-composites}{}
\DefMacro{table-pattern-stats-row18-pr}{}
\DefMacro{table-pattern-stats-row19-name}{SUB20}
\DefMacro{table-pattern-stats-row19-origin}{LLVM}
\DefMacro{table-pattern-stats-row19-type}{SubINode}
\DefMacro{table-pattern-stats-row19-shadows}{}
\DefMacro{table-pattern-stats-row19-composites}{}
\DefMacro{table-pattern-stats-row19-pr}{}
\DefMacro{table-pattern-stats-row20-name}{SUB21}
\DefMacro{table-pattern-stats-row20-origin}{LLVM}
\DefMacro{table-pattern-stats-row20-type}{SubINode}
\DefMacro{table-pattern-stats-row20-shadows}{}
\DefMacro{table-pattern-stats-row20-composites}{}
\DefMacro{table-pattern-stats-row20-pr}{7376}
\DefMacro{table-pattern-stats-row21-name}{SUB22}
\DefMacro{table-pattern-stats-row21-origin}{LLVM}
\DefMacro{table-pattern-stats-row21-type}{SubINode}
\DefMacro{table-pattern-stats-row21-shadows}{}
\DefMacro{table-pattern-stats-row21-composites}{}
\DefMacro{table-pattern-stats-row21-pr}{7376}
\DefMacro{table-pattern-stats-row22-name}{SUB23}
\DefMacro{table-pattern-stats-row22-origin}{LLVM}
\DefMacro{table-pattern-stats-row22-type}{SubINode}
\DefMacro{table-pattern-stats-row22-shadows}{}
\DefMacro{table-pattern-stats-row22-composites}{}
\DefMacro{table-pattern-stats-row22-pr}{7376}
\DefMacro{table-pattern-stats-row23-name}{SUB24}
\DefMacro{table-pattern-stats-row23-origin}{LLVM}
\DefMacro{table-pattern-stats-row23-type}{SubINode}
\DefMacro{table-pattern-stats-row23-shadows}{}
\DefMacro{table-pattern-stats-row23-composites}{}
\DefMacro{table-pattern-stats-row23-pr}{7395}
\DefMacro{table-pattern-stats-row24-name}{SUB25}
\DefMacro{table-pattern-stats-row24-origin}{LLVM}
\DefMacro{table-pattern-stats-row24-type}{SubINode}
\DefMacro{table-pattern-stats-row24-shadows}{}
\DefMacro{table-pattern-stats-row24-composites}{}
\DefMacro{table-pattern-stats-row24-pr}{7395}
\DefMacro{table-pattern-stats-row25-name}{SUB26}
\DefMacro{table-pattern-stats-row25-origin}{LLVM}
\DefMacro{table-pattern-stats-row25-type}{SubINode}
\DefMacro{table-pattern-stats-row25-shadows}{}
\DefMacro{table-pattern-stats-row25-composites}{}
\DefMacro{table-pattern-stats-row25-pr}{7395}
\DefMacro{table-pattern-stats-row26-name}{SUB27}
\DefMacro{table-pattern-stats-row26-origin}{LLVM}
\DefMacro{table-pattern-stats-row26-type}{SubINode}
\DefMacro{table-pattern-stats-row26-shadows}{}
\DefMacro{table-pattern-stats-row26-composites}{}
\DefMacro{table-pattern-stats-row26-pr}{7395}
\DefMacro{table-pattern-stats-row27-name}{SUB28}
\DefMacro{table-pattern-stats-row27-origin}{LLVM}
\DefMacro{table-pattern-stats-row27-type}{SubINode}
\DefMacro{table-pattern-stats-row27-shadows}{}
\DefMacro{table-pattern-stats-row27-composites}{}
\DefMacro{table-pattern-stats-row27-pr}{7395}
\DefMacro{table-pattern-stats-row28-name}{SUB29}
\DefMacro{table-pattern-stats-row28-origin}{LLVM}
\DefMacro{table-pattern-stats-row28-type}{SubINode}
\DefMacro{table-pattern-stats-row28-shadows}{}
\DefMacro{table-pattern-stats-row28-composites}{}
\DefMacro{table-pattern-stats-row28-pr}{7395}
\DefMacro{table-pattern-stats-row29-name}{SUB30}
\DefMacro{table-pattern-stats-row29-origin}{LLVM}
\DefMacro{table-pattern-stats-row29-type}{SubINode}
\DefMacro{table-pattern-stats-row29-shadows}{}
\DefMacro{table-pattern-stats-row29-composites}{}
\DefMacro{table-pattern-stats-row29-pr}{6441}
\DefMacro{table-pattern-stats-row30-name}{SUB31}
\DefMacro{table-pattern-stats-row30-origin}{LLVM}
\DefMacro{table-pattern-stats-row30-type}{SubINode}
\DefMacro{table-pattern-stats-row30-shadows}{}
\DefMacro{table-pattern-stats-row30-composites}{}
\DefMacro{table-pattern-stats-row30-pr}{}
\DefMacro{table-pattern-stats-row31-name}{SUB32}
\DefMacro{table-pattern-stats-row31-origin}{LLVM}
\DefMacro{table-pattern-stats-row31-type}{SubINode}
\DefMacro{table-pattern-stats-row31-shadows}{}
\DefMacro{table-pattern-stats-row31-composites}{}
\DefMacro{table-pattern-stats-row31-pr}{}
\DefMacro{table-pattern-stats-row32-name}{SUB33}
\DefMacro{table-pattern-stats-row32-origin}{LLVM}
\DefMacro{table-pattern-stats-row32-type}{SubINode}
\DefMacro{table-pattern-stats-row32-shadows}{}
\DefMacro{table-pattern-stats-row32-composites}{}
\DefMacro{table-pattern-stats-row32-pr}{}
\DefMacro{table-pattern-stats-row33-name}{SUB34}
\DefMacro{table-pattern-stats-row33-origin}{LLVM}
\DefMacro{table-pattern-stats-row33-type}{SubINode}
\DefMacro{table-pattern-stats-row33-shadows}{}
\DefMacro{table-pattern-stats-row33-composites}{}
\DefMacro{table-pattern-stats-row33-pr}{}
\DefMacro{table-pattern-stats-row34-name}{SUB35}
\DefMacro{table-pattern-stats-row34-origin}{LLVM}
\DefMacro{table-pattern-stats-row34-type}{SubINode}
\DefMacro{table-pattern-stats-row34-shadows}{}
\DefMacro{table-pattern-stats-row34-composites}{}
\DefMacro{table-pattern-stats-row34-pr}{}
\DefMacro{table-pattern-stats-row35-name}{SUB36}
\DefMacro{table-pattern-stats-row35-origin}{LLVM}
\DefMacro{table-pattern-stats-row35-type}{SubINode}
\DefMacro{table-pattern-stats-row35-shadows}{}
\DefMacro{table-pattern-stats-row35-composites}{}
\DefMacro{table-pattern-stats-row35-pr}{}
\DefMacro{table-pattern-stats-row36-name}{SUB37}
\DefMacro{table-pattern-stats-row36-origin}{LLVM}
\DefMacro{table-pattern-stats-row36-type}{SubINode}
\DefMacro{table-pattern-stats-row36-shadows}{}
\DefMacro{table-pattern-stats-row36-composites}{}
\DefMacro{table-pattern-stats-row36-pr}{}
\DefMacro{table-pattern-stats-row37-name}{SUB38}
\DefMacro{table-pattern-stats-row37-origin}{LLVM}
\DefMacro{table-pattern-stats-row37-type}{SubINode}
\DefMacro{table-pattern-stats-row37-shadows}{}
\DefMacro{table-pattern-stats-row37-composites}{}
\DefMacro{table-pattern-stats-row37-pr}{}
\DefMacro{table-pattern-stats-row38-name}{SUB39}
\DefMacro{table-pattern-stats-row38-origin}{LLVM}
\DefMacro{table-pattern-stats-row38-type}{SubINode}
\DefMacro{table-pattern-stats-row38-shadows}{}
\DefMacro{table-pattern-stats-row38-composites}{}
\DefMacro{table-pattern-stats-row38-pr}{}
\DefMacro{table-pattern-stats-row39-name}{SUB40}
\DefMacro{table-pattern-stats-row39-origin}{LLVM}
\DefMacro{table-pattern-stats-row39-type}{SubINode}
\DefMacro{table-pattern-stats-row39-shadows}{}
\DefMacro{table-pattern-stats-row39-composites}{}
\DefMacro{table-pattern-stats-row39-pr}{}
\DefMacro{table-pattern-stats-row40-name}{SUB41}
\DefMacro{table-pattern-stats-row40-origin}{LLVM}
\DefMacro{table-pattern-stats-row40-type}{SubINode}
\DefMacro{table-pattern-stats-row40-shadows}{}
\DefMacro{table-pattern-stats-row40-composites}{}
\DefMacro{table-pattern-stats-row40-pr}{}
\DefMacro{table-pattern-stats-row41-name}{SUB42}
\DefMacro{table-pattern-stats-row41-origin}{LLVM}
\DefMacro{table-pattern-stats-row41-type}{SubINode}
\DefMacro{table-pattern-stats-row41-shadows}{}
\DefMacro{table-pattern-stats-row41-composites}{}
\DefMacro{table-pattern-stats-row41-pr}{}
\DefMacro{table-pattern-stats-row42-name}{ADD1}
\DefMacro{table-pattern-stats-row42-origin}{OpenJDK}
\DefMacro{table-pattern-stats-row42-type}{AddINode}
\DefMacro{table-pattern-stats-row42-shadows}{}
\DefMacro{table-pattern-stats-row42-composites}{}
\DefMacro{table-pattern-stats-row42-pr}{}
\DefMacro{table-pattern-stats-row43-name}{ADD2}
\DefMacro{table-pattern-stats-row43-origin}{OpenJDK}
\DefMacro{table-pattern-stats-row43-type}{AddINode}
\DefMacro{table-pattern-stats-row43-shadows}{pAdd5, pAdd6, pNewAddAddSub1165}
\DefMacro{table-pattern-stats-row43-composites}{}
\DefMacro{table-pattern-stats-row43-pr}{}
\DefMacro{table-pattern-stats-row44-name}{ADD3}
\DefMacro{table-pattern-stats-row44-origin}{OpenJDK}
\DefMacro{table-pattern-stats-row44-type}{AddINode}
\DefMacro{table-pattern-stats-row44-shadows}{}
\DefMacro{table-pattern-stats-row44-composites}{}
\DefMacro{table-pattern-stats-row44-pr}{}
\DefMacro{table-pattern-stats-row45-name}{ADD4}
\DefMacro{table-pattern-stats-row45-origin}{OpenJDK}
\DefMacro{table-pattern-stats-row45-type}{AddINode}
\DefMacro{table-pattern-stats-row45-shadows}{}
\DefMacro{table-pattern-stats-row45-composites}{}
\DefMacro{table-pattern-stats-row45-pr}{}
\DefMacro{table-pattern-stats-row46-name}{ADD5}
\DefMacro{table-pattern-stats-row46-origin}{OpenJDK}
\DefMacro{table-pattern-stats-row46-type}{AddINode}
\DefMacro{table-pattern-stats-row46-shadows}{}
\DefMacro{table-pattern-stats-row46-composites}{}
\DefMacro{table-pattern-stats-row46-pr}{}
\DefMacro{table-pattern-stats-row47-name}{ADD6}
\DefMacro{table-pattern-stats-row47-origin}{OpenJDK}
\DefMacro{table-pattern-stats-row47-type}{AddINode}
\DefMacro{table-pattern-stats-row47-shadows}{}
\DefMacro{table-pattern-stats-row47-composites}{}
\DefMacro{table-pattern-stats-row47-pr}{}
\DefMacro{table-pattern-stats-row48-name}{ADD7}
\DefMacro{table-pattern-stats-row48-origin}{OpenJDK}
\DefMacro{table-pattern-stats-row48-type}{AddINode}
\DefMacro{table-pattern-stats-row48-shadows}{}
\DefMacro{table-pattern-stats-row48-composites}{}
\DefMacro{table-pattern-stats-row48-pr}{6752}
\DefMacro{table-pattern-stats-row49-name}{ADD8}
\DefMacro{table-pattern-stats-row49-origin}{OpenJDK}
\DefMacro{table-pattern-stats-row49-type}{AddINode}
\DefMacro{table-pattern-stats-row49-shadows}{}
\DefMacro{table-pattern-stats-row49-composites}{}
\DefMacro{table-pattern-stats-row49-pr}{6752}
\DefMacro{table-pattern-stats-row50-name}{ADD9}
\DefMacro{table-pattern-stats-row50-origin}{OpenJDK}
\DefMacro{table-pattern-stats-row50-type}{AddINode}
\DefMacro{table-pattern-stats-row50-shadows}{pNewAddAddSub1165}
\DefMacro{table-pattern-stats-row50-composites}{}
\DefMacro{table-pattern-stats-row50-pr}{}
\DefMacro{table-pattern-stats-row51-name}{ADD10}
\DefMacro{table-pattern-stats-row51-origin}{OpenJDK}
\DefMacro{table-pattern-stats-row51-type}{AddINode}
\DefMacro{table-pattern-stats-row51-shadows}{pNewAddAddSub1165}
\DefMacro{table-pattern-stats-row51-composites}{}
\DefMacro{table-pattern-stats-row51-pr}{}
\DefMacro{table-pattern-stats-row52-name}{ADD11}
\DefMacro{table-pattern-stats-row52-origin}{OpenJDK}
\DefMacro{table-pattern-stats-row52-type}{AddINode}
\DefMacro{table-pattern-stats-row52-shadows}{}
\DefMacro{table-pattern-stats-row52-composites}{}
\DefMacro{table-pattern-stats-row52-pr}{}
\DefMacro{table-pattern-stats-row53-name}{ADD12}
\DefMacro{table-pattern-stats-row53-origin}{OpenJDK}
\DefMacro{table-pattern-stats-row53-type}{AddINode}
\DefMacro{table-pattern-stats-row53-shadows}{}
\DefMacro{table-pattern-stats-row53-composites}{}
\DefMacro{table-pattern-stats-row53-pr}{}
\DefMacro{table-pattern-stats-row54-name}{ADD13}
\DefMacro{table-pattern-stats-row54-origin}{OpenJDK}
\DefMacro{table-pattern-stats-row54-type}{AddINode}
\DefMacro{table-pattern-stats-row54-shadows}{}
\DefMacro{table-pattern-stats-row54-composites}{}
\DefMacro{table-pattern-stats-row54-pr}{}
\DefMacro{table-pattern-stats-row55-name}{ADD14}
\DefMacro{table-pattern-stats-row55-origin}{OpenJDK}
\DefMacro{table-pattern-stats-row55-type}{AddINode}
\DefMacro{table-pattern-stats-row55-shadows}{}
\DefMacro{table-pattern-stats-row55-composites}{}
\DefMacro{table-pattern-stats-row55-pr}{}
\DefMacro{table-pattern-stats-row56-name}{ADD15}
\DefMacro{table-pattern-stats-row56-origin}{OpenJDK}
\DefMacro{table-pattern-stats-row56-type}{AddINode}
\DefMacro{table-pattern-stats-row56-shadows}{}
\DefMacro{table-pattern-stats-row56-composites}{}
\DefMacro{table-pattern-stats-row56-pr}{}
\DefMacro{table-pattern-stats-row57-name}{ADD16}
\DefMacro{table-pattern-stats-row57-origin}{OpenJDK}
\DefMacro{table-pattern-stats-row57-type}{AddINode}
\DefMacro{table-pattern-stats-row57-shadows}{}
\DefMacro{table-pattern-stats-row57-composites}{}
\DefMacro{table-pattern-stats-row57-pr}{}
\DefMacro{table-pattern-stats-row58-name}{ADD17}
\DefMacro{table-pattern-stats-row58-origin}{OpenJDK}
\DefMacro{table-pattern-stats-row58-type}{AddINode}
\DefMacro{table-pattern-stats-row58-shadows}{}
\DefMacro{table-pattern-stats-row58-composites}{}
\DefMacro{table-pattern-stats-row58-pr}{}
\DefMacro{table-pattern-stats-row59-name}{ADD18}
\DefMacro{table-pattern-stats-row59-origin}{OpenJDK}
\DefMacro{table-pattern-stats-row59-type}{AddINode}
\DefMacro{table-pattern-stats-row59-shadows}{}
\DefMacro{table-pattern-stats-row59-composites}{}
\DefMacro{table-pattern-stats-row59-pr}{}
\DefMacro{table-pattern-stats-row60-name}{ADD19}
\DefMacro{table-pattern-stats-row60-origin}{OpenJDK}
\DefMacro{table-pattern-stats-row60-type}{AddINode}
\DefMacro{table-pattern-stats-row60-shadows}{}
\DefMacro{table-pattern-stats-row60-composites}{}
\DefMacro{table-pattern-stats-row60-pr}{}
\DefMacro{table-pattern-stats-row61-name}{ADD20}
\DefMacro{table-pattern-stats-row61-origin}{OpenJDK}
\DefMacro{table-pattern-stats-row61-type}{AddINode}
\DefMacro{table-pattern-stats-row61-shadows}{}
\DefMacro{table-pattern-stats-row61-composites}{}
\DefMacro{table-pattern-stats-row61-pr}{}
\DefMacro{table-pattern-stats-row62-name}{ADD21}
\DefMacro{table-pattern-stats-row62-origin}{OpenJDK}
\DefMacro{table-pattern-stats-row62-type}{AddINode}
\DefMacro{table-pattern-stats-row62-shadows}{pAddConstantAssociative, pNewAddAddSub1040, pNewAddAddSub1043, pNewAddNotXPlusOneToNegX1, pNewAdd\_APlusC1\_Plus\_C2MinusB\_}
\DefMacro{table-pattern-stats-row62-composites}{}
\DefMacro{table-pattern-stats-row62-pr}{}
\DefMacro{table-pattern-stats-row63-name}{ADD22}
\DefMacro{table-pattern-stats-row63-origin}{OpenJDK}
\DefMacro{table-pattern-stats-row63-type}{AddINode}
\DefMacro{table-pattern-stats-row63-shadows}{pNewAddNotXPlusOneToNegX2}
\DefMacro{table-pattern-stats-row63-composites}{}
\DefMacro{table-pattern-stats-row63-pr}{}
\DefMacro{table-pattern-stats-row64-name}{ADD23}
\DefMacro{table-pattern-stats-row64-origin}{OpenJDK}
\DefMacro{table-pattern-stats-row64-type}{AddINode}
\DefMacro{table-pattern-stats-row64-shadows}{}
\DefMacro{table-pattern-stats-row64-composites}{}
\DefMacro{table-pattern-stats-row64-pr}{}
\DefMacro{table-pattern-stats-row65-name}{ADD24}
\DefMacro{table-pattern-stats-row65-origin}{OpenJDK}
\DefMacro{table-pattern-stats-row65-type}{AddINode}
\DefMacro{table-pattern-stats-row65-shadows}{}
\DefMacro{table-pattern-stats-row65-composites}{}
\DefMacro{table-pattern-stats-row65-pr}{}
\DefMacro{table-pattern-stats-row66-name}{OR1}
\DefMacro{table-pattern-stats-row66-origin}{OpenJDK}
\DefMacro{table-pattern-stats-row66-type}{OrINode}
\DefMacro{table-pattern-stats-row66-shadows}{}
\DefMacro{table-pattern-stats-row66-composites}{}
\DefMacro{table-pattern-stats-row66-pr}{}
\DefMacro{table-pattern-stats-row67-name}{OR2}
\DefMacro{table-pattern-stats-row67-origin}{OpenJDK}
\DefMacro{table-pattern-stats-row67-type}{OrINode}
\DefMacro{table-pattern-stats-row67-shadows}{}
\DefMacro{table-pattern-stats-row67-composites}{}
\DefMacro{table-pattern-stats-row67-pr}{}
\DefMacro{table-pattern-stats-row68-name}{OR3}
\DefMacro{table-pattern-stats-row68-origin}{OpenJDK}
\DefMacro{table-pattern-stats-row68-type}{OrINode}
\DefMacro{table-pattern-stats-row68-shadows}{}
\DefMacro{table-pattern-stats-row68-composites}{}
\DefMacro{table-pattern-stats-row68-pr}{}
\DefMacro{table-pattern-stats-row69-name}{OR4}
\DefMacro{table-pattern-stats-row69-origin}{OpenJDK}
\DefMacro{table-pattern-stats-row69-type}{OrINode}
\DefMacro{table-pattern-stats-row69-shadows}{}
\DefMacro{table-pattern-stats-row69-composites}{}
\DefMacro{table-pattern-stats-row69-pr}{}
\DefMacro{table-pattern-stats-row70-name}{MIN1}
\DefMacro{table-pattern-stats-row70-origin}{OpenJDK}
\DefMacro{table-pattern-stats-row70-type}{MinINode}
\DefMacro{table-pattern-stats-row70-shadows}{}
\DefMacro{table-pattern-stats-row70-composites}{}
\DefMacro{table-pattern-stats-row70-pr}{}
\DefMacro{table-pattern-stats-row71-name}{MIN2}
\DefMacro{table-pattern-stats-row71-origin}{OpenJDK}
\DefMacro{table-pattern-stats-row71-type}{MinINode}
\DefMacro{table-pattern-stats-row71-shadows}{}
\DefMacro{table-pattern-stats-row71-composites}{}
\DefMacro{table-pattern-stats-row71-pr}{}
\DefMacro{table-pattern-stats-row72-name}{MIN3}
\DefMacro{table-pattern-stats-row72-origin}{OpenJDK}
\DefMacro{table-pattern-stats-row72-type}{MinINode}
\DefMacro{table-pattern-stats-row72-shadows}{}
\DefMacro{table-pattern-stats-row72-composites}{}
\DefMacro{table-pattern-stats-row72-pr}{}
\DefMacro{table-pattern-stats-row73-name}{ADD25}
\DefMacro{table-pattern-stats-row73-origin}{LLVM}
\DefMacro{table-pattern-stats-row73-type}{AddINode}
\DefMacro{table-pattern-stats-row73-shadows}{}
\DefMacro{table-pattern-stats-row73-composites}{}
\DefMacro{table-pattern-stats-row73-pr}{7376}
\DefMacro{table-pattern-stats-row74-name}{ADD26}
\DefMacro{table-pattern-stats-row74-origin}{LLVM}
\DefMacro{table-pattern-stats-row74-type}{AddINode}
\DefMacro{table-pattern-stats-row74-shadows}{}
\DefMacro{table-pattern-stats-row74-composites}{}
\DefMacro{table-pattern-stats-row74-pr}{7376}
\DefMacro{table-pattern-stats-row75-name}{ADD27}
\DefMacro{table-pattern-stats-row75-origin}{LLVM}
\DefMacro{table-pattern-stats-row75-type}{AddINode}
\DefMacro{table-pattern-stats-row75-shadows}{}
\DefMacro{table-pattern-stats-row75-composites}{}
\DefMacro{table-pattern-stats-row75-pr}{7376}
\DefMacro{table-pattern-stats-row76-name}{ADD28}
\DefMacro{table-pattern-stats-row76-origin}{LLVM}
\DefMacro{table-pattern-stats-row76-type}{AddINode}
\DefMacro{table-pattern-stats-row76-shadows}{}
\DefMacro{table-pattern-stats-row76-composites}{}
\DefMacro{table-pattern-stats-row76-pr}{7376}
\DefMacro{table-pattern-stats-row77-name}{ADD29}
\DefMacro{table-pattern-stats-row77-origin}{LLVM}
\DefMacro{table-pattern-stats-row77-type}{AddINode}
\DefMacro{table-pattern-stats-row77-shadows}{}
\DefMacro{table-pattern-stats-row77-composites}{}
\DefMacro{table-pattern-stats-row77-pr}{}
\DefMacro{table-pattern-stats-row78-name}{ADD30}
\DefMacro{table-pattern-stats-row78-origin}{LLVM}
\DefMacro{table-pattern-stats-row78-type}{AddINode}
\DefMacro{table-pattern-stats-row78-shadows}{}
\DefMacro{table-pattern-stats-row78-composites}{}
\DefMacro{table-pattern-stats-row78-pr}{6675}
\DefMacro{table-pattern-stats-row79-name}{ADD31}
\DefMacro{table-pattern-stats-row79-origin}{LLVM}
\DefMacro{table-pattern-stats-row79-type}{AddINode}
\DefMacro{table-pattern-stats-row79-shadows}{pAddNotXPlusOne}
\DefMacro{table-pattern-stats-row79-composites}{}
\DefMacro{table-pattern-stats-row79-pr}{6858}
\DefMacro{table-pattern-stats-row80-name}{ADD32}
\DefMacro{table-pattern-stats-row80-origin}{LLVM}
\DefMacro{table-pattern-stats-row80-type}{AddINode}
\DefMacro{table-pattern-stats-row80-shadows}{}
\DefMacro{table-pattern-stats-row80-composites}{}
\DefMacro{table-pattern-stats-row80-pr}{7395}
\DefMacro{table-pattern-stats-row81-name}{ADD33}
\DefMacro{table-pattern-stats-row81-origin}{LLVM}
\DefMacro{table-pattern-stats-row81-type}{AddINode}
\DefMacro{table-pattern-stats-row81-shadows}{}
\DefMacro{table-pattern-stats-row81-composites}{}
\DefMacro{table-pattern-stats-row81-pr}{7395}
\DefMacro{table-pattern-stats-row82-name}{ADD34}
\DefMacro{table-pattern-stats-row82-origin}{LLVM}
\DefMacro{table-pattern-stats-row82-type}{AddINode}
\DefMacro{table-pattern-stats-row82-shadows}{}
\DefMacro{table-pattern-stats-row82-composites}{}
\DefMacro{table-pattern-stats-row82-pr}{7395}
\DefMacro{table-pattern-stats-row83-name}{ADD35}
\DefMacro{table-pattern-stats-row83-origin}{LLVM}
\DefMacro{table-pattern-stats-row83-type}{AddINode}
\DefMacro{table-pattern-stats-row83-shadows}{}
\DefMacro{table-pattern-stats-row83-composites}{}
\DefMacro{table-pattern-stats-row83-pr}{7395}
\DefMacro{table-pattern-stats-row84-name}{ADD36}
\DefMacro{table-pattern-stats-row84-origin}{LLVM}
\DefMacro{table-pattern-stats-row84-type}{AddINode}
\DefMacro{table-pattern-stats-row84-shadows}{}
\DefMacro{table-pattern-stats-row84-composites}{}
\DefMacro{table-pattern-stats-row84-pr}{}
\DefMacro{table-pattern-stats-row85-name}{ADD37}
\DefMacro{table-pattern-stats-row85-origin}{LLVM}
\DefMacro{table-pattern-stats-row85-type}{AddINode}
\DefMacro{table-pattern-stats-row85-shadows}{}
\DefMacro{table-pattern-stats-row85-composites}{}
\DefMacro{table-pattern-stats-row85-pr}{}
\DefMacro{table-pattern-stats-row86-name}{ADD38}
\DefMacro{table-pattern-stats-row86-origin}{LLVM}
\DefMacro{table-pattern-stats-row86-type}{AddINode}
\DefMacro{table-pattern-stats-row86-shadows}{}
\DefMacro{table-pattern-stats-row86-composites}{}
\DefMacro{table-pattern-stats-row86-pr}{}
\DefMacro{table-pattern-stats-row87-name}{ADD39}
\DefMacro{table-pattern-stats-row87-origin}{LLVM}
\DefMacro{table-pattern-stats-row87-type}{AddINode}
\DefMacro{table-pattern-stats-row87-shadows}{}
\DefMacro{table-pattern-stats-row87-composites}{}
\DefMacro{table-pattern-stats-row87-pr}{}
\DefMacro{table-pattern-stats-row88-name}{ADD40}
\DefMacro{table-pattern-stats-row88-origin}{LLVM}
\DefMacro{table-pattern-stats-row88-type}{AddINode}
\DefMacro{table-pattern-stats-row88-shadows}{}
\DefMacro{table-pattern-stats-row88-composites}{}
\DefMacro{table-pattern-stats-row88-pr}{}
\DefMacro{table-pattern-stats-row89-name}{ADD41}
\DefMacro{table-pattern-stats-row89-origin}{OWN}
\DefMacro{table-pattern-stats-row89-type}{AddINode}
\DefMacro{table-pattern-stats-row89-shadows}{pAdd7, pNewAddAddSub1165, pNewAdd\_APlusC1\_Plus\_C2MinusB\_}
\DefMacro{table-pattern-stats-row89-composites}{}
\DefMacro{table-pattern-stats-row89-pr}{7795}
\DefMacro{table-pattern-stats-row90-name}{ADD42}
\DefMacro{table-pattern-stats-row90-origin}{OWN}
\DefMacro{table-pattern-stats-row90-type}{AddINode}
\DefMacro{table-pattern-stats-row90-shadows}{pAdd1, pAdd8, pNewAddAddSub1165}
\DefMacro{table-pattern-stats-row90-composites}{}
\DefMacro{table-pattern-stats-row90-pr}{7795}
\DefMacro{table-pattern-stats-row91-name}{OR5}
\DefMacro{table-pattern-stats-row91-origin}{LLVM}
\DefMacro{table-pattern-stats-row91-type}{OrINode}
\DefMacro{table-pattern-stats-row91-shadows}{}
\DefMacro{table-pattern-stats-row91-composites}{}
\DefMacro{table-pattern-stats-row91-pr}{}
\DefMacro{table-pattern-stats-row92-name}{OR6}
\DefMacro{table-pattern-stats-row92-origin}{LLVM}
\DefMacro{table-pattern-stats-row92-type}{OrINode}
\DefMacro{table-pattern-stats-row92-shadows}{}
\DefMacro{table-pattern-stats-row92-composites}{}
\DefMacro{table-pattern-stats-row92-pr}{}
\DefMacro{table-pattern-stats-row93-name}{OR7}
\DefMacro{table-pattern-stats-row93-origin}{LLVM}
\DefMacro{table-pattern-stats-row93-type}{OrINode}
\DefMacro{table-pattern-stats-row93-shadows}{}
\DefMacro{table-pattern-stats-row93-composites}{}
\DefMacro{table-pattern-stats-row93-pr}{}
\DefMacro{table-pattern-stats-row94-name}{OR8}
\DefMacro{table-pattern-stats-row94-origin}{LLVM}
\DefMacro{table-pattern-stats-row94-type}{OrINode}
\DefMacro{table-pattern-stats-row94-shadows}{}
\DefMacro{table-pattern-stats-row94-composites}{}
\DefMacro{table-pattern-stats-row94-pr}{}
\DefMacro{table-pattern-stats-row95-name}{OR9}
\DefMacro{table-pattern-stats-row95-origin}{LLVM}
\DefMacro{table-pattern-stats-row95-type}{OrINode}
\DefMacro{table-pattern-stats-row95-shadows}{}
\DefMacro{table-pattern-stats-row95-composites}{}
\DefMacro{table-pattern-stats-row95-pr}{}
\DefMacro{table-pattern-stats-row96-name}{OR10}
\DefMacro{table-pattern-stats-row96-origin}{LLVM}
\DefMacro{table-pattern-stats-row96-type}{OrINode}
\DefMacro{table-pattern-stats-row96-shadows}{}
\DefMacro{table-pattern-stats-row96-composites}{}
\DefMacro{table-pattern-stats-row96-pr}{}
\DefMacro{table-pattern-stats-row97-name}{OR11}
\DefMacro{table-pattern-stats-row97-origin}{LLVM}
\DefMacro{table-pattern-stats-row97-type}{OrINode}
\DefMacro{table-pattern-stats-row97-shadows}{}
\DefMacro{table-pattern-stats-row97-composites}{}
\DefMacro{table-pattern-stats-row97-pr}{}
\DefMacro{table-pattern-stats-row98-name}{OR12}
\DefMacro{table-pattern-stats-row98-origin}{LLVM}
\DefMacro{table-pattern-stats-row98-type}{OrINode}
\DefMacro{table-pattern-stats-row98-shadows}{}
\DefMacro{table-pattern-stats-row98-composites}{}
\DefMacro{table-pattern-stats-row98-pr}{}
\DefMacro{table-pattern-stats-row99-name}{OR13}
\DefMacro{table-pattern-stats-row99-origin}{LLVM}
\DefMacro{table-pattern-stats-row99-type}{OrINode}
\DefMacro{table-pattern-stats-row99-shadows}{}
\DefMacro{table-pattern-stats-row99-composites}{}
\DefMacro{table-pattern-stats-row99-pr}{}
\DefMacro{table-pattern-stats-row100-name}{OR14}
\DefMacro{table-pattern-stats-row100-origin}{LLVM}
\DefMacro{table-pattern-stats-row100-type}{OrINode}
\DefMacro{table-pattern-stats-row100-shadows}{}
\DefMacro{table-pattern-stats-row100-composites}{}
\DefMacro{table-pattern-stats-row100-pr}{}
\DefMacro{table-pattern-stats-row101-name}{OR15}
\DefMacro{table-pattern-stats-row101-origin}{LLVM}
\DefMacro{table-pattern-stats-row101-type}{OrINode}
\DefMacro{table-pattern-stats-row101-shadows}{}
\DefMacro{table-pattern-stats-row101-composites}{}
\DefMacro{table-pattern-stats-row101-pr}{}
\DefMacro{table-pattern-stats-row102-name}{OR16}
\DefMacro{table-pattern-stats-row102-origin}{LLVM}
\DefMacro{table-pattern-stats-row102-type}{OrINode}
\DefMacro{table-pattern-stats-row102-shadows}{}
\DefMacro{table-pattern-stats-row102-composites}{}
\DefMacro{table-pattern-stats-row102-pr}{}
\DefMacro{table-pattern-stats-row103-name}{OR17}
\DefMacro{table-pattern-stats-row103-origin}{LLVM}
\DefMacro{table-pattern-stats-row103-type}{OrINode}
\DefMacro{table-pattern-stats-row103-shadows}{}
\DefMacro{table-pattern-stats-row103-composites}{}
\DefMacro{table-pattern-stats-row103-pr}{}
\DefMacro{table-pattern-stats-row104-name}{OR18}
\DefMacro{table-pattern-stats-row104-origin}{LLVM}
\DefMacro{table-pattern-stats-row104-type}{OrINode}
\DefMacro{table-pattern-stats-row104-shadows}{}
\DefMacro{table-pattern-stats-row104-composites}{}
\DefMacro{table-pattern-stats-row104-pr}{}
\DefMacro{table-pattern-stats-row105-name}{OR19}
\DefMacro{table-pattern-stats-row105-origin}{LLVM}
\DefMacro{table-pattern-stats-row105-type}{OrINode}
\DefMacro{table-pattern-stats-row105-shadows}{}
\DefMacro{table-pattern-stats-row105-composites}{}
\DefMacro{table-pattern-stats-row105-pr}{}
\DefMacro{table-pattern-stats-row106-name}{OR20}
\DefMacro{table-pattern-stats-row106-origin}{LLVM}
\DefMacro{table-pattern-stats-row106-type}{OrINode}
\DefMacro{table-pattern-stats-row106-shadows}{}
\DefMacro{table-pattern-stats-row106-composites}{}
\DefMacro{table-pattern-stats-row106-pr}{}
\DefMacro{table-pattern-stats-row107-name}{OR21}
\DefMacro{table-pattern-stats-row107-origin}{LLVM}
\DefMacro{table-pattern-stats-row107-type}{OrINode}
\DefMacro{table-pattern-stats-row107-shadows}{}
\DefMacro{table-pattern-stats-row107-composites}{}
\DefMacro{table-pattern-stats-row107-pr}{}
\DefMacro{table-pattern-stats-row108-name}{OR22}
\DefMacro{table-pattern-stats-row108-origin}{LLVM}
\DefMacro{table-pattern-stats-row108-type}{OrINode}
\DefMacro{table-pattern-stats-row108-shadows}{}
\DefMacro{table-pattern-stats-row108-composites}{}
\DefMacro{table-pattern-stats-row108-pr}{}
\DefMacro{table-pattern-stats-row109-name}{OR23}
\DefMacro{table-pattern-stats-row109-origin}{LLVM}
\DefMacro{table-pattern-stats-row109-type}{OrINode}
\DefMacro{table-pattern-stats-row109-shadows}{}
\DefMacro{table-pattern-stats-row109-composites}{}
\DefMacro{table-pattern-stats-row109-pr}{}
\DefMacro{table-pattern-stats-row110-name}{OR24}
\DefMacro{table-pattern-stats-row110-origin}{LLVM}
\DefMacro{table-pattern-stats-row110-type}{OrINode}
\DefMacro{table-pattern-stats-row110-shadows}{}
\DefMacro{table-pattern-stats-row110-composites}{}
\DefMacro{table-pattern-stats-row110-pr}{}
\DefMacro{table-pattern-stats-row111-name}{OR25}
\DefMacro{table-pattern-stats-row111-origin}{LLVM}
\DefMacro{table-pattern-stats-row111-type}{OrINode}
\DefMacro{table-pattern-stats-row111-shadows}{}
\DefMacro{table-pattern-stats-row111-composites}{}
\DefMacro{table-pattern-stats-row111-pr}{}
\DefMacro{table-pattern-stats-row112-name}{OR26}
\DefMacro{table-pattern-stats-row112-origin}{LLVM}
\DefMacro{table-pattern-stats-row112-type}{OrINode}
\DefMacro{table-pattern-stats-row112-shadows}{}
\DefMacro{table-pattern-stats-row112-composites}{}
\DefMacro{table-pattern-stats-row112-pr}{}
\DefMacro{table-pattern-stats-row113-name}{OR27}
\DefMacro{table-pattern-stats-row113-origin}{LLVM}
\DefMacro{table-pattern-stats-row113-type}{OrINode}
\DefMacro{table-pattern-stats-row113-shadows}{}
\DefMacro{table-pattern-stats-row113-composites}{}
\DefMacro{table-pattern-stats-row113-pr}{}
\DefMacro{table-pattern-stats-row114-name}{OR28}
\DefMacro{table-pattern-stats-row114-origin}{LLVM}
\DefMacro{table-pattern-stats-row114-type}{OrINode}
\DefMacro{table-pattern-stats-row114-shadows}{}
\DefMacro{table-pattern-stats-row114-composites}{}
\DefMacro{table-pattern-stats-row114-pr}{}
\DefMacro{table-pattern-stats-row115-name}{OR29}
\DefMacro{table-pattern-stats-row115-origin}{LLVM}
\DefMacro{table-pattern-stats-row115-type}{OrINode}
\DefMacro{table-pattern-stats-row115-shadows}{}
\DefMacro{table-pattern-stats-row115-composites}{}
\DefMacro{table-pattern-stats-row115-pr}{}
\DefMacro{table-pattern-stats-row116-name}{OR30}
\DefMacro{table-pattern-stats-row116-origin}{LLVM}
\DefMacro{table-pattern-stats-row116-type}{OrINode}
\DefMacro{table-pattern-stats-row116-shadows}{}
\DefMacro{table-pattern-stats-row116-composites}{}
\DefMacro{table-pattern-stats-row116-pr}{}
\DefMacro{table-pattern-stats-row117-name}{OR31}
\DefMacro{table-pattern-stats-row117-origin}{LLVM}
\DefMacro{table-pattern-stats-row117-type}{OrINode}
\DefMacro{table-pattern-stats-row117-shadows}{}
\DefMacro{table-pattern-stats-row117-composites}{}
\DefMacro{table-pattern-stats-row117-pr}{}
\DefMacro{table-pattern-stats-row118-name}{OR32}
\DefMacro{table-pattern-stats-row118-origin}{LLVM}
\DefMacro{table-pattern-stats-row118-type}{OrINode}
\DefMacro{table-pattern-stats-row118-shadows}{}
\DefMacro{table-pattern-stats-row118-composites}{}
\DefMacro{table-pattern-stats-row118-pr}{}
\DefMacro{table-pattern-stats-row119-name}{OR33}
\DefMacro{table-pattern-stats-row119-origin}{LLVM}
\DefMacro{table-pattern-stats-row119-type}{OrINode}
\DefMacro{table-pattern-stats-row119-shadows}{}
\DefMacro{table-pattern-stats-row119-composites}{}
\DefMacro{table-pattern-stats-row119-pr}{}
\DefMacro{table-pattern-stats-row120-name}{MUL1}
\DefMacro{table-pattern-stats-row120-origin}{OpenJDK}
\DefMacro{table-pattern-stats-row120-type}{MulINode}
\DefMacro{table-pattern-stats-row120-shadows}{}
\DefMacro{table-pattern-stats-row120-composites}{}
\DefMacro{table-pattern-stats-row120-pr}{}
\DefMacro{table-pattern-stats-row121-name}{MUL2}
\DefMacro{table-pattern-stats-row121-origin}{OpenJDK}
\DefMacro{table-pattern-stats-row121-type}{MulINode}
\DefMacro{table-pattern-stats-row121-shadows}{}
\DefMacro{table-pattern-stats-row121-composites}{}
\DefMacro{table-pattern-stats-row121-pr}{}
\DefMacro{table-pattern-stats-row122-name}{MUL3}
\DefMacro{table-pattern-stats-row122-origin}{OpenJDK}
\DefMacro{table-pattern-stats-row122-type}{MulINode}
\DefMacro{table-pattern-stats-row122-shadows}{}
\DefMacro{table-pattern-stats-row122-composites}{}
\DefMacro{table-pattern-stats-row122-pr}{}
\DefMacro{table-pattern-stats-row123-name}{MUL4}
\DefMacro{table-pattern-stats-row123-origin}{OpenJDK}
\DefMacro{table-pattern-stats-row123-type}{MulINode}
\DefMacro{table-pattern-stats-row123-shadows}{}
\DefMacro{table-pattern-stats-row123-composites}{}
\DefMacro{table-pattern-stats-row123-pr}{}
\DefMacro{table-pattern-stats-row124-name}{MUL5}
\DefMacro{table-pattern-stats-row124-origin}{OpenJDK}
\DefMacro{table-pattern-stats-row124-type}{MulINode}
\DefMacro{table-pattern-stats-row124-shadows}{}
\DefMacro{table-pattern-stats-row124-composites}{}
\DefMacro{table-pattern-stats-row124-pr}{}
\DefMacro{table-pattern-stats-row125-name}{AND1}
\DefMacro{table-pattern-stats-row125-origin}{OpenJDK}
\DefMacro{table-pattern-stats-row125-type}{AndINode}
\DefMacro{table-pattern-stats-row125-shadows}{}
\DefMacro{table-pattern-stats-row125-composites}{}
\DefMacro{table-pattern-stats-row125-pr}{}
\DefMacro{table-pattern-stats-row126-name}{LSHIFT1}
\DefMacro{table-pattern-stats-row126-origin}{OpenJDK}
\DefMacro{table-pattern-stats-row126-type}{LShiftINode}
\DefMacro{table-pattern-stats-row126-shadows}{}
\DefMacro{table-pattern-stats-row126-composites}{}
\DefMacro{table-pattern-stats-row126-pr}{}
\DefMacro{table-pattern-stats-row127-name}{LSHIFT2}
\DefMacro{table-pattern-stats-row127-origin}{OpenJDK}
\DefMacro{table-pattern-stats-row127-type}{LShiftINode}
\DefMacro{table-pattern-stats-row127-shadows}{}
\DefMacro{table-pattern-stats-row127-composites}{}
\DefMacro{table-pattern-stats-row127-pr}{}
\DefMacro{table-pattern-stats-row128-name}{LSHIFT3}
\DefMacro{table-pattern-stats-row128-origin}{OpenJDK}
\DefMacro{table-pattern-stats-row128-type}{LShiftINode}
\DefMacro{table-pattern-stats-row128-shadows}{}
\DefMacro{table-pattern-stats-row128-composites}{}
\DefMacro{table-pattern-stats-row128-pr}{}
\DefMacro{table-pattern-stats-row129-name}{LSHIFT4}
\DefMacro{table-pattern-stats-row129-origin}{OpenJDK}
\DefMacro{table-pattern-stats-row129-type}{LShiftINode}
\DefMacro{table-pattern-stats-row129-shadows}{}
\DefMacro{table-pattern-stats-row129-composites}{}
\DefMacro{table-pattern-stats-row129-pr}{}
\DefMacro{table-pattern-stats-row130-name}{LSHIFT5}
\DefMacro{table-pattern-stats-row130-origin}{OpenJDK}
\DefMacro{table-pattern-stats-row130-type}{LShiftINode}
\DefMacro{table-pattern-stats-row130-shadows}{}
\DefMacro{table-pattern-stats-row130-composites}{}
\DefMacro{table-pattern-stats-row130-pr}{}
\DefMacro{table-pattern-stats-row131-name}{LSHIFT6}
\DefMacro{table-pattern-stats-row131-origin}{OpenJDK}
\DefMacro{table-pattern-stats-row131-type}{LShiftINode}
\DefMacro{table-pattern-stats-row131-shadows}{}
\DefMacro{table-pattern-stats-row131-composites}{}
\DefMacro{table-pattern-stats-row131-pr}{}
\DefMacro{table-pattern-stats-row132-name}{RSHIFT1}
\DefMacro{table-pattern-stats-row132-origin}{OpenJDK}
\DefMacro{table-pattern-stats-row132-type}{RShiftINode}
\DefMacro{table-pattern-stats-row132-shadows}{}
\DefMacro{table-pattern-stats-row132-composites}{}
\DefMacro{table-pattern-stats-row132-pr}{}
\DefMacro{table-pattern-stats-row133-name}{URSHIFT1}
\DefMacro{table-pattern-stats-row133-origin}{OpenJDK}
\DefMacro{table-pattern-stats-row133-type}{URShiftINode}
\DefMacro{table-pattern-stats-row133-shadows}{}
\DefMacro{table-pattern-stats-row133-composites}{}
\DefMacro{table-pattern-stats-row133-pr}{}
\DefMacro{table-pattern-stats-row134-name}{URSHIFT2}
\DefMacro{table-pattern-stats-row134-origin}{OpenJDK}
\DefMacro{table-pattern-stats-row134-type}{URShiftINode}
\DefMacro{table-pattern-stats-row134-shadows}{}
\DefMacro{table-pattern-stats-row134-composites}{}
\DefMacro{table-pattern-stats-row134-pr}{}
\DefMacro{table-pattern-stats-row135-name}{URSHIFT3}
\DefMacro{table-pattern-stats-row135-origin}{OpenJDK}
\DefMacro{table-pattern-stats-row135-type}{URShiftINode}
\DefMacro{table-pattern-stats-row135-shadows}{}
\DefMacro{table-pattern-stats-row135-composites}{}
\DefMacro{table-pattern-stats-row135-pr}{}
\DefMacro{table-pattern-stats-row136-name}{URSHIFT4}
\DefMacro{table-pattern-stats-row136-origin}{OpenJDK}
\DefMacro{table-pattern-stats-row136-type}{URShiftINode}
\DefMacro{table-pattern-stats-row136-shadows}{}
\DefMacro{table-pattern-stats-row136-composites}{}
\DefMacro{table-pattern-stats-row136-pr}{}
\DefMacro{table-pattern-stats-row137-name}{URSHIFT5}
\DefMacro{table-pattern-stats-row137-origin}{OpenJDK}
\DefMacro{table-pattern-stats-row137-type}{URShiftINode}
\DefMacro{table-pattern-stats-row137-shadows}{}
\DefMacro{table-pattern-stats-row137-composites}{}
\DefMacro{table-pattern-stats-row137-pr}{}
\DefMacro{table-pattern-stats-row138-name}{ROTATELEFT1}
\DefMacro{table-pattern-stats-row138-origin}{OpenJDK}
\DefMacro{table-pattern-stats-row138-type}{RotateLeftNode}
\DefMacro{table-pattern-stats-row138-shadows}{}
\DefMacro{table-pattern-stats-row138-composites}{}
\DefMacro{table-pattern-stats-row138-pr}{}
\DefMacro{table-pattern-stats-row139-name}{AND2}
\DefMacro{table-pattern-stats-row139-origin}{LLVM}
\DefMacro{table-pattern-stats-row139-type}{AndINode}
\DefMacro{table-pattern-stats-row139-shadows}{}
\DefMacro{table-pattern-stats-row139-composites}{}
\DefMacro{table-pattern-stats-row139-pr}{}
\DefMacro{table-pattern-stats-row140-name}{AND3}
\DefMacro{table-pattern-stats-row140-origin}{LLVM}
\DefMacro{table-pattern-stats-row140-type}{AndINode}
\DefMacro{table-pattern-stats-row140-shadows}{}
\DefMacro{table-pattern-stats-row140-composites}{}
\DefMacro{table-pattern-stats-row140-pr}{}
\DefMacro{table-pattern-stats-row141-name}{AND4}
\DefMacro{table-pattern-stats-row141-origin}{LLVM}
\DefMacro{table-pattern-stats-row141-type}{AndINode}
\DefMacro{table-pattern-stats-row141-shadows}{}
\DefMacro{table-pattern-stats-row141-composites}{}
\DefMacro{table-pattern-stats-row141-pr}{}
\DefMacro{table-pattern-stats-row142-name}{AND5}
\DefMacro{table-pattern-stats-row142-origin}{LLVM}
\DefMacro{table-pattern-stats-row142-type}{AndINode}
\DefMacro{table-pattern-stats-row142-shadows}{}
\DefMacro{table-pattern-stats-row142-composites}{}
\DefMacro{table-pattern-stats-row142-pr}{}
\DefMacro{table-pattern-stats-row143-name}{AND6}
\DefMacro{table-pattern-stats-row143-origin}{LLVM}
\DefMacro{table-pattern-stats-row143-type}{AndINode}
\DefMacro{table-pattern-stats-row143-shadows}{}
\DefMacro{table-pattern-stats-row143-composites}{}
\DefMacro{table-pattern-stats-row143-pr}{}
\DefMacro{table-pattern-stats-row144-name}{AND7}
\DefMacro{table-pattern-stats-row144-origin}{LLVM}
\DefMacro{table-pattern-stats-row144-type}{AndINode}
\DefMacro{table-pattern-stats-row144-shadows}{}
\DefMacro{table-pattern-stats-row144-composites}{}
\DefMacro{table-pattern-stats-row144-pr}{}
\DefMacro{table-pattern-stats-row145-name}{AND8}
\DefMacro{table-pattern-stats-row145-origin}{LLVM}
\DefMacro{table-pattern-stats-row145-type}{AndINode}
\DefMacro{table-pattern-stats-row145-shadows}{}
\DefMacro{table-pattern-stats-row145-composites}{}
\DefMacro{table-pattern-stats-row145-pr}{}
\DefMacro{table-pattern-stats-row146-name}{AND9}
\DefMacro{table-pattern-stats-row146-origin}{LLVM}
\DefMacro{table-pattern-stats-row146-type}{AndINode}
\DefMacro{table-pattern-stats-row146-shadows}{}
\DefMacro{table-pattern-stats-row146-composites}{}
\DefMacro{table-pattern-stats-row146-pr}{}
\DefMacro{table-pattern-stats-row147-name}{AND10}
\DefMacro{table-pattern-stats-row147-origin}{LLVM}
\DefMacro{table-pattern-stats-row147-type}{AndINode}
\DefMacro{table-pattern-stats-row147-shadows}{}
\DefMacro{table-pattern-stats-row147-composites}{}
\DefMacro{table-pattern-stats-row147-pr}{}
\DefMacro{table-pattern-stats-row148-name}{AND11}
\DefMacro{table-pattern-stats-row148-origin}{LLVM}
\DefMacro{table-pattern-stats-row148-type}{AndINode}
\DefMacro{table-pattern-stats-row148-shadows}{}
\DefMacro{table-pattern-stats-row148-composites}{}
\DefMacro{table-pattern-stats-row148-pr}{}
\DefMacro{table-pattern-stats-row149-name}{AND12}
\DefMacro{table-pattern-stats-row149-origin}{LLVM}
\DefMacro{table-pattern-stats-row149-type}{AndINode}
\DefMacro{table-pattern-stats-row149-shadows}{}
\DefMacro{table-pattern-stats-row149-composites}{}
\DefMacro{table-pattern-stats-row149-pr}{}
\DefMacro{table-pattern-stats-row150-name}{AND13}
\DefMacro{table-pattern-stats-row150-origin}{LLVM}
\DefMacro{table-pattern-stats-row150-type}{AndINode}
\DefMacro{table-pattern-stats-row150-shadows}{}
\DefMacro{table-pattern-stats-row150-composites}{}
\DefMacro{table-pattern-stats-row150-pr}{}
\DefMacro{table-pattern-stats-row151-name}{AND14}
\DefMacro{table-pattern-stats-row151-origin}{LLVM}
\DefMacro{table-pattern-stats-row151-type}{AndINode}
\DefMacro{table-pattern-stats-row151-shadows}{}
\DefMacro{table-pattern-stats-row151-composites}{}
\DefMacro{table-pattern-stats-row151-pr}{}
\DefMacro{table-pattern-stats-row152-name}{AND15}
\DefMacro{table-pattern-stats-row152-origin}{LLVM}
\DefMacro{table-pattern-stats-row152-type}{AndINode}
\DefMacro{table-pattern-stats-row152-shadows}{}
\DefMacro{table-pattern-stats-row152-composites}{}
\DefMacro{table-pattern-stats-row152-pr}{}
\DefMacro{table-pattern-stats-row153-name}{AND16}
\DefMacro{table-pattern-stats-row153-origin}{LLVM}
\DefMacro{table-pattern-stats-row153-type}{AndINode}
\DefMacro{table-pattern-stats-row153-shadows}{}
\DefMacro{table-pattern-stats-row153-composites}{}
\DefMacro{table-pattern-stats-row153-pr}{}
\DefMacro{table-pattern-stats-row154-name}{AND17}
\DefMacro{table-pattern-stats-row154-origin}{LLVM}
\DefMacro{table-pattern-stats-row154-type}{AndINode}
\DefMacro{table-pattern-stats-row154-shadows}{}
\DefMacro{table-pattern-stats-row154-composites}{}
\DefMacro{table-pattern-stats-row154-pr}{}
\DefMacro{table-pattern-stats-row155-name}{AND18}
\DefMacro{table-pattern-stats-row155-origin}{LLVM}
\DefMacro{table-pattern-stats-row155-type}{AndINode}
\DefMacro{table-pattern-stats-row155-shadows}{}
\DefMacro{table-pattern-stats-row155-composites}{}
\DefMacro{table-pattern-stats-row155-pr}{}
\DefMacro{table-pattern-stats-row156-name}{AND19}
\DefMacro{table-pattern-stats-row156-origin}{LLVM}
\DefMacro{table-pattern-stats-row156-type}{AndINode}
\DefMacro{table-pattern-stats-row156-shadows}{}
\DefMacro{table-pattern-stats-row156-composites}{}
\DefMacro{table-pattern-stats-row156-pr}{}
\DefMacro{table-pattern-stats-row157-name}{AND20}
\DefMacro{table-pattern-stats-row157-origin}{LLVM}
\DefMacro{table-pattern-stats-row157-type}{AndINode}
\DefMacro{table-pattern-stats-row157-shadows}{}
\DefMacro{table-pattern-stats-row157-composites}{}
\DefMacro{table-pattern-stats-row157-pr}{}
\DefMacro{table-pattern-stats-row158-name}{AND21}
\DefMacro{table-pattern-stats-row158-origin}{LLVM}
\DefMacro{table-pattern-stats-row158-type}{AndINode}
\DefMacro{table-pattern-stats-row158-shadows}{}
\DefMacro{table-pattern-stats-row158-composites}{}
\DefMacro{table-pattern-stats-row158-pr}{}
\DefMacro{table-pattern-stats-row159-name}{AND22}
\DefMacro{table-pattern-stats-row159-origin}{LLVM}
\DefMacro{table-pattern-stats-row159-type}{AndINode}
\DefMacro{table-pattern-stats-row159-shadows}{}
\DefMacro{table-pattern-stats-row159-composites}{}
\DefMacro{table-pattern-stats-row159-pr}{}
\DefMacro{table-pattern-stats-row160-name}{AND23}
\DefMacro{table-pattern-stats-row160-origin}{LLVM}
\DefMacro{table-pattern-stats-row160-type}{AndINode}
\DefMacro{table-pattern-stats-row160-shadows}{}
\DefMacro{table-pattern-stats-row160-composites}{}
\DefMacro{table-pattern-stats-row160-pr}{}
\DefMacro{table-pattern-stats-row161-name}{AND24}
\DefMacro{table-pattern-stats-row161-origin}{LLVM}
\DefMacro{table-pattern-stats-row161-type}{AndINode}
\DefMacro{table-pattern-stats-row161-shadows}{}
\DefMacro{table-pattern-stats-row161-composites}{}
\DefMacro{table-pattern-stats-row161-pr}{}

\DefMacro{table-pattern-stats-float-env}{table*}
\DefMacro{table-pattern-stats-caption}{Statistics of \ptns.\label{tab:pattern-stats}}
\DefMacro{table-pattern-stats-fontsize}{tiny}
\DefMacro{table-pattern-stats-col-name}{Name}
\DefMacro{table-pattern-stats-col-before}{Before}
\DefMacro{table-pattern-stats-col-after}{After}
\DefMacro{table-pattern-stats-col-origin}{Origin}
\DefMacro{table-pattern-stats-col-type}{Type}
\DefMacro{table-pattern-stats-col-shadows}{\Shadows}
\DefMacro{table-pattern-stats-col-concatenates}{Concatenates}
\DefMacro{table-pattern-stats-col-pr}{PR}

\DefMacro{table-shadow-stats-num-rows}{9}
\DefMacro{table-shadow-stats-row0-shadowing-before}{\texttt{x - (y + C0)}}
\DefMacro{table-shadow-stats-row0-shadowing-precondition}{\makecell{\texttt{okToConvert(}\\\hfill\texttt{y + C0, x)}}}
\DefMacro{table-shadow-stats-row0-shadowed-before-0}{\texttt{C0 - (x + C1)}}
\DefMacro{table-shadow-stats-row0-shadowed-precondition-0}{\makecell{\texttt{okToConvert(}\\\hfill\texttt{x + C1, C0)}}}
\DefMacro{table-shadow-stats-row1-shadowing-before}{\texttt{(a - b) + (c - d)}}
\DefMacro{table-shadow-stats-row1-shadowing-precondition}{$\top$}
\DefMacro{table-shadow-stats-row1-shadowed-before-0}{\texttt{(a - b) + (b - c)}}
\DefMacro{table-shadow-stats-row1-shadowed-precondition-0}{$\top$}
\DefMacro{table-shadow-stats-row1-shadowed-before-1}{\texttt{(a - b) + (c - a)}}
\DefMacro{table-shadow-stats-row1-shadowed-precondition-1}{$\top$}
\DefMacro{table-shadow-stats-row1-shadowed-before-2}{\texttt{(0 - a) + (0 - b)}}
\DefMacro{table-shadow-stats-row1-shadowed-precondition-2}{$\top$}
\DefMacro{table-shadow-stats-row2-shadowing-before}{\texttt{x + (0 - y)}}
\DefMacro{table-shadow-stats-row2-shadowing-precondition}{$\top$}
\DefMacro{table-shadow-stats-row2-shadowed-before-0}{\texttt{(0 - a) + (0 - b)}}
\DefMacro{table-shadow-stats-row2-shadowed-precondition-0}{$\top$}
\DefMacro{table-shadow-stats-row3-shadowing-before}{\texttt{(0 - y) + x}}
\DefMacro{table-shadow-stats-row3-shadowing-precondition}{$\top$}
\DefMacro{table-shadow-stats-row3-shadowed-before-0}{\texttt{(0 - a) + (0 - b)}}
\DefMacro{table-shadow-stats-row3-shadowed-precondition-0}{$\top$}
\DefMacro{table-shadow-stats-row4-shadowing-before}{\texttt{(x + CON) + y}}
\DefMacro{table-shadow-stats-row4-shadowing-precondition}{$\top$}
\DefMacro{table-shadow-stats-row4-shadowed-before-0}{\texttt{(x + CON1) + CON2}}
\DefMacro{table-shadow-stats-row4-shadowed-precondition-0}{$\top$}
\DefMacro{table-shadow-stats-row4-shadowed-before-1}{\texttt{(((z | C2) \char`\^{} C1) + 1) + rhs}}
\DefMacro{table-shadow-stats-row4-shadowed-precondition-1}{\texttt{C2 == \char`\~{}C1}}
\DefMacro{table-shadow-stats-row4-shadowed-before-2}{\texttt{(((z \& C2) \char`\^{} C1) + 1) + rhs}}
\DefMacro{table-shadow-stats-row4-shadowed-precondition-2}{\texttt{C2 == C1}}
\DefMacro{table-shadow-stats-row4-shadowed-before-3}{\texttt{(x + 1) + (y \char`\^{} -1)}}
\DefMacro{table-shadow-stats-row4-shadowed-precondition-3}{$\top$}
\DefMacro{table-shadow-stats-row4-shadowed-before-4}{\texttt{(a + C1) + (C2 - b)}}
\DefMacro{table-shadow-stats-row4-shadowed-precondition-4}{$\top$}
\DefMacro{table-shadow-stats-row5-shadowing-before}{\texttt{x + (y + CON)}}
\DefMacro{table-shadow-stats-row5-shadowing-precondition}{$\top$}
\DefMacro{table-shadow-stats-row5-shadowed-before-0}{\texttt{(y \char`\^{} -1) + (x + 1)}}
\DefMacro{table-shadow-stats-row5-shadowed-precondition-0}{$\top$}
\DefMacro{table-shadow-stats-row6-shadowing-before}{\texttt{(x \char`\^{} -1) + C}}
\DefMacro{table-shadow-stats-row6-shadowing-precondition}{$\top$}
\DefMacro{table-shadow-stats-row6-shadowed-before-0}{\texttt{(x \char`\^{} -1) + 1}}
\DefMacro{table-shadow-stats-row6-shadowed-precondition-0}{$\top$}
\DefMacro{table-shadow-stats-row7-shadowing-before}{\texttt{x + (CON - y)}}
\DefMacro{table-shadow-stats-row7-shadowing-precondition}{$\top$}
\DefMacro{table-shadow-stats-row7-shadowed-before-0}{\texttt{x + (0 - y)}}
\DefMacro{table-shadow-stats-row7-shadowed-precondition-0}{$\top$}
\DefMacro{table-shadow-stats-row7-shadowed-before-1}{\texttt{(0 - a) + (0 - b)}}
\DefMacro{table-shadow-stats-row7-shadowed-precondition-1}{$\top$}
\DefMacro{table-shadow-stats-row7-shadowed-before-2}{\texttt{(a + C1) + (C2 - b)}}
\DefMacro{table-shadow-stats-row7-shadowed-precondition-2}{$\top$}
\DefMacro{table-shadow-stats-row8-shadowing-before}{\texttt{(CON - y) + x}}
\DefMacro{table-shadow-stats-row8-shadowing-precondition}{$\top$}
\DefMacro{table-shadow-stats-row8-shadowed-before-0}{\texttt{(CON1 - x) + CON2}}
\DefMacro{table-shadow-stats-row8-shadowed-precondition-0}{$\top$}
\DefMacro{table-shadow-stats-row8-shadowed-before-1}{\texttt{(0 - y) + x}}
\DefMacro{table-shadow-stats-row8-shadowed-precondition-1}{$\top$}
\DefMacro{table-shadow-stats-row8-shadowed-before-2}{\texttt{(0 - a) + (0 - b)}}
\DefMacro{table-shadow-stats-row8-shadowed-precondition-2}{$\top$}

\DefMacro{table-shadow-stats-float-env}{table}
\DefMacro{table-shadow-stats-caption}{
\Shadow between \ptns. Constants are in upper-case and free variables
are in lower-case. \label{tab:shadow-stats}}
\DefMacro{table-shadow-stats-fontsize}{scriptsize}
\DefMacro{table-shadow-stats-col-shadowing}{\Shadowing}
\DefMacro{table-shadow-stats-col-shadowed}{\Shadowed}
\DefMacro{table-shadow-stats-col-name}{Name}
\DefMacro{table-shadow-stats-col-before}{Before}
\DefMacro{table-shadow-stats-col-precondition}{Precondition}

\DefMacro{table-composite-stats-num-rows}{109}
\DefMacro{table-composite-stats-row0-compositing-after}{\texttt{x + -C0}}
\DefMacro{table-composite-stats-row0-compositing-precondition}{$\top$}
\DefMacro{table-composite-stats-row0-composited-before-0}{\texttt{(CON1 - x) + CON2}}
\DefMacro{table-composite-stats-row0-composited-precondition-0}{$\top$}
\DefMacro{table-composite-stats-row0-composited-before-1}{\texttt{(0 - y) + x}}
\DefMacro{table-composite-stats-row0-composited-precondition-1}{$\top$}
\DefMacro{table-composite-stats-row0-composited-before-2}{\texttt{(x >>> Z) + Y}}
\DefMacro{table-composite-stats-row0-composited-precondition-2}{\texttt{Z < 5 \&\& -5 < Y \&\& Y < 0 \&\& x >= -(Y << Z)}}
\DefMacro{table-composite-stats-row0-composited-before-3}{\texttt{(x + CON1) + CON2}}
\DefMacro{table-composite-stats-row0-composited-precondition-3}{$\top$}
\DefMacro{table-composite-stats-row0-composited-before-4}{\texttt{CON + x}}
\DefMacro{table-composite-stats-row0-composited-precondition-4}{$\top$}
\DefMacro{table-composite-stats-row0-composited-before-5}{\texttt{(x \^{} -1) + 1}}
\DefMacro{table-composite-stats-row0-composited-precondition-5}{$\top$}
\DefMacro{table-composite-stats-row0-composited-before-6}{\texttt{(x + CON) + y}}
\DefMacro{table-composite-stats-row0-composited-precondition-6}{$\top$}
\DefMacro{table-composite-stats-row0-composited-before-7}{\texttt{(((z | C2) \^{} C1) + 1) + rhs}}
\DefMacro{table-composite-stats-row0-composited-precondition-7}{\texttt{C2 == \~{}C1}}
\DefMacro{table-composite-stats-row0-composited-before-8}{\texttt{(((z \& C2) \^{} C1) + 1) + rhs}}
\DefMacro{table-composite-stats-row0-composited-precondition-8}{\texttt{C2 == C1}}
\DefMacro{table-composite-stats-row0-composited-before-9}{\texttt{x + x}}
\DefMacro{table-composite-stats-row0-composited-precondition-9}{$\top$}
\DefMacro{table-composite-stats-row0-composited-before-10}{\texttt{(x \^{} -1) + C}}
\DefMacro{table-composite-stats-row0-composited-precondition-10}{$\top$}
\DefMacro{table-composite-stats-row0-composited-before-11}{\texttt{((y \^{} -1) + x) + 1}}
\DefMacro{table-composite-stats-row0-composited-precondition-11}{$\top$}
\DefMacro{table-composite-stats-row0-composited-before-12}{\texttt{(x + (y \^{} -1)) + 1}}
\DefMacro{table-composite-stats-row0-composited-precondition-12}{$\top$}
\DefMacro{table-composite-stats-row0-composited-before-13}{\texttt{(CON - y) + x}}
\DefMacro{table-composite-stats-row0-composited-precondition-13}{$\top$}
\DefMacro{table-composite-stats-row0-composited-before-14}{\texttt{(x | C2) + C}}
\DefMacro{table-composite-stats-row0-composited-precondition-14}{\texttt{C2 == -C}}
\DefMacro{table-composite-stats-row1-compositing-after}{\texttt{(x - y) + C0}}
\DefMacro{table-composite-stats-row1-compositing-precondition}{\texttt{Lib.okToConvert(x + C0, y)}}
\DefMacro{table-composite-stats-row1-composited-before-0}{\texttt{(CON1 - x) + CON2}}
\DefMacro{table-composite-stats-row1-composited-precondition-0}{$\top$}
\DefMacro{table-composite-stats-row1-composited-before-1}{\texttt{(0 - y) + x}}
\DefMacro{table-composite-stats-row1-composited-precondition-1}{$\top$}
\DefMacro{table-composite-stats-row1-composited-before-2}{\texttt{(CON - y) + x}}
\DefMacro{table-composite-stats-row1-composited-precondition-2}{$\top$}
\DefMacro{table-composite-stats-row2-compositing-after}{\texttt{(x - y) + -C0}}
\DefMacro{table-composite-stats-row2-compositing-precondition}{\texttt{Lib.okToConvert(y + C0, x)}}
\DefMacro{table-composite-stats-row2-composited-before-0}{\texttt{(CON1 - x) + CON2}}
\DefMacro{table-composite-stats-row2-composited-precondition-0}{$\top$}
\DefMacro{table-composite-stats-row2-composited-before-1}{\texttt{(0 - y) + x}}
\DefMacro{table-composite-stats-row2-composited-precondition-1}{$\top$}
\DefMacro{table-composite-stats-row2-composited-before-2}{\texttt{(CON - y) + x}}
\DefMacro{table-composite-stats-row2-composited-precondition-2}{$\top$}
\DefMacro{table-composite-stats-row3-compositing-after}{\texttt{0 - y}}
\DefMacro{table-composite-stats-row3-compositing-precondition}{$\top$}
\DefMacro{table-composite-stats-row3-composited-before-0}{\texttt{x - (0 - y)}}
\DefMacro{table-composite-stats-row3-composited-precondition-0}{$\top$}
\DefMacro{table-composite-stats-row3-composited-before-1}{\texttt{C - (x \^{} -1)}}
\DefMacro{table-composite-stats-row3-composited-precondition-1}{$\top$}
\DefMacro{table-composite-stats-row3-composited-before-2}{\texttt{C0 - (x + C1)}}
\DefMacro{table-composite-stats-row3-composited-precondition-2}{\texttt{Lib.okToConvert(x + C1, C0)}}
\DefMacro{table-composite-stats-row3-composited-before-3}{\texttt{C - (C2 - x)}}
\DefMacro{table-composite-stats-row3-composited-precondition-3}{$\top$}
\DefMacro{table-composite-stats-row3-composited-before-4}{\texttt{x - (x \& y)}}
\DefMacro{table-composite-stats-row3-composited-precondition-4}{$\top$}
\DefMacro{table-composite-stats-row3-composited-before-5}{\texttt{x - C0}}
\DefMacro{table-composite-stats-row3-composited-precondition-5}{$\top$}
\DefMacro{table-composite-stats-row3-composited-before-6}{\texttt{a - (b - c)}}
\DefMacro{table-composite-stats-row3-composited-precondition-6}{\texttt{Lib.outcnt(b - c) == 1}}
\DefMacro{table-composite-stats-row3-composited-before-7}{\texttt{x - (y + C0)}}
\DefMacro{table-composite-stats-row3-composited-precondition-7}{\texttt{Lib.okToConvert(y + C0, x)}}
\DefMacro{table-composite-stats-row3-composited-before-8}{\texttt{0 - (x - y)}}
\DefMacro{table-composite-stats-row3-composited-precondition-8}{\texttt{x != 0}}
\DefMacro{table-composite-stats-row3-composited-before-9}{\texttt{0 - (x + CON)}}
\DefMacro{table-composite-stats-row3-composited-precondition-9}{\texttt{CON != 0}}
\DefMacro{table-composite-stats-row3-composited-before-10}{\texttt{0 - (a >> 31)}}
\DefMacro{table-composite-stats-row3-composited-precondition-10}{$\top$}
\DefMacro{table-composite-stats-row3-composited-before-11}{\texttt{x - (y + x)}}
\DefMacro{table-composite-stats-row3-composited-precondition-11}{$\top$}
\DefMacro{table-composite-stats-row4-compositing-after}{\texttt{0 - y}}
\DefMacro{table-composite-stats-row4-compositing-precondition}{$\top$}
\DefMacro{table-composite-stats-row4-composited-before-0}{\texttt{x - (0 - y)}}
\DefMacro{table-composite-stats-row4-composited-precondition-0}{$\top$}
\DefMacro{table-composite-stats-row4-composited-before-1}{\texttt{C - (x \^{} -1)}}
\DefMacro{table-composite-stats-row4-composited-precondition-1}{$\top$}
\DefMacro{table-composite-stats-row4-composited-before-2}{\texttt{C0 - (x + C1)}}
\DefMacro{table-composite-stats-row4-composited-precondition-2}{\texttt{Lib.okToConvert(x + C1, C0)}}
\DefMacro{table-composite-stats-row4-composited-before-3}{\texttt{C - (C2 - x)}}
\DefMacro{table-composite-stats-row4-composited-precondition-3}{$\top$}
\DefMacro{table-composite-stats-row4-composited-before-4}{\texttt{x - (x \& y)}}
\DefMacro{table-composite-stats-row4-composited-precondition-4}{$\top$}
\DefMacro{table-composite-stats-row4-composited-before-5}{\texttt{x - C0}}
\DefMacro{table-composite-stats-row4-composited-precondition-5}{$\top$}
\DefMacro{table-composite-stats-row4-composited-before-6}{\texttt{a - (b - c)}}
\DefMacro{table-composite-stats-row4-composited-precondition-6}{\texttt{Lib.outcnt(b - c) == 1}}
\DefMacro{table-composite-stats-row4-composited-before-7}{\texttt{x - (y + C0)}}
\DefMacro{table-composite-stats-row4-composited-precondition-7}{\texttt{Lib.okToConvert(y + C0, x)}}
\DefMacro{table-composite-stats-row4-composited-before-8}{\texttt{0 - (x - y)}}
\DefMacro{table-composite-stats-row4-composited-precondition-8}{\texttt{x != 0}}
\DefMacro{table-composite-stats-row4-composited-before-9}{\texttt{0 - (x + CON)}}
\DefMacro{table-composite-stats-row4-composited-precondition-9}{\texttt{CON != 0}}
\DefMacro{table-composite-stats-row4-composited-before-10}{\texttt{0 - (a >> 31)}}
\DefMacro{table-composite-stats-row4-composited-precondition-10}{$\top$}
\DefMacro{table-composite-stats-row4-composited-before-11}{\texttt{x - (x + y)}}
\DefMacro{table-composite-stats-row4-composited-precondition-11}{$\top$}
\DefMacro{table-composite-stats-row4-composited-before-12}{\texttt{x - (y + x)}}
\DefMacro{table-composite-stats-row4-composited-precondition-12}{$\top$}
\DefMacro{table-composite-stats-row5-compositing-after}{\texttt{0 - y}}
\DefMacro{table-composite-stats-row5-compositing-precondition}{$\top$}
\DefMacro{table-composite-stats-row5-composited-before-0}{\texttt{x - (0 - y)}}
\DefMacro{table-composite-stats-row5-composited-precondition-0}{$\top$}
\DefMacro{table-composite-stats-row5-composited-before-1}{\texttt{C - (x \^{} -1)}}
\DefMacro{table-composite-stats-row5-composited-precondition-1}{$\top$}
\DefMacro{table-composite-stats-row5-composited-before-2}{\texttt{C0 - (x + C1)}}
\DefMacro{table-composite-stats-row5-composited-precondition-2}{\texttt{Lib.okToConvert(x + C1, C0)}}
\DefMacro{table-composite-stats-row5-composited-before-3}{\texttt{C - (C2 - x)}}
\DefMacro{table-composite-stats-row5-composited-precondition-3}{$\top$}
\DefMacro{table-composite-stats-row5-composited-before-4}{\texttt{x - (x \& y)}}
\DefMacro{table-composite-stats-row5-composited-precondition-4}{$\top$}
\DefMacro{table-composite-stats-row5-composited-before-5}{\texttt{x - C0}}
\DefMacro{table-composite-stats-row5-composited-precondition-5}{$\top$}
\DefMacro{table-composite-stats-row5-composited-before-6}{\texttt{a - (b - c)}}
\DefMacro{table-composite-stats-row5-composited-precondition-6}{\texttt{Lib.outcnt(b - c) == 1}}
\DefMacro{table-composite-stats-row5-composited-before-7}{\texttt{x - (y + C0)}}
\DefMacro{table-composite-stats-row5-composited-precondition-7}{\texttt{Lib.okToConvert(y + C0, x)}}
\DefMacro{table-composite-stats-row5-composited-before-8}{\texttt{0 - (x - y)}}
\DefMacro{table-composite-stats-row5-composited-precondition-8}{\texttt{x != 0}}
\DefMacro{table-composite-stats-row5-composited-before-9}{\texttt{0 - (x + CON)}}
\DefMacro{table-composite-stats-row5-composited-precondition-9}{\texttt{CON != 0}}
\DefMacro{table-composite-stats-row5-composited-before-10}{\texttt{0 - (a >> 31)}}
\DefMacro{table-composite-stats-row5-composited-precondition-10}{$\top$}
\DefMacro{table-composite-stats-row5-composited-before-11}{\texttt{x - (x + y)}}
\DefMacro{table-composite-stats-row5-composited-precondition-11}{$\top$}

\DefMacro{table-composite-stats-float-env}{table*}
\DefMacro{table-composite-stats-caption}{\Composite between \ptns. Identifiers in upper-case letters are constants. \label{tab:composite-stats}}
\DefMacro{table-composite-stats-fontsize}{tiny}
\DefMacro{table-composite-stats-col-compositing}{\Compositing}
\DefMacro{table-composite-stats-col-composited}{\Composited}
\DefMacro{table-composite-stats-col-name}{Name}
\DefMacro{table-composite-stats-col-before}{Before}
\DefMacro{table-composite-stats-col-after}{After}
\DefMacro{table-composite-stats-col-precondition}{Precondition}

\DefMacro{table-lcoc-stats-row0-names-0}{SUB1}
\DefMacro{table-lcoc-stats-row0-manual-coc}{167}
\DefMacro{table-lcoc-stats-row0-manual-ioc}{24}
\DefMacro{table-lcoc-stats-row0-generated-coc}{171}
\DefMacro{table-lcoc-stats-row0-generated-ioc}{21}
\DefMacro{table-lcoc-stats-row0-pattern-coc}{47}
\DefMacro{table-lcoc-stats-row0-pattern-ioc}{9}
\DefMacro{table-lcoc-stats-row0-reduction-coc}{71.86}
\DefMacro{table-lcoc-stats-row0-reduction-ioc}{62.50}
\DefMacro{table-lcoc-stats-row1-names-0}{SUB2}
\DefMacro{table-lcoc-stats-row1-manual-coc}{264}
\DefMacro{table-lcoc-stats-row1-manual-ioc}{38}
\DefMacro{table-lcoc-stats-row1-generated-coc}{412}
\DefMacro{table-lcoc-stats-row1-generated-ioc}{43}
\DefMacro{table-lcoc-stats-row1-pattern-coc}{88}
\DefMacro{table-lcoc-stats-row1-pattern-ioc}{17}
\DefMacro{table-lcoc-stats-row1-reduction-coc}{66.67}
\DefMacro{table-lcoc-stats-row1-reduction-ioc}{55.26}
\DefMacro{table-lcoc-stats-row2-names-0}{SUB3}
\DefMacro{table-lcoc-stats-row2-manual-coc}{340}
\DefMacro{table-lcoc-stats-row2-manual-ioc}{48}
\DefMacro{table-lcoc-stats-row2-generated-coc}{421}
\DefMacro{table-lcoc-stats-row2-generated-ioc}{45}
\DefMacro{table-lcoc-stats-row2-pattern-coc}{89}
\DefMacro{table-lcoc-stats-row2-pattern-ioc}{17}
\DefMacro{table-lcoc-stats-row2-reduction-coc}{73.82}
\DefMacro{table-lcoc-stats-row2-reduction-ioc}{64.58}
\DefMacro{table-lcoc-stats-row3-names-0}{SUB4}
\DefMacro{table-lcoc-stats-row3-manual-coc}{134}
\DefMacro{table-lcoc-stats-row3-manual-ioc}{20}
\DefMacro{table-lcoc-stats-row3-generated-coc}{280}
\DefMacro{table-lcoc-stats-row3-generated-ioc}{29}
\DefMacro{table-lcoc-stats-row3-pattern-coc}{38}
\DefMacro{table-lcoc-stats-row3-pattern-ioc}{8}
\DefMacro{table-lcoc-stats-row3-reduction-coc}{71.64}
\DefMacro{table-lcoc-stats-row3-reduction-ioc}{60.00}
\DefMacro{table-lcoc-stats-row4-names-0}{SUB5}
\DefMacro{table-lcoc-stats-row4-manual-coc}{134}
\DefMacro{table-lcoc-stats-row4-manual-ioc}{20}
\DefMacro{table-lcoc-stats-row4-generated-coc}{280}
\DefMacro{table-lcoc-stats-row4-generated-ioc}{29}
\DefMacro{table-lcoc-stats-row4-pattern-coc}{38}
\DefMacro{table-lcoc-stats-row4-pattern-ioc}{8}
\DefMacro{table-lcoc-stats-row4-reduction-coc}{71.64}
\DefMacro{table-lcoc-stats-row4-reduction-ioc}{60.00}
\DefMacro{table-lcoc-stats-row5-names-0}{SUB6}
\DefMacro{table-lcoc-stats-row5-manual-coc}{134}
\DefMacro{table-lcoc-stats-row5-manual-ioc}{20}
\DefMacro{table-lcoc-stats-row5-generated-coc}{280}
\DefMacro{table-lcoc-stats-row5-generated-ioc}{29}
\DefMacro{table-lcoc-stats-row5-pattern-coc}{38}
\DefMacro{table-lcoc-stats-row5-pattern-ioc}{8}
\DefMacro{table-lcoc-stats-row5-reduction-coc}{71.64}
\DefMacro{table-lcoc-stats-row5-reduction-ioc}{60.00}
\DefMacro{table-lcoc-stats-row6-names-0}{SUB7}
\DefMacro{table-lcoc-stats-row6-manual-coc}{200}
\DefMacro{table-lcoc-stats-row6-manual-ioc}{31}
\DefMacro{table-lcoc-stats-row6-generated-coc}{333}
\DefMacro{table-lcoc-stats-row6-generated-ioc}{36}
\DefMacro{table-lcoc-stats-row6-pattern-coc}{48}
\DefMacro{table-lcoc-stats-row6-pattern-ioc}{9}
\DefMacro{table-lcoc-stats-row6-reduction-coc}{76.00}
\DefMacro{table-lcoc-stats-row6-reduction-ioc}{70.97}
\DefMacro{table-lcoc-stats-row7-names-0}{SUB8}
\DefMacro{table-lcoc-stats-row7-manual-coc}{213}
\DefMacro{table-lcoc-stats-row7-manual-ioc}{32}
\DefMacro{table-lcoc-stats-row7-generated-coc}{395}
\DefMacro{table-lcoc-stats-row7-generated-ioc}{44}
\DefMacro{table-lcoc-stats-row7-pattern-coc}{66}
\DefMacro{table-lcoc-stats-row7-pattern-ioc}{10}
\DefMacro{table-lcoc-stats-row7-reduction-coc}{69.01}
\DefMacro{table-lcoc-stats-row7-reduction-ioc}{68.75}
\DefMacro{table-lcoc-stats-row8-names-0}{SUB9}
\DefMacro{table-lcoc-stats-row8-manual-coc}{169}
\DefMacro{table-lcoc-stats-row8-manual-ioc}{27}
\DefMacro{table-lcoc-stats-row8-generated-coc}{443}
\DefMacro{table-lcoc-stats-row8-generated-ioc}{45}
\DefMacro{table-lcoc-stats-row8-pattern-coc}{47}
\DefMacro{table-lcoc-stats-row8-pattern-ioc}{11}
\DefMacro{table-lcoc-stats-row8-reduction-coc}{72.19}
\DefMacro{table-lcoc-stats-row8-reduction-ioc}{59.26}
\DefMacro{table-lcoc-stats-row9-names-0}{SUB10}
\DefMacro{table-lcoc-stats-row9-manual-coc}{169}
\DefMacro{table-lcoc-stats-row9-manual-ioc}{27}
\DefMacro{table-lcoc-stats-row9-generated-coc}{443}
\DefMacro{table-lcoc-stats-row9-generated-ioc}{45}
\DefMacro{table-lcoc-stats-row9-pattern-coc}{47}
\DefMacro{table-lcoc-stats-row9-pattern-ioc}{11}
\DefMacro{table-lcoc-stats-row9-reduction-coc}{72.19}
\DefMacro{table-lcoc-stats-row9-reduction-ioc}{59.26}
\DefMacro{table-lcoc-stats-row10-names-0}{SUB11}
\DefMacro{table-lcoc-stats-row10-manual-coc}{169}
\DefMacro{table-lcoc-stats-row10-manual-ioc}{27}
\DefMacro{table-lcoc-stats-row10-generated-coc}{443}
\DefMacro{table-lcoc-stats-row10-generated-ioc}{45}
\DefMacro{table-lcoc-stats-row10-pattern-coc}{47}
\DefMacro{table-lcoc-stats-row10-pattern-ioc}{11}
\DefMacro{table-lcoc-stats-row10-reduction-coc}{72.19}
\DefMacro{table-lcoc-stats-row10-reduction-ioc}{59.26}
\DefMacro{table-lcoc-stats-row11-names-0}{SUB12}
\DefMacro{table-lcoc-stats-row11-manual-coc}{169}
\DefMacro{table-lcoc-stats-row11-manual-ioc}{27}
\DefMacro{table-lcoc-stats-row11-generated-coc}{443}
\DefMacro{table-lcoc-stats-row11-generated-ioc}{45}
\DefMacro{table-lcoc-stats-row11-pattern-coc}{47}
\DefMacro{table-lcoc-stats-row11-pattern-ioc}{11}
\DefMacro{table-lcoc-stats-row11-reduction-coc}{72.19}
\DefMacro{table-lcoc-stats-row11-reduction-ioc}{59.26}
\DefMacro{table-lcoc-stats-row12-names-0}{SUB13}
\DefMacro{table-lcoc-stats-row12-manual-coc}{179}
\DefMacro{table-lcoc-stats-row12-manual-ioc}{26}
\DefMacro{table-lcoc-stats-row12-generated-coc}{319}
\DefMacro{table-lcoc-stats-row12-generated-ioc}{32}
\DefMacro{table-lcoc-stats-row12-pattern-coc}{71}
\DefMacro{table-lcoc-stats-row12-pattern-ioc}{15}
\DefMacro{table-lcoc-stats-row12-reduction-coc}{60.34}
\DefMacro{table-lcoc-stats-row12-reduction-ioc}{42.31}
\DefMacro{table-lcoc-stats-row13-names-0}{SUB14}
\DefMacro{table-lcoc-stats-row13-manual-coc}{611}
\DefMacro{table-lcoc-stats-row13-manual-ioc}{87}
\DefMacro{table-lcoc-stats-row13-generated-coc}{1,942}
\DefMacro{table-lcoc-stats-row13-generated-ioc}{196}
\DefMacro{table-lcoc-stats-row13-pattern-coc}{188}
\DefMacro{table-lcoc-stats-row13-pattern-ioc}{48}
\DefMacro{table-lcoc-stats-row13-reduction-coc}{69.23}
\DefMacro{table-lcoc-stats-row13-reduction-ioc}{44.83}
\DefMacro{table-lcoc-stats-row13-names-1}{SUB15}
\DefMacro{table-lcoc-stats-row13-names-2}{SUB16}
\DefMacro{table-lcoc-stats-row13-names-3}{SUB17}
\DefMacro{table-lcoc-stats-row14-names-0}{SUB18}
\DefMacro{table-lcoc-stats-row14-manual-coc}{338}
\DefMacro{table-lcoc-stats-row14-manual-ioc}{48}
\DefMacro{table-lcoc-stats-row14-generated-coc}{344}
\DefMacro{table-lcoc-stats-row14-generated-ioc}{36}
\DefMacro{table-lcoc-stats-row14-pattern-coc}{38}
\DefMacro{table-lcoc-stats-row14-pattern-ioc}{5}
\DefMacro{table-lcoc-stats-row14-reduction-coc}{88.76}
\DefMacro{table-lcoc-stats-row14-reduction-ioc}{89.58}
\DefMacro{table-lcoc-stats-row15-names-0}{ADD1}
\DefMacro{table-lcoc-stats-row15-manual-coc}{284}
\DefMacro{table-lcoc-stats-row15-manual-ioc}{40}
\DefMacro{table-lcoc-stats-row15-generated-coc}{443}
\DefMacro{table-lcoc-stats-row15-generated-ioc}{50}
\DefMacro{table-lcoc-stats-row15-pattern-coc}{83}
\DefMacro{table-lcoc-stats-row15-pattern-ioc}{13}
\DefMacro{table-lcoc-stats-row15-reduction-coc}{70.77}
\DefMacro{table-lcoc-stats-row15-reduction-ioc}{67.50}
\DefMacro{table-lcoc-stats-row16-names-0}{ADD2}
\DefMacro{table-lcoc-stats-row16-manual-coc}{290}
\DefMacro{table-lcoc-stats-row16-manual-ioc}{38}
\DefMacro{table-lcoc-stats-row16-generated-coc}{503}
\DefMacro{table-lcoc-stats-row16-generated-ioc}{51}
\DefMacro{table-lcoc-stats-row16-pattern-coc}{60}
\DefMacro{table-lcoc-stats-row16-pattern-ioc}{14}
\DefMacro{table-lcoc-stats-row16-reduction-coc}{79.31}
\DefMacro{table-lcoc-stats-row16-reduction-ioc}{63.16}
\DefMacro{table-lcoc-stats-row17-names-0}{ADD3}
\DefMacro{table-lcoc-stats-row17-manual-coc}{173}
\DefMacro{table-lcoc-stats-row17-manual-ioc}{25}
\DefMacro{table-lcoc-stats-row17-generated-coc}{888}
\DefMacro{table-lcoc-stats-row17-generated-ioc}{90}
\DefMacro{table-lcoc-stats-row17-pattern-coc}{94}
\DefMacro{table-lcoc-stats-row17-pattern-ioc}{22}
\DefMacro{table-lcoc-stats-row17-reduction-coc}{45.66}
\DefMacro{table-lcoc-stats-row17-reduction-ioc}{12.00}
\DefMacro{table-lcoc-stats-row17-names-1}{ADD4}
\DefMacro{table-lcoc-stats-row18-names-0}{ADD5}
\DefMacro{table-lcoc-stats-row18-manual-coc}{173}
\DefMacro{table-lcoc-stats-row18-manual-ioc}{25}
\DefMacro{table-lcoc-stats-row18-generated-coc}{888}
\DefMacro{table-lcoc-stats-row18-generated-ioc}{90}
\DefMacro{table-lcoc-stats-row18-pattern-coc}{94}
\DefMacro{table-lcoc-stats-row18-pattern-ioc}{22}
\DefMacro{table-lcoc-stats-row18-reduction-coc}{45.66}
\DefMacro{table-lcoc-stats-row18-reduction-ioc}{12.00}
\DefMacro{table-lcoc-stats-row18-names-1}{ADD6}
\DefMacro{table-lcoc-stats-row19-names-0}{ADD7}
\DefMacro{table-lcoc-stats-row19-manual-coc}{173}
\DefMacro{table-lcoc-stats-row19-manual-ioc}{25}
\DefMacro{table-lcoc-stats-row19-generated-coc}{443}
\DefMacro{table-lcoc-stats-row19-generated-ioc}{45}
\DefMacro{table-lcoc-stats-row19-pattern-coc}{47}
\DefMacro{table-lcoc-stats-row19-pattern-ioc}{11}
\DefMacro{table-lcoc-stats-row19-reduction-coc}{72.83}
\DefMacro{table-lcoc-stats-row19-reduction-ioc}{56.00}
\DefMacro{table-lcoc-stats-row20-names-0}{ADD8}
\DefMacro{table-lcoc-stats-row20-manual-coc}{173}
\DefMacro{table-lcoc-stats-row20-manual-ioc}{25}
\DefMacro{table-lcoc-stats-row20-generated-coc}{443}
\DefMacro{table-lcoc-stats-row20-generated-ioc}{45}
\DefMacro{table-lcoc-stats-row20-pattern-coc}{47}
\DefMacro{table-lcoc-stats-row20-pattern-ioc}{11}
\DefMacro{table-lcoc-stats-row20-reduction-coc}{72.83}
\DefMacro{table-lcoc-stats-row20-reduction-ioc}{56.00}
\DefMacro{table-lcoc-stats-row21-names-0}{ADD9}
\DefMacro{table-lcoc-stats-row21-manual-coc}{141}
\DefMacro{table-lcoc-stats-row21-manual-ioc}{21}
\DefMacro{table-lcoc-stats-row21-generated-coc}{290}
\DefMacro{table-lcoc-stats-row21-generated-ioc}{31}
\DefMacro{table-lcoc-stats-row21-pattern-coc}{38}
\DefMacro{table-lcoc-stats-row21-pattern-ioc}{8}
\DefMacro{table-lcoc-stats-row21-reduction-coc}{73.05}
\DefMacro{table-lcoc-stats-row21-reduction-ioc}{61.90}
\DefMacro{table-lcoc-stats-row22-names-0}{ADD10}
\DefMacro{table-lcoc-stats-row22-manual-coc}{141}
\DefMacro{table-lcoc-stats-row22-manual-ioc}{21}
\DefMacro{table-lcoc-stats-row22-generated-coc}{290}
\DefMacro{table-lcoc-stats-row22-generated-ioc}{31}
\DefMacro{table-lcoc-stats-row22-pattern-coc}{38}
\DefMacro{table-lcoc-stats-row22-pattern-ioc}{8}
\DefMacro{table-lcoc-stats-row22-reduction-coc}{73.05}
\DefMacro{table-lcoc-stats-row22-reduction-ioc}{61.90}
\DefMacro{table-lcoc-stats-row23-names-0}{ADD11}
\DefMacro{table-lcoc-stats-row23-manual-coc}{462}
\DefMacro{table-lcoc-stats-row23-manual-ioc}{67}
\DefMacro{table-lcoc-stats-row23-generated-coc}{563}
\DefMacro{table-lcoc-stats-row23-generated-ioc}{65}
\DefMacro{table-lcoc-stats-row23-pattern-coc}{115}
\DefMacro{table-lcoc-stats-row23-pattern-ioc}{22}
\DefMacro{table-lcoc-stats-row23-reduction-coc}{75.11}
\DefMacro{table-lcoc-stats-row23-reduction-ioc}{67.16}
\DefMacro{table-lcoc-stats-row24-names-0}{ADD12}
\DefMacro{table-lcoc-stats-row24-manual-coc}{609}
\DefMacro{table-lcoc-stats-row24-manual-ioc}{85}
\DefMacro{table-lcoc-stats-row24-generated-coc}{1,942}
\DefMacro{table-lcoc-stats-row24-generated-ioc}{196}
\DefMacro{table-lcoc-stats-row24-pattern-coc}{188}
\DefMacro{table-lcoc-stats-row24-pattern-ioc}{48}
\DefMacro{table-lcoc-stats-row24-reduction-coc}{69.13}
\DefMacro{table-lcoc-stats-row24-reduction-ioc}{43.53}
\DefMacro{table-lcoc-stats-row24-names-1}{ADD13}
\DefMacro{table-lcoc-stats-row24-names-2}{ADD14}
\DefMacro{table-lcoc-stats-row24-names-3}{ADD15}
\DefMacro{table-lcoc-stats-row25-names-0}{ADD16}
\DefMacro{table-lcoc-stats-row25-manual-coc}{710}
\DefMacro{table-lcoc-stats-row25-manual-ioc}{78}
\DefMacro{table-lcoc-stats-row25-generated-coc}{1,530}
\DefMacro{table-lcoc-stats-row25-generated-ioc}{152}
\DefMacro{table-lcoc-stats-row25-pattern-coc}{416}
\DefMacro{table-lcoc-stats-row25-pattern-ioc}{44}
\DefMacro{table-lcoc-stats-row25-reduction-coc}{41.41}
\DefMacro{table-lcoc-stats-row25-reduction-ioc}{43.59}
\DefMacro{table-lcoc-stats-row25-names-1}{ADD17}
\DefMacro{table-lcoc-stats-row26-names-0}{ADD18}
\DefMacro{table-lcoc-stats-row26-manual-coc}{204}
\DefMacro{table-lcoc-stats-row26-manual-ioc}{29}
\DefMacro{table-lcoc-stats-row26-generated-coc}{338}
\DefMacro{table-lcoc-stats-row26-generated-ioc}{37}
\DefMacro{table-lcoc-stats-row26-pattern-coc}{34}
\DefMacro{table-lcoc-stats-row26-pattern-ioc}{5}
\DefMacro{table-lcoc-stats-row26-reduction-coc}{83.33}
\DefMacro{table-lcoc-stats-row26-reduction-ioc}{82.76}
\DefMacro{table-lcoc-stats-row27-names-0}{OR1}
\DefMacro{table-lcoc-stats-row27-manual-coc}{360}
\DefMacro{table-lcoc-stats-row27-manual-ioc}{38}
\DefMacro{table-lcoc-stats-row27-generated-coc}{1,594}
\DefMacro{table-lcoc-stats-row27-generated-ioc}{157}
\DefMacro{table-lcoc-stats-row27-pattern-coc}{371}
\DefMacro{table-lcoc-stats-row27-pattern-ioc}{44}
\DefMacro{table-lcoc-stats-row27-reduction-coc}{-3.06}
\DefMacro{table-lcoc-stats-row27-reduction-ioc}{-15.79}
\DefMacro{table-lcoc-stats-row27-names-1}{OR3}
\DefMacro{table-lcoc-stats-row28-names-0}{OR2}
\DefMacro{table-lcoc-stats-row28-manual-coc}{351}
\DefMacro{table-lcoc-stats-row28-manual-ioc}{38}
\DefMacro{table-lcoc-stats-row28-generated-coc}{1,598}
\DefMacro{table-lcoc-stats-row28-generated-ioc}{157}
\DefMacro{table-lcoc-stats-row28-pattern-coc}{375}
\DefMacro{table-lcoc-stats-row28-pattern-ioc}{44}
\DefMacro{table-lcoc-stats-row28-reduction-coc}{-6.84}
\DefMacro{table-lcoc-stats-row28-reduction-ioc}{-15.79}
\DefMacro{table-lcoc-stats-row28-names-1}{OR4}
\DefMacro{table-lcoc-stats-row29-names-0}{MIN1}
\DefMacro{table-lcoc-stats-row29-manual-coc}{160}
\DefMacro{table-lcoc-stats-row29-manual-ioc}{25}
\DefMacro{table-lcoc-stats-row29-generated-coc}{292}
\DefMacro{table-lcoc-stats-row29-generated-ioc}{30}
\DefMacro{table-lcoc-stats-row29-pattern-coc}{83}
\DefMacro{table-lcoc-stats-row29-pattern-ioc}{19}
\DefMacro{table-lcoc-stats-row29-reduction-coc}{48.12}
\DefMacro{table-lcoc-stats-row29-reduction-ioc}{24.00}
\DefMacro{table-lcoc-stats-row30-names-0}{AND1}
\DefMacro{table-lcoc-stats-row30-manual-coc}{182}
\DefMacro{table-lcoc-stats-row30-manual-ioc}{25}
\DefMacro{table-lcoc-stats-row30-generated-coc}{327}
\DefMacro{table-lcoc-stats-row30-generated-ioc}{36}
\DefMacro{table-lcoc-stats-row30-pattern-coc}{33}
\DefMacro{table-lcoc-stats-row30-pattern-ioc}{5}
\DefMacro{table-lcoc-stats-row30-reduction-coc}{81.87}
\DefMacro{table-lcoc-stats-row30-reduction-ioc}{80.00}
\DefMacro{table-lcoc-stats-row31-names-0}{LSHIFT1}
\DefMacro{table-lcoc-stats-row31-manual-coc}{326}
\DefMacro{table-lcoc-stats-row31-manual-ioc}{38}
\DefMacro{table-lcoc-stats-row31-generated-coc}{490}
\DefMacro{table-lcoc-stats-row31-generated-ioc}{54}
\DefMacro{table-lcoc-stats-row31-pattern-coc}{106}
\DefMacro{table-lcoc-stats-row31-pattern-ioc}{15}
\DefMacro{table-lcoc-stats-row31-reduction-coc}{67.48}
\DefMacro{table-lcoc-stats-row31-reduction-ioc}{60.53}
\DefMacro{table-lcoc-stats-row32-names-0}{LSHIFT2}
\DefMacro{table-lcoc-stats-row32-manual-coc}{165}
\DefMacro{table-lcoc-stats-row32-manual-ioc}{19}
\DefMacro{table-lcoc-stats-row32-generated-coc}{743}
\DefMacro{table-lcoc-stats-row32-generated-ioc}{80}
\DefMacro{table-lcoc-stats-row32-pattern-coc}{119}
\DefMacro{table-lcoc-stats-row32-pattern-ioc}{20}
\DefMacro{table-lcoc-stats-row32-reduction-coc}{27.88}
\DefMacro{table-lcoc-stats-row32-reduction-ioc}{-5.26}
\DefMacro{table-lcoc-stats-row32-names-1}{LSHIFT3}
\DefMacro{table-lcoc-stats-row33-names-0}{LSHIFT4}
\DefMacro{table-lcoc-stats-row33-manual-coc}{279}
\DefMacro{table-lcoc-stats-row33-manual-ioc}{34}
\DefMacro{table-lcoc-stats-row33-generated-coc}{1,059}
\DefMacro{table-lcoc-stats-row33-generated-ioc}{104}
\DefMacro{table-lcoc-stats-row33-pattern-coc}{139}
\DefMacro{table-lcoc-stats-row33-pattern-ioc}{26}
\DefMacro{table-lcoc-stats-row33-reduction-coc}{50.18}
\DefMacro{table-lcoc-stats-row33-reduction-ioc}{23.53}
\DefMacro{table-lcoc-stats-row33-names-1}{LSHIFT5}
\DefMacro{table-lcoc-stats-row34-names-0}{LSHIFT6}
\DefMacro{table-lcoc-stats-row34-manual-coc}{264}
\DefMacro{table-lcoc-stats-row34-manual-ioc}{28}
\DefMacro{table-lcoc-stats-row34-generated-coc}{484}
\DefMacro{table-lcoc-stats-row34-generated-ioc}{52}
\DefMacro{table-lcoc-stats-row34-pattern-coc}{92}
\DefMacro{table-lcoc-stats-row34-pattern-ioc}{14}
\DefMacro{table-lcoc-stats-row34-reduction-coc}{65.15}
\DefMacro{table-lcoc-stats-row34-reduction-ioc}{50.00}
\DefMacro{table-lcoc-stats-row35-names-0}{RSHIFT1}
\DefMacro{table-lcoc-stats-row35-manual-coc}{353}
\DefMacro{table-lcoc-stats-row35-manual-ioc}{42}
\DefMacro{table-lcoc-stats-row35-generated-coc}{479}
\DefMacro{table-lcoc-stats-row35-generated-ioc}{53}
\DefMacro{table-lcoc-stats-row35-pattern-coc}{97}
\DefMacro{table-lcoc-stats-row35-pattern-ioc}{14}
\DefMacro{table-lcoc-stats-row35-reduction-coc}{72.52}
\DefMacro{table-lcoc-stats-row35-reduction-ioc}{66.67}
\DefMacro{table-lcoc-stats-row36-names-0}{URSHIFT1}
\DefMacro{table-lcoc-stats-row36-manual-coc}{303}
\DefMacro{table-lcoc-stats-row36-manual-ioc}{32}
\DefMacro{table-lcoc-stats-row36-generated-coc}{470}
\DefMacro{table-lcoc-stats-row36-generated-ioc}{55}
\DefMacro{table-lcoc-stats-row36-pattern-coc}{90}
\DefMacro{table-lcoc-stats-row36-pattern-ioc}{17}
\DefMacro{table-lcoc-stats-row36-reduction-coc}{70.30}
\DefMacro{table-lcoc-stats-row36-reduction-ioc}{46.88}
\DefMacro{table-lcoc-stats-row37-names-0}{URSHIFT2}
\DefMacro{table-lcoc-stats-row37-manual-coc}{455}
\DefMacro{table-lcoc-stats-row37-manual-ioc}{55}
\DefMacro{table-lcoc-stats-row37-generated-coc}{677}
\DefMacro{table-lcoc-stats-row37-generated-ioc}{69}
\DefMacro{table-lcoc-stats-row37-pattern-coc}{90}
\DefMacro{table-lcoc-stats-row37-pattern-ioc}{16}
\DefMacro{table-lcoc-stats-row37-reduction-coc}{80.22}
\DefMacro{table-lcoc-stats-row37-reduction-ioc}{70.91}
\DefMacro{table-lcoc-stats-row38-names-0}{URSHIFT3}
\DefMacro{table-lcoc-stats-row38-manual-coc}{339}
\DefMacro{table-lcoc-stats-row38-manual-ioc}{41}
\DefMacro{table-lcoc-stats-row38-generated-coc}{472}
\DefMacro{table-lcoc-stats-row38-generated-ioc}{53}
\DefMacro{table-lcoc-stats-row38-pattern-coc}{83}
\DefMacro{table-lcoc-stats-row38-pattern-ioc}{14}
\DefMacro{table-lcoc-stats-row38-reduction-coc}{75.52}
\DefMacro{table-lcoc-stats-row38-reduction-ioc}{65.85}
\DefMacro{table-lcoc-stats-row39-names-0}{URSHIFT4}
\DefMacro{table-lcoc-stats-row39-manual-coc}{246}
\DefMacro{table-lcoc-stats-row39-manual-ioc}{27}
\DefMacro{table-lcoc-stats-row39-generated-coc}{412}
\DefMacro{table-lcoc-stats-row39-generated-ioc}{44}
\DefMacro{table-lcoc-stats-row39-pattern-coc}{64}
\DefMacro{table-lcoc-stats-row39-pattern-ioc}{10}
\DefMacro{table-lcoc-stats-row39-reduction-coc}{73.98}
\DefMacro{table-lcoc-stats-row39-reduction-ioc}{62.96}
\DefMacro{table-lcoc-stats-row40-names-0}{URSHIFT5}
\DefMacro{table-lcoc-stats-row40-manual-coc}{294}
\DefMacro{table-lcoc-stats-row40-manual-ioc}{39}
\DefMacro{table-lcoc-stats-row40-generated-coc}{334}
\DefMacro{table-lcoc-stats-row40-generated-ioc}{34}
\DefMacro{table-lcoc-stats-row40-pattern-coc}{54}
\DefMacro{table-lcoc-stats-row40-pattern-ioc}{8}
\DefMacro{table-lcoc-stats-row40-reduction-coc}{81.63}
\DefMacro{table-lcoc-stats-row40-reduction-ioc}{79.49}
\DefMacro{table-lcoc-stats-row-sum-manual-coc}{11,000}
\DefMacro{table-lcoc-stats-row-sum-manual-ioc}{1,462}
\DefMacro{table-lcoc-stats-row-sum-generated-coc}{24,941}
\DefMacro{table-lcoc-stats-row-sum-generated-ioc}{2,581}
\DefMacro{table-lcoc-stats-row-sum-pattern-coc}{3,987}
\DefMacro{table-lcoc-stats-row-sum-pattern-ioc}{692}
\DefMacro{table-lcoc-stats-row-sum-reduction-coc}{63.75}
\DefMacro{table-lcoc-stats-row-sum-reduction-ioc}{52.67}

\DefMacro{table-lcoc-stats-float-env}{table}
\DefMacro{table-lcoc-stats-caption}{%
Comparison of number of characters and number of identifiers between
hand-written C/C++ code in \OpenJDK and corresponding \ptns written
using \Tool.\label{tab:lcoc-stats}}
\DefMacro{table-lcoc-stats-fontsize}{footnotesize}
\DefMacro{table-lcoc-stats-col-names}{Names}
\DefMacro{table-lcoc-stats-col-manual}{Hand-written}
\DefMacro{table-lcoc-stats-col-generated}{\Tool Generated}
\DefMacro{table-lcoc-stats-col-pattern}{\Tool \Ptn}
\DefMacro{table-lcoc-stats-col-coc}{\#Chars}
\DefMacro{table-lcoc-stats-col-ioc}{\#Ids}
\DefMacro{table-lcoc-stats-col-reduction}{Reduction (\%)}

\DefMacro{figure-performance-speedup-manual-relative-to-none-caption}{Performance speedup of hand-written code relative to none.\label{fig:performance-speedup-manual-relative-to-none}}
\DefMacro{figure-performance-speedup-generated-relative-to-manual-caption}{Performance comparison of generated code relative to hand-written code of \OpenJDK \optimizations on \Renaissance benchmarks.\label{fig:performance-speedup-generated-relative-to-manual}\vspace{-15pt}}

\DefMacro{figure-example}{An example of a peephole optimization as implemented in \OpenJDK and \Tool, and associated test.\label{fig:example}}
\DefMacro{figure-example-test-code}{An example \JIT test available in \OpenJDK.\label{fig:example:test-code}}
\DefMacro{figure-example-pattern-code}{\Ptn written using \Tool.\label{fig:example:pattern-code}} %
\DefMacro{figure-example-manual-code}{Hand-written code (with bug) in OpenJDK.\label{fig:example:manual-code}}
\DefMacro{figure-example-generated-code}{Code generated from \Tool.\label{fig:example:generated-code}}

\DefMacro{figure-syntax}{\Tool Syntax. The non-terminals that are not defined in the figure share the same definition as Java~\cite{JavaLangSpecification}.\label{fig:syntax}\vspace{-5pt}}

\DefMacro{figure-api-list}{List of API annotations and methods.\label{fig:api-list}}

\DefMacro{figure-example-asts}{\eASTs for \ptn \UseMacro{ptn-pAdd6} in Figure \ref{fig:example:pattern-code}.\label{fig:example-asts}}
\DefMacro{figure-example-asts-before}{\eAST of \CodeIn{before} expression.\label{fig:example-asts:before}}
\DefMacro{figure-example-asts-after}{\eAST of \CodeIn{after} expression.\label{fig:example-asts:after}}

\DefMacro{figure-example-shadow}{Illustration of \shadow detecting algorithm on \ptn \UseMacro{ptn-pAdd2} ($X$) \shadowing \UseMacro{ptn-pAdd5} ($Y$). Every \eAST node is represented by a variable, and the dashed lines connect equivalent nodes (e.g., $x_1$ is equivalent to $y_1$). The formula $F$ shows the final SMT formula that specifies the shadow relation. \label{fig:example-shadow}\vspace{-5pt}}

\DefMacro{figure-count-ids}{Example of identifier counting.\label{fig:count-ids}}
\DefMacro{figure-count-ids-pattern}{\Ptn code written using \Tool (Figure~\ref{fig:example:pattern-code}).\label{fig:count-ids-pattern}}
\DefMacro{figure-count-ids-manual}{Hand-written code in OpenJDK (Figure~\ref{fig:example:manual-code}).\label{fig:count-ids-manual}}

\DefMacro{figure-pr-patterns}{Associated \ptns in pull requests.\label{fig:pr-patterns}}
\DefMacro{figure-pr-patterns-pNewSubAddSub1574}{\Ptn \UseMacro{ptn-pNewSubAddSub1574}.\label{fig:pr-patterns:pNewSubAddSub1574}}
\DefMacro{figure-pr-patterns-pNewAddAddSub1156}{\Ptn \UseMacro{ptn-pNewAddAddSub1156}.\label{fig:pr-patterns:pNewAddAddSub1156}}
\DefMacro{figure-pr-patterns-pNewAddAddSub1202}{\Ptn \UseMacro{ptn-pNewAddAddSub1202}.\label{fig:pr-patterns:pNewAddAddSub1202}}
\DefMacro{figure-pr-patterns-pNewSubAddSub1564}{\Ptn \UseMacro{ptn-pNewSubAddSub1564}.\label{fig:pr-patterns:pNewSubAddSub1564}}
\DefMacro{figure-pr-patterns-pNewSub_XOrY_Minus_XXorY_}{\Ptn \UseMacro{ptn-pNewSub_XOrY_Minus_XXorY_}.\label{fig:pr-patterns:pNewSub_XOrY_Minus_XXorY_}}
\DefMacro{figure-pr-patterns-pNewXPlus_ConMinusY_}{\Ptn \UseMacro{ptn-pNewXPlus_ConMinusY_}.\label{fig:pr-patterns:pNewXPlus_ConMinusY_}}
\DefMacro{figure-pr-patterns-pNewXPlus_ConMinusY_Sym}{\Ptn \UseMacro{ptn-pNewXPlus_ConMinusY_Sym}.\label{fig:pr-patterns:pNewXPlus_ConMinusY_Sym}}
\DefMacro{figure-pr-patterns-pAdd2}{\Ptn \UseMacro{ptn-pAdd2}.\label{fig:pr-patterns:pAdd2}}
\DefMacro{figure-pr-patterns-pAdd5}{\Ptn \UseMacro{ptn-pAdd5}.\label{fig:pr-patterns:pAdd5}}
\DefMacro{figure-pr-patterns-pAdd6}{\Ptn \UseMacro{ptn-pAdd6}.\label{fig:pr-patterns:pAdd6}}

\DefMacro{algo-shadow-deciding-caption}{\Shadow determining algorithm.\label{algo:shadow-deciding}}
\DefMacro{algo-shadow-deciding-fontsize}{small}
\DefMacro{algo-shadow-deciding-shape-caption}{Auxiliary \shadow determining algorithm (shape).\label{algo:shadow-deciding-shape}}
\DefMacro{algo-shadow-deciding-shape-fontsize}{footnotesize}
\DefMacro{algo-shadow-deciding-value-caption}{Auxiliary \shadow determining algorithm (value).\label{algo:shadow-deciding-value}}
\DefMacro{algo-shadow-deciding-value-fontsize}{footnotesize}
\DefMacro{algo-shadow-deciding-aux-caption}{Auxiliary \shadow determining algorithm (helper).\label{algo:shadow-deciding-aux}}
\DefMacro{algo-shadow-deciding-aux-fontsize}{footnotesize}
\DefMacro{algo-shadow-deciding-full-caption}{Full \shadow determining algorithm.\label{algo:shadow-deciding-full}}
\DefMacro{algo-shadow-deciding-full-fontsize}{footnotesize}

\newcommand{\RQEasierWriting}{RQ1\xspace}
\newcommand{\RQPerformance}{RQ2\xspace}
\newcommand{\RQShadow}{RQ3\xspace}
\newcommand{\RQTestGenerationAndPullRequests}{RQ4\xspace}

\DefMacro{total-character-reduction}{\num[round-mode=places,round-precision=0]{\UseMacro{table-lcoc-stats-row-sum-reduction-coc}}}
\DefMacro{total-identifier-reduction}{\num[round-mode=places,round-precision=0]{\UseMacro{table-lcoc-stats-row-sum-reduction-ioc}}}

\DefMacro{shadow-timeout-per-pair}{2} %
\DefMacro{total-shadows}{\UseMacro{table-shadow-stats-num-rows}}

\DefMacro{total-measured-repetitions}{50}
\DefMacro{num-end-to-end-runs}{5}
\DefMacro{total-repetitions-per-run}{100}
\DefMacro{measured-repetitions-per-run}{10}

\DefMacro{total-num-patterns}{\UseMacro{table-pattern-stats-summary-row0-num-patterns}}
\DefMacro{total-num-patterns-OpenJDK}{\UseMacro{table-pattern-stats-summary-row0-num-OpenJDK}}
\DefMacro{total-num-patterns-LLVM}{\UseMacro{table-pattern-stats-summary-row0-num-LLVM}}
\DefMacro{total-num-patterns-OWN}{\numberstringnum{\UseMacro{table-pattern-stats-summary-row0-num-OWN}}}

\DefMacro{total-num-PRs}{eight} %
\DefMacro{total-num-non-test-PRs}{\numberstringnum{\UseMacro{table-pattern-stats-summary-row0-num-prs}}}
\DefMacro{total-num-integrated-PRs}{seven}
\DefMacro{Total-num-integrated-PRs}{Seven}
\DefMacro{total-num-new-optimization-PRs}{six}
\DefMacro{num-integrated-new-optimization-PRs}{five}
\DefMacro{total-num-shadow-PRs}{one}
\DefMacro{num-integrated-shadow-PRs}{one}
\DefMacro{total-num-test-PRs}{one}

\DefMacro{num-tests-in-test-PR}{10}

\newcommand{\DataURL}{\url{https://github.com/EngineeringSoftware/jog}}

\begin{document}

\title{\Title}

\author{Zhiqiang Zang}
\affiliation{%
\institution{The University of Texas at Austin}
\city{Austin}
\state{Texas}
\country{USA}
}
\email{zhiqiang.zang@utexas.edu}

\author{Aditya Thimmaiah}
\affiliation{%
\institution{The University of Texas at Austin}
\city{Austin}
\state{Texas}
\country{USA}
}
\email{auditt@utexas.edu}

\author{Milos Gligoric}
\affiliation{%
\institution{The University of Texas at Austin}
\city{Austin}
\state{Texas}
\country{USA}
}
\email{gligoric@utexas.edu}

\begin{abstract}
We present \Tool, a framework that facilitates developing Java \JIT
peephole \optimizations alongside \JIT tests.
\Tool enables developers to
write a \ptn, in Java itself, that specifies desired code
transformations by writing code before and after the \optimization,
as well as any necessary preconditions. Such \ptns can be written in
the same way that tests of the \optimization are already written in OpenJDK.
\Tool translates each \ptn into C/C++ code that can be integrated as
a \JIT \optimization pass. \Tool also generates Java tests for
\optimizations from \ptns. Furthermore, \Tool can
automatically detect possible \shadow relation between a pair of
\optimizations where the effect of the \shadowed \optimization is
overridden by another. Our evaluation shows that \Tool makes it
easier to write readable \JIT \optimizations alongside tests without
decreasing the effectiveness of \JIT \optimizations. We wrote
\UseMacro{total-num-patterns} \ptns, including
\UseMacro{total-num-patterns-OpenJDK} existing \optimizations in
\OpenJDK, \UseMacro{total-num-patterns-LLVM} new \optimizations
adapted from \LLVM, and \UseMacro{total-num-patterns-OWN} new
\optimizations that we proposed.
We opened \UseMacro{total-num-PRs} pull requests (PRs) for \OpenJDK,
including \UseMacro{total-num-new-optimization-PRs} for new
\optimizations, \UseMacro{total-num-shadow-PRs} on removing
\shadowed \optimizations, and \UseMacro{total-num-test-PRs} for
newly generated \JIT tests;
\UseMacro{total-num-integrated-PRs} PRs have already been integrated
into the master branch of \OpenJDK.
\end{abstract}



\maketitle

\section{Introduction}
\label{sec:intro}

\emph{Peephole optimization}~\cite{Muchnick97Advanced,
McKeeman65Peephole} is an optimization technique performed on a
small set of instructions (known as a \emph{window}), e.g., \CodeIn{A
+ A} is transformed to \CodeIn{A \char`\<{}\char`\<{} 1}. Popular compilers such as
GCC, LLVM, and Java \JIT, include dozens if not hundreds of peephole
\optimizations~\cite{LinkToGCCPeephole, LinkToLLVMPeephole,
LinkToOpenJDKPeephole}.

Traditionally, each peephole \optimization is implemented as a
\emph{compiler pass}.  Each compiler pass detects windows, i.e., a
sequence of instructions that can be \optimized, and replaces each
window with an equivalent, albeit more efficient, sequence of
instructions.  These implementations are written in the language in
which the compiler is implemented (C/C++ for Java \JIT) and they
leverage compiler infrastructure to detect instructions of interest.
Representation of these instructions inside the compiler
infrastructure is substantially different from code written in the
programming language itself~\cite{Lopes15Alive}.
This disconnect introduces a burden on compiler developers to perform
proper reasoning to detect windows of interest, to do the instruction
mapping from high-level code (what developers write) to low-level
code, and to document their intention.  The process is tedious and error
prone.

\Alive~\cite{Lopes15Alive} was an improvement over the traditional
approach: a developer writes \ptns in a domain specific language (DSL)
over the \emph{intermediate representation} (IR) of the program (LLVM
bitcode) which are then translated into compiler passes.
The DSL used in \Alive is still very much disconnected from code
written in the programming language being optimized (C++).
This disconnect introduces a steep learning curve and lacks most of
common software tools, e.g., an IDE.  \Alive also focused on C++
intricacies and undefined behavior.

Our insight is that \emph{many peephole \optimizations can be
expressed in the programming language that is being \optimized} (e.g.,
Java).
We found the motivation in existing tests for Java \JIT.
Most tests for \JIT \optimizations in \OpenJDK are written in Java and
some of the tests contain Java code that follows specific \emph{\ptns}
so as to trigger the \optimizations under
test~\cite{LinkToOpenJDKPeepholeTests}.
Figure~\ref{fig:example:test-code} shows such a \JIT test
from \OpenJDK, which triggers the peephole \optimization that
transforms \CodeIn{(a - b) + (c - a)} into \CodeIn{c - b}, by
returning \CodeIn{(a - b) + (c - a)} (line
\ref{line:example:before-in-test}).
Such a pattern expresses, in Java code, the window to be recognized by a
specific \optimization.
\emph{We propose to extend this concept to use the \ptns not only to
write tests but to express the entire \optimization, including code
before and after the \optimization.}

\begin{figure}[!t]
\begin{lstlisting}[language=java-pretty]
@Test
@IR(failOn = {IRNode.ADD}) (*@ \label{line:example:IR:start} @*) (*@ \label{line:example:IR:failOn} @*)
@IR(counts = {IRNode.SUB, "1"}) (*@ \label{line:example:IR:end} @*) (*@ \label{line:example:IR:counts} @*)
// Checks (a - b) + (c - a) => (c - b) (*@ \label{line:example:after-in-test} @*)
public long test8(long a, long b, long c) {
  return (a - b) + (c - a); (*@ \label{line:example:before-in-test} @*)
}
\end{lstlisting}
\vspace{-5pt}
\caption{\UseMacro{figure-example-test-code}}
\vspace{-10pt}
\end{figure}

We present \Tool, the first framework that enables developers to write
\optimization \ptns in a high-level language (Java).
Namely, using \Tool, a Java \JIT compiler developer writes
\optimization \ptns as Java statements. \Ptns are type-checked (by the
Java compiler) and automatically translated into compiler passes (by
\Tool). Additionally, Java tests for the \optimizations can be
automatically generated from the \ptns.
Writing \ptns in Java for the Java \JIT compiler ensures that
sequences of statements are meaningful, i.e., windows can indeed
appear in programs (which is not necessarily the case when matching
intermediate representation or compiler abstractions).
Next, writing \ptns in Java simplifies the reasoning behind each
peephole \optimization: what used to be comments documenting an
\optimization inside the Java \JIT compiler for dozens of lines of
code, or what used to be a test that describes how to trigger an
\optimization, becomes a self-documenting \ptn.
Finally, while writing \ptns in \Tool, a developer can use software
engineering tools available for the language (e.g., IDE, linter).
Having \ptns in Java also enables future application of program
equivalence checkers that work on either Java code or bytecode (which
can be easily obtained by compiling \Tool \ptns).

Furthermore, conciseness of \ptns makes it easier to analyze relations
between \optimizations. \Tool automatically detects possible shadow
relation between a pair of \optimizations where the effect of the
shadowed \optimization is overridden by another.
Consider two \optimizations $X$ and $Y$: $X$ transforms \CodeIn{(a -
b) + (c - d)} into \CodeIn{(a + c) - (b + d)} and $Y$ transforms
\CodeIn{(a - b) + (b - c)} into \CodeIn{a - c}, where \CodeIn{a},
\CodeIn{b}, \CodeIn{c}, \CodeIn{d} are all free variables.
Note that any expression matching \CodeIn{(a - b) + (b - c)} ($X$)
also matches \CodeIn{(a - b) + (c - d)} ($Y$), which means $X$ can be
applied wherever $Y$ can be applied, so the effect of $X$ will shadow
$Y$ if $X$ is always applied before $Y$ in a compiler pass. \Tool can
automatically report the shadow relations.

Using \Tool, we wrote a total of \UseMacro{total-num-patterns}
\optimization \ptns, including \UseMacro{total-num-patterns-OpenJDK}
existing \optimizations in \OpenJDK, \UseMacro{total-num-patterns-LLVM}
new \optimizations adapted from LLVM, and
\UseMacro{total-num-patterns-OWN} new \optimizations. Most of the
patterns that we extracted from \OpenJDK were existing tests of the
\optimizations, or they were hand-written as examples in the comments
documenting the C/C++ implementation.
Our most complex \ptn has only
\UseMacro{table-lcoc-stats-row23-pattern-coc} characters in contrast
to the \UseMacro{table-lcoc-stats-row23-manual-coc} characters of its
C/C++ counterpart. Our evaluation shows that generating code from
\ptns using \Tool does not reduce the effectiveness of \JIT
\optimizations. We also identified a bug in existing Java \JIT as one
\optimization was unreachable as a consequence of being shadowed by
another.

Recently, we have opened a group of \UseMacro{total-num-PRs} pull
requests (PRs) for \OpenJDK (\UseMacro{total-num-new-optimization-PRs}
for new \optimizations, \UseMacro{total-num-shadow-PRs} for fixing the
aforementioned shadowed \optimizations, and
\UseMacro{total-num-test-PRs} for new \Tool generated \JIT tests of
existing \optimizations). \UseMacro{Total-num-integrated-PRs} of the
PRs were already accepted and integrated into the master branch; the
remaining PR is under review. We intend on opening PRs on the
remaining \optimizations in the future.

\vspace{5pt}
\noindent
The main contributions of this paper include the following:

\begin{itemize}[topsep=5pt,itemsep=5pt,partopsep=0ex,parsep=0ex,leftmargin=*]
\item We present \Tool, the first framework that allows developers to
specify a Java \JIT peephole \optimization as a \ptn written in Java
itself, extending the existing approach to writing tests for \JIT.
The \ptn is automatically translated into C/C++ code as a \JIT
\optimization pass, and a Java test for the \optimization is
generated from the \ptn as needed.

\item \JIT \optimizations written in \Tool is easier to read and
understand. We translated \UseMacro{total-num-patterns-OpenJDK}
existing \ptns in \OpenJDK. The evaluation shows a
\UseMacro{total-character-reduction}\% reduction in characters of
code and a \UseMacro{total-identifier-reduction}\% reduction in the
number of identifiers in code when writing \optimizations in \Tool
relative to existing hand-written code in \OpenJDK.

\item Code generated from \Tool maintains the effectiveness of \JIT
\optimizations. The evaluation shows that the impact on performance
is minimal on replacing existing hand-written code in \OpenJDK with
\Tool generated code for existing \ptns. We also wrote
\UseMacro{total-num-patterns-LLVM} new \ptns adapted from \LLVM. A
total of \UseMacro{total-num-new-optimization-PRs} PRs on the new
\ptns were opened, of which
\UseMacro{num-integrated-new-optimization-PRs} PRs have been
integrated into the master branch of \OpenJDK.

\item We present an algorithm to determine if one \optimization
\shadows another written in \ptns using \Tool. We ran the algorithm
on all the translated \ptns to detect \shadows between \ptns. We
opened \UseMacro{total-num-shadow-PRs} PR on removing \shadowed
\ptns that has been integrated into the master branch of \OpenJDK.

\item \JIT tests generated from \Tool complements existing test suites
in \OpenJDK. We generated tests for existing \optimizations in
\OpenJDK and opened \UseMacro{total-num-test-PRs} PR to add
\UseMacro{num-tests-in-test-PR} new tests for existing untested
\optimizations in \OpenJDK.
\end{itemize}

\vspace{3pt}
\noindent
We believe that \Tool enables developers to quickly write and evaluate
a large number of peephole \optimizations by writing \ptns in a
familiar programming language and very much similar to the way the
existing tests for \JIT are written.
\Tool is publicly available at \DataURL.

\section{Example}
\label{sec:example}

The IR test, written in Java using IR test
framework~\cite{LinkToIRTestFramework}, is a recommended approach in
\OpenJDK to testing \JIT peephole optimizations.
We already showed such a test in Figure~\ref{fig:example:test-code}.
While the test runs, the method annotated by \CodeIn{@Test}
(\CodeIn{test8}) is compiled by \JIT, with the expression \CodeIn{(a -
b) + (c - a)} \optimized to \CodeIn{c - b}. Then the IR shape of the
compiled method is checked against certain rules specified in
\CodeIn{@IR}
(line~\ref{line:example:IR:start}--\ref{line:example:IR:end}). The
rules verify that the \optimization from \CodeIn{(a - b) + (c - a)} to
\CodeIn{c - b} indeed happens, by checking that the compiled method
must not contain \CodeIn{ADD} node (line~\ref{line:example:IR:failOn})
and must have exactly one \CodeIn{SUB} node
(line~\ref{line:example:IR:counts}).

\begin{figure}[t]
\begin{subfigure}[b]{\linewidth}
\begin{lstlisting}[language=jog]
@Pattern
public void (*@\UseMacro{ptn-pAdd6}@*)(long a, long b, long c) {(*@\label{line:example:pattern:var-decl}@*)
  before((a - b) + (c - a));(*@\label{line:example:pattern:before}@*)
  after(c - b);(*@\label{line:example:pattern:after}@*)
}
\end{lstlisting}
\caption{\UseMacro{figure-example-pattern-code}}
\end{subfigure}
\\
\begin{subfigure}[b]{\linewidth}
\begin{lstlisting}[language=cpp-pretty-diff]
 Node *AddLNode::Ideal(PhaseGVN *phase, bool can_reshape) {... (*@ \label{line:example:cond1} @*)
   Node* in1 = in(1);
   Node* in2 = in(2);
   int op1 = in1->Opcode();
   int op2 = in2->Opcode();
   if (op1 == Op_SubL) {... (*@ \label{line:example:cond2} @*)
     // Convert "(a-b)+(c-a)" into "(c-b)"
-    if (op2 == Op_SubL && in1->in(1) == in1->in(2)) {(*@ \label{line:example:incorrect} @*)
+    if (op2 == Op_SubL && in1->in(1) == in2->in(2)) {(*@ \label{line:example:correct} @*)(*@ \label{line:example:cond3-4} @*)
       return new SubLNode(in2->in(1), in1->in(2));
     }
   }...
 }
\end{lstlisting}
\caption{\UseMacro{figure-example-manual-code}}
\end{subfigure}
\\
\begin{subfigure}[b]{\linewidth}
\begin{lstlisting}[language=cpp-pretty]
Node *AddLNode::Ideal(PhaseGVN *phase, bool can_reshape) {...
  Node* _JOG_in1 = in(1); (*@ \label{line:example:generated:decl:start} @*)
  Node* _JOG_in11 = _JOG_in1 != NULL && 1 < _JOG_in1->req() ?
                      _JOG_in1->in(1) : NULL;
  Node* _JOG_in12 = _JOG_in1 != NULL && 2 < _JOG_in1->req() ?
                      _JOG_in1->in(2) : NULL;
  Node* _JOG_in2 = in(2);
  Node* _JOG_in21 = _JOG_in2 != NULL && 1 < _JOG_in2->req() ?
                      _JOG_in2->in(1) : NULL;
  Node* _JOG_in22 = _JOG_in2 != NULL && 2 < _JOG_in2->req() ?
                      _JOG_in2->in(2) : NULL; (*@ \label{line:example:generated:decl:end} @*)
  if (_JOG_in1->Opcode() == Op_SubL (*@ \label{line:example:generated:operator-check:left} @*) (*@ \label{line:example:generated:if:start} @*)
      && _JOG_in2->Opcode() == Op_SubL (*@ \label{line:example:generated:operator-check:right} @*)
      && _JOG_in11 == _JOG_in22) { (*@ \label{line:example:generated:sharing-child-check} @*)
    return new SubLNode(_JOG_in21, _JOG_in12); (*@ \label{line:example:generated:return} @*)
  }... (*@ \label{line:example:generated:if:end} @*)
}
\end{lstlisting}
\caption{\UseMacro{figure-example-generated-code}}
\end{subfigure}
\caption{\UseMacro{figure-example}}
\end{figure}

Using \Tool, developers can write the \optimization under test in the
same way as in the already existing test.
Figure~\ref{fig:example:pattern-code} shows a \ptn written using \Tool
that expresses the \optimization, which is a Java method annotated
with \CodeIn{@Pattern}.
The parameters of the method declare all the variables
(line~\ref{line:example:pattern:var-decl} in
Figure~\ref{fig:example:pattern-code}), \CodeIn{a}, \CodeIn{b}, and
\CodeIn{c}, that are used in the method body.
Parameter type \CodeIn{long} indicates the data type involved in the
\optimization.
Two API calls inside the method body, \CodeIn{before((a - b) + (c -
a))} (line~\ref{line:example:pattern:before} in
Figure~\ref{fig:example:pattern-code}) and \CodeIn{after(c - b)}
(line~\ref{line:example:pattern:after} in
Figure~\ref{fig:example:pattern-code}), specify the matched expression
before the \optimization and the transformed expression after the
\optimization, respectively. Both \CodeIn{before} and \CodeIn{after}
API call are written in the same way as the existing test is written.
\CodeIn{before((a - b) + (c - a))} directly uses existing test code
from \CodeIn{return (a - b) + (c - a);}
(line~\ref{line:example:before-in-test} in
Figure~\ref{fig:example:test-code}), and \CodeIn{after(c - b)} is
extracted from the comment \CodeIn{// Check (a - b) + (c - a) => (c -
b)} (line~\ref{line:example:after-in-test} in
Figure~\ref{fig:example:test-code}).

Because the \ptn and the test are written in the same way, not only
does \Tool provide an intuitive way to express an \optimization,
without writing any extra code, but also it can automatically generate
the test from the \ptn. First the {\CodeIn{@Test}} method declares
exactly the same free variables as the \ptn (\CodeIn{long a, long b,
long c}), and returns exactly \CodeIn{before}'s expression in the
\ptn (\CodeIn{return (a - b) + (c - a)};).
Next \Tool analyzes \CodeIn{before((a - b) + (c - a))} and
\CodeIn{after(c - b)} in the \ptn, (1)~to find in \CodeIn{after} the
numbers of operators (one \CodeIn{SUB}) and (2)~to find which
operators disappear from \CodeIn{before} to \CodeIn{after}
(\CodeIn{ADD}). \Tool then maps the operators to the corresponding IR
node types used in IR tests and makes \CodeIn{@IR} annotations
(\CodeIn{@IR({counts = {IRNode.SUB, ``1''}})} and \CodeIn{@IR({failOn
= {IRNode.ADD}})}). Eventually the exactly same test as shown in
Figure~\ref{fig:example:test-code} can be generated from the \Tool \ptn.

More importantly, \Tool automatically translates a \ptn into the C/C++
code that can be directly included in a \JIT optimization pass.
Figure~\ref{fig:example:generated-code} shows the C/C++ code
translated by \Tool from the \ptn, and
Figure~\ref{fig:example:manual-code} shows the hand-written code
extracted from \OpenJDK,
that implements the same \JIT peephole \optimization that transforms
\CodeIn{(a - b) + (c - a)} into \CodeIn{c - b}.
The implementation contains two steps: (A)~match any expression that
is of interest to the \optimization and (B)~return a new \optimized
equivalent expression. In this example, any matched expression
satisfies the following four conditions: (1)~the expression is an
addition expression (implicitly line~\ref{line:example:cond1} in
Figure~\ref{fig:example:manual-code} because the method works only
inside an additive expression); (2)~the left operand (\CodeIn{a - b})
is a subtraction expression (line~\ref{line:example:cond2} in
Figure~\ref{fig:example:manual-code}); (3)~the right operand
(\CodeIn{c - a}) is also a subtraction expression
(line~\ref{line:example:cond3-4} in
Figure~\ref{fig:example:manual-code}); (4)~the left operand of the
left sub-expression (\CodeIn{a}) is equal to the right operand of the
right sub-expression (\CodeIn{a} again)
(line~\ref{line:example:cond3-4} in
Figure~\ref{fig:example:manual-code}). After a match is found, the
code constructs a new subtraction expression (\CodeIn{c - b}) using
\CodeIn{b} and \CodeIn{c}, and returns it.
The transformation reduces the cost of evaluating the expression by
two operations, from two subtractions and one addition to only one
subtraction. Interestingly, this code has a bug (in \OpenJDK) because
of wrong access to the right operand of the right sub-expression,
which is supposed to be \CodeIn{in2->in(2)} while developers wrote it
as \CodeIn{in1->in(2)}.
It took 13 years to discover and fix the bug;
line~\ref{line:example:incorrect} was inserted in 2008 and had been
not touched until 2021~\cite{BugInExample}.
If the optimization had rather been implemented using \Tool, the bug could
have been avoided.

\Tool reads from \CodeIn{before} and \CodeIn{after} APIs the
expressions to match and return, respectively. \Tool analyzes the
expressions to infer the conditions to check and to infer the new
expression to construct, and eventually assembles everything in C/C++
code as the output.
Figure~\ref{fig:example:generated-code} shows the code generated from
the \ptn in Figure~\ref{fig:example:pattern-code}.
The generated code keeps the
same functionality while avoiding the bug in the hand-written code of
Figure~\ref{fig:example:manual-code}.

\section{\Tool Framework}
\label{sec:framework}

This section describes the \Tool framework in detail. We describe the
syntax for writing \ptns, semantics of the statements, translation
details, and test generation from \ptns.

\subsection{Syntax}

Figure~\ref{fig:syntax} defines the syntax of \Tool, which is a subset
of Java (for non-terminals that are not defined in the figure, please
refer to the Java grammar~\cite{JavaLangSpecification}).
Every \optimization is written as a method in Java, which we call a
\emph{\ptn}. The method body contains several statements. Each
statement can be \CodeIn{BeforeStmt}, \CodeIn{AfterStmt}, conditional
or assignment. We introduce \CodeIn{BeforeStmt} to specify the
expression that a \ptn has to match, and we introduce
\CodeIn{AfterStmt} to specify the \optimized expression as a result of
applying the \optimization.

\begin{figure}[t]
\begin{lstlisting}[language=syntax]
Pattern := MethodModifier* MethodHeader MethodBody
MethodHeader := "void" Identifier "(" FormalParameterList ")"
MethodBody := "{" Stmt* "}"
Stmt := BeforeStmt | AfterStmt | IfStmt | AssignStmt
BeforeStmt := before "(" expression ")" ;
AfterStmt := after "(" expression ")" ;
\end{lstlisting}
\caption{\UseMacro{figure-syntax}}
\end{figure}

\subsection{Semantics}
The parameters of the method declare the variables used in the \ptn.
There are two types of variables: \emph{constant} and \emph{free}.
A constant variable represents a literal (e.g., 42); a free variable
represents any expression, including literals. A parameter declares a
free variable unless explicitly declared as a constant variable.

The semantics of a \ptn is a peephole \optimization that transforms a
certain set of instructions into another set of instructions. Thus, a
\ptn must have one \CodeIn{BeforeStmt} and one \CodeIn{AfterStmt}.
Both statements contain an expression. The expression inside
\CodeIn{BeforeStmt} defines the set of instructions that can be
transformed by the \optimization, and the expression inside
\CodeIn{AfterStmt} defines the set of instructions as a result of the
\optimization. It is possible that the \optimization is supposed to be
applied only under certain preconditions. Any necessary precondition
can be specified as the condition of an \CodeIn{IfStmt}, and either
\CodeIn{BeforeStmt} or \CodeIn{AfterStmt} can be included in the
``then'' branch of such \CodeIn{IfStmt}. To ensure every pair of
\CodeIn{BeforeStmt} and \CodeIn{AfterStmt} match with each other,
\CodeIn{AfterStmt} must be either a sibling node of
\CodeIn{BeforeStmt} after it (in sequential order) or a descendant of
such a sibling node, e.g., \CodeIn{if ([COND]) \{BeforeStmt\}
AfterStmt} is not a valid \ptn because the \CodeIn{AfterStmt} is
neither a sibling node of \CodeIn{BeforeStmt}, nor a descendant of
such a sibling node.

\subsection{Translation}
\label{sec:framework:translation}

We implement \Tool in the Java programming language and provide two
API annotations, \CodeIn{@Pattern} and \CodeIn{@Constant}, and two API
methods, \CodeIn{void before(int expression)} and \CodeIn{void
after(int expression)} (\CodeIn{int} can also be \CodeIn{long}), to
express a \ptn. We also reuse Java constructs to make it easier to
write a \ptn, such as \CodeIn{if} statements and assignments.

A \ptn is recognized by a method annotated with \CodeIn{@Pattern}. All
variables used in the body of the method must be declared as
parameters of the method. A parameter can be annotated with
\CodeIn{@Constant} to indicate that the parameter represents a
constant variable rather than a free variable. A valid \ptn requires a
\CodeIn{before} method call and an \CodeIn{after} method call in the
method body and it may contain \CodeIn{if} statements for
preconditions or assignments for local re-assignment of variables.

\Tool starts translating a \ptn by parsing the expression from
\CodeIn{before} \API and constructing an \emph{\eAST} (extended
abstract syntax tree, which is strictly a directed acyclic graph) for
the expression. During the construction, \Tool maintains a map from
identifiers in the expression, such as variables or number literals,
to leaf nodes in the \eAST. This map is then used to construct the
\eAST for the expression from \CodeIn{after} API or any preconditions,
because \Tool reuses the same node in \CodeIn{before} when seeing the
same identifier in \CodeIn{after} \API or preconditions, to ensure the
correct transformation from \CodeIn{before} to \CodeIn{after}.
Figure~\ref{fig:example-asts} shows the \eASTs for the
\CodeIn{before}'s and \CodeIn{after}'s expression of \ptn
\UseMacro{ptn-pAdd6} (Figure~\ref{fig:example:pattern-code}).

\begin{figure}[t]
\tikzstyle{every node} = [font=\scriptsize] %
\tikzstyle{labelnode} = [rectangle, inner sep=1pt, text centered, text depth=.25ex, font=\footnotesize] %
\tikzstyle{dagnode} = [circle, draw, thick, minimum size=6mm, font=\footnotesize]
\tikzstyle{invisiblenode} = [circle, draw, minimum size=6mm, font=\footnotesize]
\tikzstyle{ra} = [->, shorten >=1pt, >=stealth, semithick]
\begin{subfigure}[b]{0.49\linewidth}
\centering

\begin{tikzpicture}[scale=1.0]

\node[dagnode] (root) at (0,0) { \CodeIn{+} };
\node[dagnode] (minus1) [below left=5mm and 8mm of root] { \CodeIn{-} };
\node[dagnode] (minus2) [below right=5mm and 8mm of root] { \CodeIn{-} };
\node[dagnode] (a) [below left=5mm and 2mm of minus1] { \CodeIn{a} };
\node[dagnode] (b) [below right=5mm and 2mm of minus1] { \CodeIn{b} };
\node[dagnode] (c) [below left=5mm and 2mm of minus2] { \CodeIn{c} };

\node[labelnode] (label-r) [right=1mm of root] { $r_b$ };
\node[labelnode] (label-p) [right=1mm of minus1] { $p$ };
\node[labelnode] (label-q) [right=1mm of minus2] { $q$ };
\node[labelnode] (label-a) [right=1mm of a] { $a$ };
\node[labelnode] (label-b) [right=1mm of b] { $b$ };
\node[labelnode] (label-c) [right=1mm of c] { $c$ };

\draw [ra] (root.south west) -- (minus1.north);
\draw [ra] (root.south east) -- (minus2.north);
\draw [ra] (minus1.south west) -- (a.north);
\draw [ra] (minus1.south east) -- (b.north);
\draw [ra] (minus2.south west) -- (c.north);
\draw [ra] (minus2.south east) to [out=-45, in=45] (a);
\end{tikzpicture}%

\caption{\UseMacro{figure-example-asts-before}}
\end{subfigure}
\hfill
\begin{subfigure}[b]{0.49\linewidth}
\centering

\begin{tikzpicture}[scale=1.0]

\node[dagnode] (root) at (0,0) { \CodeIn{-} };
\node[dagnode] (c) [below left=5mm and 2mm of root] { \CodeIn{c} };
\node[dagnode] (b) [below right=5mm and 2mm of root] { \CodeIn{b} };

\node[labelnode] (label-r) [right=1mm of root] { $r_a$ };
\node[labelnode] (label-c) [right=1mm of c] { $c$ };
\node[labelnode] (label-b) [right=1mm of b] { $b$ };

\draw [ra] (root.south west) -- (c.north);
\draw [ra] (root.south east) -- (b.north);
\end{tikzpicture}%

\caption{\UseMacro{figure-example-asts-after}}
\end{subfigure}
\caption{\UseMacro{figure-example-asts}}
\end{figure}

\Tool next translates \eASTs into C/C++ code that can be included in a
\JIT \optimization pass. As we have seen in Section~\ref{sec:example},
the generated C/C++ code consists of an \CodeIn{if} statement. The
condition of the \CodeIn{if} statement is the conjunction of all the
conditions that have to be satisfied for any expression to be matched
by the \ptn. The \CodeIn{then} branch of the \CodeIn{if} statement
ends with a \CodeIn{return} statement that returns an optimized
expression. \Tool first traverses \CodeIn{before}'s \eAST, and for
every node in the \eAST \Tool translates the path from the root to the
node into a pointer access chain in C/C++ (line
\ref{line:example:generated:decl:start}--\ref{line:example:generated:decl:end}
in Figure~\ref{fig:example:generated-code}). For example, node $b$ in
Figure~\ref{fig:example-asts:before} can be accessed by
\CodeIn{in(1)->in(2)}. Note one node could be accessed in more than
one way, and \Tool always picks the smallest one in lexicographic
order. Considering node $a$ in in
Figure~\ref{fig:example-asts:before}, which is both the left child of
node $p$ (\CodeIn{in(1)}) and the right child of node $q$
(\CodeIn{in(2)}), this node can be accessed by both
\CodeIn{in(1)->in(1)} and \CodeIn{in(2)->in(2)}, \Tool translates the
node into \CodeIn{in(1)->in(1)}.
Next, \Tool generates the conditions. \Tool traverses
\CodeIn{before}'s \eAST again to generate operator check and possible
constant check, for example, checking subtraction operators for node
$p$, \CodeIn{\_JOG\_in1->Opcode() == Op\_SubL}
(line~\ref{line:example:generated:operator-check:left} in
Figure~\ref{fig:example:generated-code}), and $q$,
\CodeIn{\_JOG\_in2->Opcode() == Op\_SubL}
(line~\ref{line:example:generated:operator-check:right} in
Figure~\ref{fig:example:generated-code}), where \CodeIn{\_JOG\_in1 =
in(1)} and \CodeIn{\_JOG\_in2 = in(2)}. Also, \Tool generates
same-node check for any node that can be accessed in more than one way
from the root. For instance, node $a$ in
Figure~\ref{fig:example-asts:before} results in the condition
\CodeIn{\_JOG\_in11 == \_JOG\_in22}
(line~\ref{line:example:generated:sharing-child-check} in
Figure~\ref{fig:example:generated-code}), where \CodeIn{\_JOG\_in11 =
in(1)->in(1)} and \CodeIn{\_JOG\_in22 = in(2)->in(2)}. Additionally,
if the \ptn provided contains any \CodeIn{if} conditions, i.e., the
specified \optimization requires preconditions, \Tool translates the
\eASTs of the preconditions into conditions in C/C++ code in the same
way.

\begin{figure*}[t]
\centering

\tikzstyle{every node} = [font=\scriptsize] %
\tikzstyle{labelnode} = [rectangle, inner sep=1pt, text centered, text depth=.25ex, minimum height=2.4ex, font=\scriptsize, anchor=center] %
\tikzstyle{dagnode} = [circle, draw, thick, minimum size=6mm, font=\footnotesize]
\tikzstyle{point} = [circle, inner sep=0pt, minimum size=1pt, fill=white] %
\tikzstyle{ra} = [->, shorten >=1pt, >=stealth, semithick]
\tikzstyle{dl} = [dashed, shorten >=1pt, shorten <=1pt]
\tikzstyle{textnode} = [rectangle, inner sep=2pt, text centered, text width=24ex, font=\footnotesize]
\tikzstyle{textbox} = [rectangle, draw, align=left, text width=35mm, font=\footnotesize]

\begin{tikzpicture}[scale=1.0]
\node[dagnode] (x-root) at (-60mm,0) { \CodeIn{+} };
\node[dagnode] (x-p) [below left=5mm and 8mm of x-root] { \CodeIn{-} };
\node[dagnode] (x-q) [below right=5mm and 8mm of x-root] { \CodeIn{-} };
\node[dagnode] (x-a) [below left=5mm and 2mm of x-p] { \CodeIn{a} };
\node[dagnode] (x-b) [below right=5mm and 2mm of x-p] { \CodeIn{b} };
\node[dagnode] (x-c) [below left=5mm and 2mm of x-q] { \CodeIn{c} };
\node[dagnode] (x-d) [below right=5mm and 2mm of x-q] { \CodeIn{d} };

\node[labelnode] (label-r) [below=1mm of x-root] { $x_1$ };
\node[labelnode] (label-p) [below=1mm of x-p] { $x_2$ };
\node[labelnode] (label-q) [below=1mm of x-q] { $x_5$ };
\node[labelnode] (label-a) [below=1mm of x-a] { $x_3$ };
\node[labelnode] (label-b) [below=1mm of x-b] { $x_4$ };
\node[labelnode] (label-c) [below=1mm of x-c] { $x_6$ };
\node[labelnode] (label-d) [below=1mm of x-d] { $x_7$ };

\draw [ra] (x-root.south west) -- (x-p.north);
\draw [ra] (x-root.south east) -- (x-q.north);
\draw [ra] (x-p.south west) -- (x-a.north);
\draw [ra] (x-p.south east) -- (x-b.north);
\draw [ra] (x-q.south west) -- (x-c.north);
\draw [ra] (x-q.south east) -- (x-d.north);

\node[dagnode] (y-root) at (0,0) { \CodeIn{+} };
\node[dagnode] (y-p) [below left=5mm and 2mm of y-root] { \CodeIn{-} };
\node[dagnode] (y-q) [below right=5mm and 2mm of y-root] { \CodeIn{-} };
\node[dagnode] (y-a) [below left=5mm and 2mm of y-p] { \CodeIn{a} };
\node[dagnode] (y-b) [below right=5mm and 2mm of y-p] { \CodeIn{b} };
\node[dagnode] (y-c) [below right=5mm and 2mm of y-q] { \CodeIn{c} };

\node[labelnode] (label-r) [below=0.5mm of y-root] { $ y_1$ };
\node[labelnode] (label-p) [below=0.5mm of y-p] { $y_2$ };
\node[labelnode] (label-q) [below=0.5mm of y-q] { $y_5$ };
\node[labelnode] (label-a) [below=0.5mm of y-a] { $y_3$ };
\node[labelnode] (label-b) [below=0.5mm of y-b] { $y_4$ };
\node[labelnode] (label-c) [below=0.5mm of y-c] { $y_6$ };

\draw [ra] (y-root.south west) -- (y-p.north);
\draw [ra] (y-root.south east) -- (y-q.north);
\draw [ra] (y-p.south west) -- (y-a.north);
\draw [ra] (y-p.south east) -- (y-b.north);
\draw [ra] (y-q.south west) -- (y-b.north);
\draw [ra] (y-q.south east) -- (y-c.north);
pp
\draw [dl] (x-root) to [out=45, in=135] (y-root);
\draw [dl] (x-p) to [out=90, in=135] (y-p);
\draw [dl] (x-q) to [out=45, in=90] (y-q);
\draw [dl] (x-a) to [out=-135, in=-135] (y-a);
\draw [dl] (x-b) to [out=-135, in=-135] (y-b);
\draw [dl] (x-c) to [out=-45, in=-45] (y-b);
\draw [dl] (x-d) to [out=-45, in=-45] (y-c);

\node[textbox] (ptn-x) [left=15mm of x-root] {
\begin{lstlisting}[language=jog-in-tikz]
@Pattern
public void (*@\UseMacro{ptn-pAdd2}@*)(int a, int b,
                   int c, int d) {
  before((a - b) + (c - d));
  after((a + c) - (b + d));
}
\end{lstlisting}
};
\node[textbox] (ptn-y) [right=15mm of y-root] {
\begin{lstlisting}[language=jog-in-tikz]
@Pattern
public void (*@\UseMacro{ptn-pAdd5}@*)(int a, int b,
                   int c) {
  before((a - b) + (b - c));
  after(a - c);
}
\end{lstlisting}
};

\node[textnode] (before-x) [above=7mm of x-root] {
$B^x:$ \CodeIn{(a - b) + (c - d)}
};
\node[textnode] (before-y) [above=7mm of y-root] {
$B^y:$ \CodeIn{(a - b) + (b - c)}
};

\node[textnode, text width=27ex] (phi-x) [below left=10mm and 20mm of x-root] {
$\Phi^x$ (Constraints on shape of \eAST $B^x$):
\begin{align*}
& x_1 = \mathrm{tree}\ (+)\ x_2\ x_5\\
\land\ & x_2 = \mathrm{tree}\ (+)\ x_3\ x_4\\
\land\ & x_5 = \mathrm{tree}\ (+)\ x_6\ x_7
\end{align*}
};
\node[textnode, text width=27ex] (phi-y) [below right=10mm and 20mm of y-root] {
$\Phi^y$ (Constraints on shape of \eAST $B^y$):
\begin{align*}
& y_1 = \mathrm{tree}\ (+)\ y_2\ y_5\\
\land\ & y_2 = \mathrm{tree}\ (+)\ y_3\ y_4\\
\land\ & y_5 = \mathrm{tree}\ (+)\ y_4\ x_6
\end{align*}
};
\node[textnode, text width=100ex] (psi) [anchor=north, below =23mm of $(phi-x.north)!0.5!(phi-y.north)$] {
$\Psi$ (Equivalence between \eAST $B_x$ and $B_y$):
\[
x_1 = y_1 \land x_2 = y_2 \land x_3 = y_3 \land x_4 = y_4 \land x_5 = y_5 \land x_6 = y_4 \land x_7 = y_6
\]
};
\node[textnode, text width=100ex] (F) [below=1mm of psi] {
$F$ (Final SMT formula to specify the relation of $X$ \shadowing $Y$):
\[
\lqall{y_1, y_2, y_3, y_4, y_5, y_6}{\Phi^y \llimply \lqexist{x_1, x_2, x_3, x_4, x_5, x_6, x_7}{\Phi^x \land \Psi}}
\]
};
\end{tikzpicture}%

\caption{\UseMacro{figure-example-shadow}}
\end{figure*}

To translate \CodeIn{after}'s \eAST, \Tool performs a Depth-First
Search (DFS). Every leaf node in \CodeIn{after}'s \eAST is shared with
\CodeIn{before}'s \eAST so \Tool reuses the pointer access chain for
the node, i.e., \CodeIn{\_JOG\_in21} for node $c$ and
\CodeIn{\_JOG\_in22} for node $b$ in
Figure~\ref{fig:example-asts:after}. For an internal node in
\CodeIn{after}'s \eAST, \Tool instantiates a new expression according
to the operator of the node. For example, node $r_a$ in
Figure~\ref{fig:example-asts:after} leads to \CodeIn{new
SubLNode(\_JOG\_in21, \_JOG\_in12)}
(line~\ref{line:example:generated:return} in
Figure~\ref{fig:example:generated-code}). Finally \Tool generates a
return statement that returns the instantiation generated for the root
node as translation of \CodeIn{after}'s \eAST.

With \CodeIn{before}'s and preconditions' \eASTs translated into
conditions and \CodeIn{after}'s \eAST translated into a return
statement, \Tool encapsulates them in an \CodeIn{if} statement
(line~\ref{line:example:generated:if:start}--\ref{line:example:generated:if:end}
in Figure~\ref{fig:example:generated-code}) and prepend proper
variable declarations
(line~\ref{line:example:generated:decl:start}--\ref{line:example:generated:decl:end}
in Figure~\ref{fig:example:generated-code}). This concludes the
translation of one \ptn. When there are multiple \ptns, \Tool
translate them in the order of the \ptns written in the file provided.

\subsection{Test Generation}

Writing the \ptn in the same way that the existing test is written allows
\Tool to generate an IR test from the \ptn. We next describe the
process of test generation using the example in
Figure~\ref{fig:example:test-code}. Although the test is an already
existing IR test in \OpenJDK, \Tool can generate exactly the same test
from the \ptn (Figure~\ref{fig:example:pattern-code}).

The {\CodeIn{@Test}} method first declares exactly the same
free variables as the \ptn (\CodeIn{long a, long b, long c}), and
returns exactly \CodeIn{before}'s expression in the \ptn
(\CodeIn{return (a - b) + (c - a)};). One exception is that when the
\ptn has a constant variable
(Section~\ref{sec:framework:translation}), \Tool uses a random number
to substitute the constant variable.
Next \Tool analyzes \CodeIn{before} and \CodeIn{after} in the \ptn.
\Tool searches in \CodeIn{after}'s \eAST (\CodeIn{c - b}) to count the
number of operators (one \CodeIn{SUB}), and compares \CodeIn{before}'s
and \CodeIn{after}'s \eASTs to obtain the operators that exist in
\CodeIn{before} but not in \CodeIn{after} (\CodeIn{ADD}). \Tool then
maps the operators to the corresponding IR node types used in IR tests
and makes \CodeIn{@IR} annotations (\CodeIn{@IR({counts = {IRNode.SUB,
``1''}})} and \CodeIn{@IR({failOn = {IRNode.ADD}})}).

Our current implementation does not generate tests for the \ptns with
preconditions that specify invariants between variables. For example,
a \ptn rewritten from OpenJDK
~\cite{OpenJDKOptimizationWithComplicatedPrecondition} that transforms
\CodeIn{(x \char`\>{}\char`\>{}> C0) + C1} to \CodeIn{(x + (C1 \char`\<{}\char`\<{} C0)) \char`\>{}\char`\>{}> C0)}
requires a precondition \CodeIn{C0 < 5 \&\& -5 < C1 \&\& C1 < 0 \&\& x
>= -(y \char`\<{}\char`\<{} C0)}.
A random integer number would not be a good test input for constant
variable \CodeIn{C0} or \CodeIn{C1} in the \ptn because it cannot
satisfy the precondition so as to trigger the \optimization. We plan
to leverage constraint solvers~\cite{DeMoura08Z3} to obtain valid test
inputs for such tests in future work.

\section{\Shadowing \Optimizations}
\label{sec:shadow}

Java \JIT compilers contain a large number of peephole \optimizations.
The maintenance becomes difficult as new \optimizations are included.
When developers want to add a new \optimization, they have to be
careful that this \optimization's effect is not overridden by some
existing \optimization. Consider two \optimizations X and Y in an
optimization pass, which are sequentially placed, i.e., X followed by
Y. If the set of instructions that Y matches is a subset of the set of
instructions that X matches, then Y will never be invoked because X is
always invoked before Y for any matched instructions. In this case, we
say X \emph{\shadows} Y or Y is \emph{\shadowed} by X. For example,
Figure~\ref{fig:example-shadow} shows such a pair of \optimizations
written in \ptns, where \ptn \UseMacro{ptn-pAdd2} \shadows \ptn
\UseMacro{ptn-pAdd5}.

The \shadow problem between two arbitrary \optimizations written in
\ptns X and Y can be rewritten as: for any expression matched by Y,
does X match the expression.
\Tool encodes the problem into an SMT formula and solves it using an SMT
solver (Z3\cite{DeMoura08Z3}). Figure~\ref{algo:shadow-deciding} shows
the overall algorithm, which we explain using a running example in
Figure~\ref{fig:example-shadow}.

\begin{figure}
\begin{\UseMacro{algo-shadow-deciding-fontsize}}

\begin{algorithmic}[1]
\Input $X$, $Y$: \Ptn
\Output $res \in \left\{\mathit{YES}, \mathit{NO}, \mathit{UNKNOWN}\right\}$ if $X$ shadows $Y$
\Function{Determine}{$X$, $Y$}
\State $B^x$ $\gets$ $\mathrm{before}(X)$ \label{line:shadow-deciding:beforex}
\State $B^y$ $\gets$ $\mathrm{before}(Y)$ \label{line:shadow-deciding:beforey}
\If{not \Call{SameShape}{$B^x$, $B^y$}} \label{line:shadow-deciding:matchshape}
\State \Return $\mathit{NO}$ \label{line:shadow-deciding:no}
\EndIf
\State Define a recursive data type $T$ with two constructors: \label{line:shadow-deciding:defineT}
\Statex[3] $\mathrm{nil}: \mathrm{int} \rightarrow T$
\Statex[3] $\mathrm{tree}: \mathrm{opcode}\ T\ T\ \rightarrow T$
\State $V^x$, $M^x$ $\gets$ \Call{CreateNewVariables}{$B^x$, $T$} \label{line:shadow-deciding:createNewVariables:x}
\State $V^y$, $M^y$ $\gets$ \Call{CreateNewVariables}{$B^y$, $T$} \label{line:shadow-deciding:createNewVariables:y}
\State $\Phi^x$ $\gets$ \Call{ConstrainShape}{$B^x$, $M^x$} \label{line:shadow-deciding:constrainShape:x}
\State $\Phi^y$ $\gets$ \Call{ConstrainShape}{$B^y$, $M^y$} \label{line:shadow-deciding:constrainShape:y}
\State $\Psi$ $\gets$ \Call{ConstrainEquivalence}{$B^x$, $B^y$, $M^x$, $M^y$} \label{line:shadow-deciding:constrainEquivalence}
\State $F$ $\gets$ $\lqAll{v^y \in V^y}{} \Phi^y \llimply \lqExist{v^x \in V^x}{} \Phi^x \land \Psi$ \label{line:shadow-deciding:formula}
\State \Return \Call{Prove}{$F$} \label{line:shadow-deciding:result}
\EndFunction
\end{algorithmic}

\end{\UseMacro{algo-shadow-deciding-fontsize}}
\caption{\UseMacro{algo-shadow-deciding-caption}}
\end{figure}

The algorithm first extracts \CodeIn{before}'s \eASTs ($B^x$ and
$B^y$) from \ptn $X$ and $Y$, respectively
(line~\ref{line:shadow-deciding:beforex}--\ref{line:shadow-deciding:beforey}),
and then checks if $B^x$ and $B^y$ share the same shape
(line~\ref{line:shadow-deciding:matchshape}). In the running example
from Figure~\ref{fig:example-shadow}, $B^x$ matches $B^y$ node-by-node,
except node $b$ in $B^y$ corresponds to both nodes $b$ and $c$ in
$B^x$.
Note that function \CodeIn{SameShape} performs a weak instead of exact
matching on node types, which allows a leaf node to match with an
internal node
because a leaf node may represent an expression as well as a variable.
Consider an expression \CodeIn{((e + f) - b) + (c - d)}, \ptn
(\CodeIn{(a - b) + (c - d) => (a + c) - (b + d)}) can still match the
expression if we replace \CodeIn{a} with \CodeIn{e + f}.

If $B^x$ and $B^y$ have different shapes, the algorithm will
immediately return \textit{NO} for the final result
(line~\ref{line:shadow-deciding:no}), i.e, $X$ does not \shadow $Y$.
To have such a rough shape check helps the algorithm more efficiently
determine the \shadow relation for two totally different \ptns, which
is common in practice. However, having the same shape does not
necessarily mean $X$ \shadows $Y$, i.e., any expression matched by
$B^y$ can also be matched by $B^x$.
Consider two \ptns $U$, \CodeIn{a + a => $\cdots$}, and $V$, \CodeIn{a
+ b => $\cdots$}, $U$ and $V$ share the same shape but $U$ does not
\shadow $V$. A counterexample is expression $1 + 2$ which is matched
by $V$ but not $U$.

To further solve the \shadow problem, we describe it formally as: for
all expression $E_y $ matched by $Y$, can we always construct another
expression $E_x$ matched by $X$ and ensure that the two expressions
are equivalent?
If the answer is yes, then $X$ \shadows $Y$; otherwise $X$ does not
\shadow $Y$. Note that we say two expressions are equivalent iff they
have exactly the same \eAST. We make such definition because \JIT
checks the structure of an expression, rather than evaluate the
expression, to determine if an \optimization can be applied on the
expression. If two expressions are equivalent, they are evaluated to
the same value, but the converse does not hold. For example,
expression \CodeIn{a + b} and \CodeIn{a + (b + 0)} are always
evaluated to the same value but they are not equivalent in our
definition.
Thus, with this definition of equivalence, the target SMT formula we
want to construct is:
\[
\lqall{E_y}{\left(Y\ \text{matches}\ E_y\right) \llimply \lqexist{E_x}{\left(X\ \mathrm{matches}\ E_x\right) \land \left(E_x = E_y\right)}}.
\]

First we construct the formulas for $Y(X)$ matching $E_y(E_x)$. We
need to encode $B^y$ into a list of constraints that $E_y$ needs to
satisfy in order to be matched. We define a recursive data type $T$
with two constructors: (1)~terminal constructor \CodeIn{nil} with no
argument, and (2)~non-terminal constructor \CodeIn{tree} with the
opcode and all the operands as arguments
(line~\ref{line:shadow-deciding:defineT}). We also create a variable
with type $T$ for every node in the \eAST
(line~\ref{line:shadow-deciding:createNewVariables:x}
and~\ref{line:shadow-deciding:createNewVariables:y}). In our example,
the nodes in \eAST $B^x$ are represented by variables $x_1$ to $x_7$.
Next, we encode the shape of the \eAST into several constraints
(line~\ref{line:shadow-deciding:constrainShape:x}
and~\ref{line:shadow-deciding:constrainShape:y}). For example, the
root node of $B^x$ and its two children in
Figure~\ref{fig:example-shadow} satisfies the constraint
$x_1 = \mathrm{tree}\ (+)\ x_2\ x_5$,
where $x_1$, $x_2$, $x_5$ is the variable mapped to the root node, the
left child, and the right child, respectively. We traverse the entire
\eAST to add one such constraint for every internal node.
Specifically, for a node that represents a constant or number
literals, we include an extra constraint on the value contained using
the terminal constructor \CodeIn{nil}.
Figure~\ref{fig:example-shadow} lists the constraints encoded from
\eAST $B^x$ and $B^y$, resulting in $\Phi^x$ and $\Phi^y$,
respectively.

Next, we encode equivalence between $B^x$ and $B^y$ into formulas
(line~\ref{line:shadow-deciding:constrainEquivalence}). We perform a
DFS on both \eASTs at the same time and add one equivalence relation,
in terms of the variables of type $T$ mapped, for every pair of nodes
visited, e.g., that the root nodes of $B^x$ and $B^y$ are equivalent
is encoded into $x_1 = y_1$.
For our running example, every dashed line in
Figure~\ref{fig:example-shadow} connects two equivalent nodes, and
$\Psi$ conjoins all the equivalence constraints between $B^x$ and
$B^y$.

With all the constraints ready, we now assemble them into the complete
SMT formula (line~\ref{line:shadow-deciding:formula}). The final
formula $F$ for our running example is
\[
\lqall{y_1, y_2, y_3, y_4, y_5, y_6}{\Phi^y \llimply \lqexist{x_1, x_2, x_3, x_4, x_5, x_6, x_7}{\Phi^x \land \Psi}},
\]
where
$\Phi^x$ is
\[
\left(x_1 = \mathrm{tree}\ (+)\ x_2\ x_5\right) \land \left(x_2 = \mathrm{tree}\ (+)\ x_3\ x_4\right) \land \left(x_5 = \mathrm{tree}\ (+)\ x_6\ x_7\right),
\]
$\Phi^y$ is
\[
\left(y_1 = \mathrm{tree}\ (+)\ y_2\ y_5\right) \land \left(y_2 = \mathrm{tree}\ (+)\ y_3\ y_4\right) \land \left(y_5 = \mathrm{tree}\ (+)\ y_4\ x_6\right),
\]
and $\Psi$ is $\left(x_1 = y_1\right) \land \left(x_2 = y_2\right) \land \left(x_3 = y_3\right) \land \left(x_4 = y_4\right) \land \left(x_5 = y_5\right) \land \left(x_6 = y_4\right) \land \left(x_7 = y_6\right)$.
Proving the formula then answers whether $X$ \shadows $Y$
(line~\ref{line:shadow-deciding:result}). If the formula is valid,
then function \CodeIn{Prove} returns \textit{YES}, i.e., $X$ \shadows
$Y$; if the formula is not valid, then \CodeIn{Prove} returns
\textit{NO}, i.e., $X$ does not \shadow $Y$; if the SMT solver is not
able to determine the outcome before timeout, then \CodeIn{Prove} returns
\textit{UNKNOWN}.
In our example, for any set of $y$ variables, there always exist a set
of $x$ variables such that the entire formula holds, e.g., $x_1 \gets
y_1$, $x_2 \gets y_2$, $x_3 \gets y_3$, $x_4 \gets y_4$, $x_5 \gets
y_5$, $x_6 \gets y_4$, $x_7 \gets y_6$. Thus, the formula $F$ is
proven, so $YES$ is returned, i.e., $X$ shadows $Y$.

Neither $X$ nor $Y$ has any preconditions in the example from
Figure~\ref{fig:example-shadow}, but \Tool can solve the shadow
problem for \ptns with preconditions. We encode \eASTs of
preconditions in the same way as \eASTs of \CodeIn{before}.
Preconditions may contain equivalence on values as well as shapes
(e.g., a pattern to match \CodeIn{0 - (x + C)} with precondition
\CodeIn{C != 0} where \CodeIn{C} is a
constant~\cite{OpenJDKOptimizationWithValuePrecondition}), so we
introduce another set of variables to encode constraints on values.
Then we encode both shape and value constraints, and both shape and
value equivalence. We construct a target SMT formula involved with
both shape and value constraints and equivalence.

\section{Evaluation}
\label{sec:eval}

We describe the setup of our experiments, quantify code complexity of
\ptns written using \Tool, show performance comparison with
hand-written optimizations, and describe test generation and our
contributions to OpenJDK.

\subsection{Setup}
\label{sec:eval:setup}

Table~\ref{tab:pattern-stats-summary} is the summary of our work to
write \UseMacro{total-num-patterns} \ptns using \Tool. For the first
category of \ptns from \OpenJDK, we selected \CodeIn{addnode.cpp},
\CodeIn{subnode.cpp} and \CodeIn{mulnode.cpp} in
\CodeIn{src/hotspot/}-\CodeIn{share/opto/} and we studied
\CodeIn{Ideal} methods defined in these files. The \CodeIn{Ideal}
method reshapes the IR graph rooted at \CodeIn{this} node and returns
the reshaped node as an \optimized node. Every \CodeIn{Ideal} method
may contain many peephole \optimizations. We identified and rewrote
\UseMacro{total-num-patterns-OpenJDK} \optimizations into \ptns using
\Tool. For the second category of \ptns, we studied \LLVM's
InstCombine pass that performs numerous algebraic simplifications that
improve efficiency, referring to \Alive's
approach~\cite{Lopes15Alive}. We translated
\UseMacro{total-num-patterns-LLVM} \ptns from
\CodeIn{InstCombineAddSub.cpp}, \CodeIn{InstCombineAndOrXor.cpp} in
\CodeIn{llvm/lib/Tra}\-\CodeIn{nsforms}\-\CodeIn{/InstCombine/}. When
we studied \optimizations from source code files in \OpenJDK and
\LLVM, we followed the order in which \optimizations appear in the
files to rewrite them in \ptns using \Tool, such that the generated
C/C++ code from these \ptns will be in a proper order.
Additionally, we proposed \UseMacro{total-num-patterns-OWN}
\optimizations
and we wrote them as \ptns.

\begin{\UseMacro{table-pattern-stats-summary-float-env}}[t]
\centering
\caption{\UseMacro{table-pattern-stats-summary-caption}}
\begin{\UseMacro{table-pattern-stats-summary-fontsize}}
\begin{tabular}{rrrrr}
\toprule
\textbf{\UseMacro{table-pattern-stats-summary-col-num-patterns}} & \textbf{\UseMacro{table-pattern-stats-summary-col-num-OpenJDK}} & \textbf{\UseMacro{table-pattern-stats-summary-col-num-LLVM}} & \textbf{\UseMacro{table-pattern-stats-summary-col-num-OWN}} & \textbf{\UseMacro{table-pattern-stats-summary-col-num-prs}}\\
\midrule
\UseMacro{table-pattern-stats-summary-row0-num-patterns} & \UseMacro{table-pattern-stats-summary-row0-num-OpenJDK} & \UseMacro{table-pattern-stats-summary-row0-num-LLVM} & \UseMacro{table-pattern-stats-summary-row0-num-OWN} & \UseMacro{table-pattern-stats-summary-row0-num-prs}\\
\bottomrule
\end{tabular}
\end{\UseMacro{table-pattern-stats-summary-fontsize}}
\end{\UseMacro{table-pattern-stats-summary-float-env}}

We ran all experiments on a 64-bit Ubuntu 18.04.1 desktop with an
Intel(R) Core(TM) i7-8700 CPU @3.20GHz and 64GB RAM. The SHA of
\OpenJDK repository~\cite{OpenJDKRepo} we used is \CodeIn{b334d96} and
the SHA of \LLVM repository~\cite{LLVMRepo} is \CodeIn{103e1d9}.

\vspace{5pt}
\noindent
We evaluate \Tool by answering the following research questions:
\begin{enumerate}[topsep=5pt,itemsep=1.5ex,partopsep=0ex,parsep=0ex,leftmargin=25pt]
\item[\textbf{\RQEasierWriting:}] How does \Tool compare to
hand-written \optimizations in terms of code complexity?

\item[\textbf{\RQPerformance:}] How does the code generated from
\Tool compare in performance to existing hand-written code in
\OpenJDK ?

\item[\textbf{\RQShadow:}] How effective is \Tool at detecting
\shadows between \ptns?

\item[\textbf{\RQTestGenerationAndPullRequests:}] How is \Tool used
to generate tests from \ptns and how does it contribute to
\OpenJDK?
\end{enumerate}

\noindent
We address \RQEasierWriting as to better understand the benefit of
using \Tool to write \optimizations for Java \JIT compilers; we use
code reduction (in terms of the number of characters and identifiers)
as a proxy when answering this question.
We address \RQPerformance to understand the performance of \Tool's
generated code compared to hand-written code; namely, we wanted to
understand the impact on the effectiveness of \JIT \optimizations. We
address \RQShadow to study the effectiveness of \Tool for detecting
shadowing \optimizations. We address \RQTestGenerationAndPullRequests
to evaluate \Tool's test generation from \ptns and describe pull
requests we opened for \OpenJDK.

\subsection{Code Complexity}
\label{sec:eval:lcoc}

We select two code features,
number of
characters and number of identifiers, as a metric to quantify code
complexity of \ptns written using \Tool~\cite{BuseLearningMetricCodeReadability}. We count the number of
characters and the number of identifiers for every \ptn written using
\Tool and its counterpart hand-written in OpenJDK. We exclude any
white-spaces or newlines when counting characters. We exclude any
reserved words in Java or C/C++ languages when counting identifiers,
and Figure~\ref{fig:count-ids} illustrates the way we count the
identifiers using an earlier example (see section~\ref{sec:example}).
Table~\ref{tab:lcoc-stats} compares these numbers between hand-written
C/C++ code in \OpenJDK and corresponding \ptns written using \Tool.
Namely, ``\UseMacro{table-lcoc-stats-col-manual}'' means the
hand-written C/C++ code in \OpenJDK, and
``\UseMacro{table-lcoc-stats-col-pattern}'' means Java code in \Tool.
The columns of ``\UseMacro{table-lcoc-stats-col-reduction}'' show the
percentage of characters and identifiers reduction from hand-written
C/C++ code to \Tool \ptns.
Additionally, we provide characters and identifiers of generated C/C++
code from \Tool as a reference, which is shown in the columns of
``\UseMacro{table-lcoc-stats-col-generated}''.

\newcommand{\FigNumberAboveLineShrink}{1.2mm}
\newcommand{\FigNumberTokenNoScale}{1.4}
\newcommand{\FigNumberTokenVSpace}{-0.1mm}
\newcommand{\FigNumberTokenHSpace}{-0.8mm}
\newcommand{\FigNumberTokenNoHSpace}{-1.7mm}
\newcommand{\LineVSep}{4mm}
\newcommand{\Indent}{2mm}
\begin{figure}[t]
\begin{subfigure}{\linewidth}

\begin{tikzpicture}[
thick,
node distance=\FigNumberTokenVSpace and \FigNumberTokenHSpace,
every node/.style={scale=0.50},
token/.style={rectangle, scale=1.6, text height=1.5ex, text depth=.25ex, text centered},
marker/.style={circle, inner sep=0mm},
key/.style={rectangle}
]

\node at (0,0) (l1t5) [token] {\CodeIn{long}};
\node [right = \FigNumberTokenHSpace of l1t5] (l1a) [token] {\CodeIn{a}};
\node [right = \FigNumberTokenNoHSpace of l1a] (l1t7) [token] {\CodeIn{, long}};
\node [right = \FigNumberTokenHSpace of l1t7] (l1b) [token] {\CodeIn{b}};
\node [right = \FigNumberTokenNoHSpace of l1b] (l1t9) [token] {\CodeIn{, long}};
\node [right = \FigNumberTokenHSpace of l1t9] (l1c) [token] {\CodeIn{c}};
\node [right = \FigNumberTokenHSpace of l1c] (l1tB) [token] {\CodeIn{\{}};

\draw ($ (l1a.north west) + (\FigNumberAboveLineShrink,0) $) to node[above, scale=\FigNumberTokenNoScale] {$1$} ($ (l1a.north east) - (\FigNumberAboveLineShrink,0) $);
\draw ($ (l1b.north west) + (\FigNumberAboveLineShrink,0) $) to node[above, scale=\FigNumberTokenNoScale] {$2$} ($ (l1b.north east) - (\FigNumberAboveLineShrink,0) $);
\draw ($ (l1c.north west) + (\FigNumberAboveLineShrink,0) $) to node[above, scale=\FigNumberTokenNoScale] {$3$} ($ (l1c.north east) - (\FigNumberAboveLineShrink,0) $);

\node [right = \FigNumberTokenHSpace of l1tB] (l2before) [token] {\CodeIn{before}};
\node [right = \FigNumberTokenNoHSpace of l2before] (l2t2) [token] {\CodeIn{((}};
\node [right = \FigNumberTokenNoHSpace of l2t2] (l2a-1) [token] {\CodeIn{a}};
\node [right = \FigNumberTokenHSpace of l2a-1] (l2t4) [token] {\CodeIn{-}};
\node [right = \FigNumberTokenHSpace of l2t4] (l2b) [token] {\CodeIn{b}};
\node [right = \FigNumberTokenNoHSpace of l2b] (l2t6) [token] {\CodeIn{) + (}};
\node [right = \FigNumberTokenNoHSpace of l2t6] (l2c) [token] {\CodeIn{c}};
\node [right = \FigNumberTokenHSpace of l2c] (l2t8) [token] {\CodeIn{-}};
\node [right = \FigNumberTokenHSpace of l2t8] (l2a-2) [token] {\CodeIn{a}};
\node [right = \FigNumberTokenNoHSpace of l2a-2] (l2tA) [token] {\CodeIn{));}};

\draw ($ (l2before.north west) + (\FigNumberAboveLineShrink,0) $) to node[above, scale=\FigNumberTokenNoScale] {$4$} ($ (l2before.north east) - (\FigNumberAboveLineShrink,0) $);
\draw ($ (l2a-1.north west) + (\FigNumberAboveLineShrink,0) $) to node[above, scale=\FigNumberTokenNoScale] {$5$} ($ (l2a-1.north east) - (\FigNumberAboveLineShrink,0) $);
\draw ($ (l2b.north west) + (\FigNumberAboveLineShrink,0) $) to node[above, scale=\FigNumberTokenNoScale] {$6$} ($ (l2b.north east) - (\FigNumberAboveLineShrink,0) $);
\draw ($ (l2c.north west) + (\FigNumberAboveLineShrink,0) $) to node[above, scale=\FigNumberTokenNoScale] {$7$} ($ (l2c.north east) - (\FigNumberAboveLineShrink,0) $);
\draw ($ (l2a-2.north west) + (\FigNumberAboveLineShrink,0) $) to node[above, scale=\FigNumberTokenNoScale] {$8$} ($ (l2a-2.north east) - (\FigNumberAboveLineShrink,0) $);

\node [right = \FigNumberTokenHSpace of l2tA] (l3after) [token] {\CodeIn{after}};
\node [right = \FigNumberTokenNoHSpace of l3after] (l3t2) [token] {\CodeIn{(}};
\node [right = \FigNumberTokenNoHSpace of l3t2] (l3c) [token] {\CodeIn{c}};
\node [right = \FigNumberTokenHSpace of l3c] (l3t4) [token] {\CodeIn{-}};
\node [right = \FigNumberTokenHSpace of l3t4] (l3b) [token] {\CodeIn{b}};
\node [right = \FigNumberTokenNoHSpace of l3b] (l3t6) [token] {\CodeIn{); \}}};

\draw ($ (l3after.north west) + (\FigNumberAboveLineShrink,0) $) to node[above, scale=\FigNumberTokenNoScale] {$9$} ($ (l3after.north east) - (\FigNumberAboveLineShrink,0) $);
\draw ($ (l3c.north west) + (\FigNumberAboveLineShrink,0) $) to node[above, scale=\FigNumberTokenNoScale] {$10$} ($ (l3c.north east) - (\FigNumberAboveLineShrink,0) $);
\draw ($ (l3b.north west) + (\FigNumberAboveLineShrink,0) $) to node[above, scale=\FigNumberTokenNoScale] {$11$} ($ (l3b.north east) - (\FigNumberAboveLineShrink,0) $);

\end{tikzpicture}

\caption{\UseMacro{figure-count-ids-pattern}}
\end{subfigure}
\begin{subfigure}{\linewidth}

\newcounter{idcounter}
\setcounter{idcounter}{0}
\begin{tikzpicture}[
thick,
node distance=\FigNumberTokenVSpace and \FigNumberTokenHSpace,
every node/.style={scale=0.50},
token/.style={rectangle, scale=1.6, text height=1.5ex, text depth=.25ex, text centered},
marker/.style={circle, inner sep=0mm},
key/.style={rectangle}
]

\node at (0, 0) (l1node) [token] {\CodeIn{Node}};
\node [right = \FigNumberTokenNoHSpace of l1node] (l1t7) [token] {\CodeIn{*}};
\node [right = \FigNumberTokenHSpace of l1t7] (l1in1) [token] {\CodeIn{in1}};
\node [right = \FigNumberTokenNoHSpace of l1in1] (l1t9) [token] {\CodeIn{ = }};
\node [right = \FigNumberTokenHSpace of l1t9] (l1in) [token] {\CodeIn{in}};
\node [right = \FigNumberTokenNoHSpace of l1in] (l1tB) [token] {\CodeIn{(1);}};

\stepcounter{idcounter}
\draw ($ (l1node.north west) + (\FigNumberAboveLineShrink,0) $) to node[above, scale=\FigNumberTokenNoScale] {$\theidcounter$} ($ (l1node.north east) - (\FigNumberAboveLineShrink,0) $);
\stepcounter{idcounter}
\draw ($ (l1in1.north west) + (\FigNumberAboveLineShrink,0) $) to node[above, scale=\FigNumberTokenNoScale] {$\theidcounter$} ($ (l1in1.north east) - (\FigNumberAboveLineShrink,0) $);
\stepcounter{idcounter}
\draw ($ (l1in.north west) + (\FigNumberAboveLineShrink,0) $) to node[above, scale=\FigNumberTokenNoScale] {$\theidcounter$} ($ (l1in.north east) - (\FigNumberAboveLineShrink,0) $);

\node [right = \FigNumberTokenHSpace of l1tB] (l2node) [token] {\CodeIn{Node}};
\node [right = \FigNumberTokenNoHSpace of l2node] (l2t7) [token] {\CodeIn{*}};
\node [right = \FigNumberTokenHSpace of l2t7] (l2in2) [token] {\CodeIn{in2}};
\node [right = \FigNumberTokenNoHSpace of l2in2] (l2t9) [token] {\CodeIn{ = }};
\node [right = \FigNumberTokenHSpace of l2t9] (l2in) [token] {\CodeIn{in}};
\node [right = \FigNumberTokenNoHSpace of l2in] (l2tB) [token] {\CodeIn{(2);}};

\stepcounter{idcounter}
\draw ($ (l2node.north west) + (\FigNumberAboveLineShrink,0) $) to node[above, scale=\FigNumberTokenNoScale] {$\theidcounter$} ($ (l2node.north east) - (\FigNumberAboveLineShrink,0) $);
\stepcounter{idcounter}
\draw ($ (l2in2.north west) + (\FigNumberAboveLineShrink,0) $) to node[above, scale=\FigNumberTokenNoScale] {$\theidcounter$} ($ (l2in2.north east) - (\FigNumberAboveLineShrink,0) $);
\stepcounter{idcounter}
\draw ($ (l2in.north west) + (\FigNumberAboveLineShrink,0) $) to node[above, scale=\FigNumberTokenNoScale] {$\theidcounter$} ($ (l2in.north east) - (\FigNumberAboveLineShrink,0) $);

\node [below = \LineVSep of l1node.south west, anchor=west] (l3int) [token] {\CodeIn{int}};
\node [right = \FigNumberTokenHSpace of l3int] (l3op1) [token] {\CodeIn{op1}};
\node [right = \FigNumberTokenHSpace of l3op1] (l3t1) [token] {\CodeIn{ = }};
\node [right = \FigNumberTokenHSpace of l3t1] (l3in1) [token] {\CodeIn{in1}};
\node [right = \FigNumberTokenNoHSpace of l3in1] (l3t2) [token] {\CodeIn{->}};
\node [right = \FigNumberTokenNoHSpace of l3t2] (l3opcode) [token] {\CodeIn{Opcode}};
\node [right = \FigNumberTokenNoHSpace of l3opcode] (l3t3) [token] {\CodeIn{();}};

\stepcounter{idcounter}
\draw ($ (l3op1.north west) + (\FigNumberAboveLineShrink,0) $) to node[above, scale=\FigNumberTokenNoScale] {$\theidcounter$} ($ (l3op1.north east) - (\FigNumberAboveLineShrink,0) $);
\stepcounter{idcounter}
\draw ($ (l3in1.north west) + (\FigNumberAboveLineShrink,0) $) to node[above, scale=\FigNumberTokenNoScale] {$\theidcounter$} ($ (l3in1.north east) - (\FigNumberAboveLineShrink,0) $);
\stepcounter{idcounter}
\draw ($ (l3opcode.north west) + (\FigNumberAboveLineShrink,0) $) to node[above, scale=\FigNumberTokenNoScale] {$\theidcounter$} ($ (l3opcode.north east) - (\FigNumberAboveLineShrink,0) $);

\node [right = \FigNumberTokenHSpace of l3t3] (l4int) [token] {\CodeIn{int}};
\node [right = \FigNumberTokenHSpace of l4int] (l4op1) [token] {\CodeIn{op2}};
\node [right = \FigNumberTokenHSpace of l4op1] (l4t1) [token] {\CodeIn{=}};
\node [right = \FigNumberTokenHSpace of l4t1] (l4in2) [token] {\CodeIn{in2}};
\node [right = \FigNumberTokenNoHSpace of l4in2] (l4t2) [token] {\CodeIn{->}};
\node [right = \FigNumberTokenNoHSpace of l4t2] (l4opcode) [token] {\CodeIn{Opcode}};
\node [right = \FigNumberTokenNoHSpace of l4opcode] (l4t3) [token] {\CodeIn{();}};

\stepcounter{idcounter}
\draw ($ (l4op1.north west) + (\FigNumberAboveLineShrink,0) $) to node[above, scale=\FigNumberTokenNoScale] {$\theidcounter$} ($ (l4op1.north east) - (\FigNumberAboveLineShrink,0) $);
\stepcounter{idcounter}
\draw ($ (l4in2.north west) + (\FigNumberAboveLineShrink,0) $) to node[above, scale=\FigNumberTokenNoScale] {$\theidcounter$} ($ (l4in2.north east) - (\FigNumberAboveLineShrink,0) $);
\stepcounter{idcounter}
\draw ($ (l4opcode.north west) + (\FigNumberAboveLineShrink,0) $) to node[above, scale=\FigNumberTokenNoScale] {$\theidcounter$} ($ (l4opcode.north east) - (\FigNumberAboveLineShrink,0) $);

\node [below right = \LineVSep and \Indent of l3int.south west, anchor=west] (l5t1) [token] {\CodeIn{if (}};
\node [right = \FigNumberTokenNoHSpace of l5t1] (l5op1) [token] {\CodeIn{op1}};
\node [right = \FigNumberTokenHSpace of l5op1] (l5t2) [token] {\CodeIn{==}};
\node [right = \FigNumberTokenHSpace of l5t2] (l5opsubl) [token] {\CodeIn{Op\_SubL}};
\node [right = \FigNumberTokenNoHSpace of l5opsubl] (l5t3) [token] {\CodeIn{) \{}};

\stepcounter{idcounter}
\draw ($ (l5op1.north west) + (\FigNumberAboveLineShrink,0) $) to node[above, scale=\FigNumberTokenNoScale] {$\theidcounter$} ($ (l5op1.north east) - (\FigNumberAboveLineShrink,0) $);
\stepcounter{idcounter}
\draw ($ (l5opsubl.north west) + (\FigNumberAboveLineShrink,0) $) to node[above, scale=\FigNumberTokenNoScale] {$\theidcounter$} ($ (l5opsubl.north east) - (\FigNumberAboveLineShrink,0) $);

\node [right = \FigNumberTokenHSpace of l5t3] (l6t1) [token] {\CodeIn{if (}};
\node [right = \FigNumberTokenNoHSpace of l6t1] (l6op2) [token] {\CodeIn{op2}};
\node [right = \FigNumberTokenHSpace of l6op2] (l6t2) [token] {\CodeIn{==}};
\node [right = \FigNumberTokenHSpace of l6t2] (l6opsubl) [token] {\CodeIn{Op\_SubL}};
\node [right = \FigNumberTokenHSpace of l6opsubl] (l6t3) [token] {\CodeIn{\&\&}};
\node [right = \FigNumberTokenHSpace of l6t3] (l6in1) [token] {\CodeIn{in1}};
\node [right = \FigNumberTokenNoHSpace of l6in1] (l6t4) [token] {\CodeIn{->}};
\node [right = \FigNumberTokenNoHSpace of l6t4] (l6in-1) [token] {\CodeIn{in}};
\node [right = \FigNumberTokenNoHSpace of l6in-1] (l6t5) [token] {\CodeIn{(1) ==}};
\node [right = \FigNumberTokenHSpace of l6t5] (l6in2) [token] {\CodeIn{in2}};
\node [right = \FigNumberTokenNoHSpace of l6in2] (l6t6) [token] {\CodeIn{->}};
\node [right = \FigNumberTokenNoHSpace of l6t6] (l6in-2) [token] {\CodeIn{in}};
\node [right = \FigNumberTokenNoHSpace of l6in-2] (l6t5) [token] {\CodeIn{(2) \{}};

\stepcounter{idcounter}
\draw ($ (l6op2.north west) + (\FigNumberAboveLineShrink,0) $) to node[above, scale=\FigNumberTokenNoScale] {$\theidcounter$} ($ (l6op2.north east) - (\FigNumberAboveLineShrink,0) $);
\stepcounter{idcounter}
\draw ($ (l6opsubl.north west) + (\FigNumberAboveLineShrink,0) $) to node[above, scale=\FigNumberTokenNoScale] {$\theidcounter$} ($ (l6opsubl.north east) - (\FigNumberAboveLineShrink,0) $);
\stepcounter{idcounter}
\draw ($ (l6in1.north west) + (\FigNumberAboveLineShrink,0) $) to node[above, scale=\FigNumberTokenNoScale] {$\theidcounter$} ($ (l6in1.north east) - (\FigNumberAboveLineShrink,0) $);
\stepcounter{idcounter}
\draw ($ (l6in-1.north west) + (\FigNumberAboveLineShrink,0) $) to node[above, scale=\FigNumberTokenNoScale] {$\theidcounter$} ($ (l6in-1.north east) - (\FigNumberAboveLineShrink,0) $);
\stepcounter{idcounter}
\draw ($ (l6in2.north west) + (\FigNumberAboveLineShrink,0) $) to node[above, scale=\FigNumberTokenNoScale] {$\theidcounter$} ($ (l6in2.north east) - (\FigNumberAboveLineShrink,0) $);
\stepcounter{idcounter}
\draw ($ (l6in-2.north west) + (\FigNumberAboveLineShrink,0) $) to node[above, scale=\FigNumberTokenNoScale] {$\theidcounter$} ($ (l6in-2.north east) - (\FigNumberAboveLineShrink,0) $);

\node [below right = \LineVSep and \Indent of l5t1.south west, anchor=west] (l7t1) [token] {\CodeIn{return new}};
\node [right = \FigNumberTokenHSpace of l7t1] (l7sublnode) [token] {\CodeIn{SubLNode}};
\node [right = \FigNumberTokenNoHSpace of l7sublnode] (l7t2) [token] {\CodeIn{(}};
\node [right = \FigNumberTokenNoHSpace of l7t2] (l7in2) [token] {\CodeIn{in2}};
\node [right = \FigNumberTokenNoHSpace of l7in2] (l7t3) [token] {\CodeIn{->}};
\node [right = \FigNumberTokenNoHSpace of l7t3] (l7in-1) [token] {\CodeIn{in}};
\node [right = \FigNumberTokenNoHSpace of l7in-1] (l7t4) [token] {\CodeIn{(1),}};
\node [right = \FigNumberTokenHSpace of l7t4] (l7in1) [token] {\CodeIn{in1}};
\node [right = \FigNumberTokenNoHSpace of l7in1] (l7t5) [token] {\CodeIn{->}};
\node [right = \FigNumberTokenNoHSpace of l7t5] (l7in-2) [token] {\CodeIn{in}};
\node [right = \FigNumberTokenNoHSpace of l7in-2] (l7t6) [token] {\CodeIn{(2)); \} \} \}}};

\stepcounter{idcounter}
\draw ($ (l7sublnode.north west) + (\FigNumberAboveLineShrink,0) $) to node[above, scale=\FigNumberTokenNoScale] {$\theidcounter$} ($ (l7sublnode.north east) - (\FigNumberAboveLineShrink,0) $);
\stepcounter{idcounter}
\draw ($ (l7in2.north west) + (\FigNumberAboveLineShrink,0) $) to node[above, scale=\FigNumberTokenNoScale] {$\theidcounter$} ($ (l7in2.north east) - (\FigNumberAboveLineShrink,0) $);
\stepcounter{idcounter}
\draw ($ (l7in-1.north west) + (\FigNumberAboveLineShrink,0) $) to node[above, scale=\FigNumberTokenNoScale] {$\theidcounter$} ($ (l7in-1.north east) - (\FigNumberAboveLineShrink,0) $);
\stepcounter{idcounter}
\draw ($ (l7in1.north west) + (\FigNumberAboveLineShrink,0) $) to node[above, scale=\FigNumberTokenNoScale] {$\theidcounter$} ($ (l7in1.north east) - (\FigNumberAboveLineShrink,0) $);
\stepcounter{idcounter}
\draw ($ (l7in-2.north west) + (\FigNumberAboveLineShrink,0) $) to node[above, scale=\FigNumberTokenNoScale] {$\theidcounter$} ($ (l7in-2.north east) - (\FigNumberAboveLineShrink,0) $);
\end{tikzpicture}

\caption{\UseMacro{figure-count-ids-manual}}
\end{subfigure}
\caption{\UseMacro{figure-count-ids}}
\end{figure}

\begin{\UseMacro{table-lcoc-stats-float-env}}[t]
\centering
\caption{\UseMacro{table-lcoc-stats-caption}}
\begin{\UseMacro{table-lcoc-stats-fontsize}}
\begin{tabular}{lrrrrrrrr}
\toprule
\multirow{2}{*}{\textbf{\UseMacro{table-lcoc-stats-col-names}}} & \multicolumn{2}{c}{\textbf{\UseMacro{table-lcoc-stats-col-manual}}} & \multicolumn{2}{c}{\textbf{\UseMacro{table-lcoc-stats-col-generated}}} & \multicolumn{2}{c}{\textbf{\UseMacro{table-lcoc-stats-col-pattern}}} & \multicolumn{2}{c}{\textbf{\UseMacro{table-lcoc-stats-col-reduction}}}\\
\cmidrule(r){2-3}\cmidrule(lr){4-5}\cmidrule(l){6-7}\cmidrule(l){8-9}
& \textbf{\UseMacro{table-lcoc-stats-col-coc}} & \textbf{\UseMacro{table-lcoc-stats-col-ioc}} & \textbf{\UseMacro{table-lcoc-stats-col-coc}} & \textbf{\UseMacro{table-lcoc-stats-col-ioc}} & \textbf{\UseMacro{table-lcoc-stats-col-coc}} & \textbf{\UseMacro{table-lcoc-stats-col-ioc}} & \textbf{\UseMacro{table-lcoc-stats-col-coc}} & \textbf{\UseMacro{table-lcoc-stats-col-ioc}}\\
\midrule
\UseMacro{table-lcoc-stats-row0-names-0} & \UseMacro{table-lcoc-stats-row0-manual-coc} & \UseMacro{table-lcoc-stats-row0-manual-ioc} & \UseMacro{table-lcoc-stats-row0-generated-coc} & \UseMacro{table-lcoc-stats-row0-generated-ioc} & \UseMacro{table-lcoc-stats-row0-pattern-coc} & \UseMacro{table-lcoc-stats-row0-pattern-ioc} & \UseMacro{table-lcoc-stats-row0-reduction-coc} & \UseMacro{table-lcoc-stats-row0-reduction-ioc}\\
\UseMacro{table-lcoc-stats-row1-names-0} & \UseMacro{table-lcoc-stats-row1-manual-coc} & \UseMacro{table-lcoc-stats-row1-manual-ioc} & \UseMacro{table-lcoc-stats-row1-generated-coc} & \UseMacro{table-lcoc-stats-row1-generated-ioc} & \UseMacro{table-lcoc-stats-row1-pattern-coc} & \UseMacro{table-lcoc-stats-row1-pattern-ioc} & \UseMacro{table-lcoc-stats-row1-reduction-coc} & \UseMacro{table-lcoc-stats-row1-reduction-ioc}\\
\UseMacro{table-lcoc-stats-row2-names-0} & \UseMacro{table-lcoc-stats-row2-manual-coc} & \UseMacro{table-lcoc-stats-row2-manual-ioc} & \UseMacro{table-lcoc-stats-row2-generated-coc} & \UseMacro{table-lcoc-stats-row2-generated-ioc} & \UseMacro{table-lcoc-stats-row2-pattern-coc} & \UseMacro{table-lcoc-stats-row2-pattern-ioc} & \UseMacro{table-lcoc-stats-row2-reduction-coc} & \UseMacro{table-lcoc-stats-row2-reduction-ioc}\\
\UseMacro{table-lcoc-stats-row3-names-0} & \UseMacro{table-lcoc-stats-row3-manual-coc} & \UseMacro{table-lcoc-stats-row3-manual-ioc} & \UseMacro{table-lcoc-stats-row3-generated-coc} & \UseMacro{table-lcoc-stats-row3-generated-ioc} & \UseMacro{table-lcoc-stats-row3-pattern-coc} & \UseMacro{table-lcoc-stats-row3-pattern-ioc} & \UseMacro{table-lcoc-stats-row3-reduction-coc} & \UseMacro{table-lcoc-stats-row3-reduction-ioc}\\
\UseMacro{table-lcoc-stats-row4-names-0} & \UseMacro{table-lcoc-stats-row4-manual-coc} & \UseMacro{table-lcoc-stats-row4-manual-ioc} & \UseMacro{table-lcoc-stats-row4-generated-coc} & \UseMacro{table-lcoc-stats-row4-generated-ioc} & \UseMacro{table-lcoc-stats-row4-pattern-coc} & \UseMacro{table-lcoc-stats-row4-pattern-ioc} & \UseMacro{table-lcoc-stats-row4-reduction-coc} & \UseMacro{table-lcoc-stats-row4-reduction-ioc}\\
\UseMacro{table-lcoc-stats-row5-names-0} & \UseMacro{table-lcoc-stats-row5-manual-coc} & \UseMacro{table-lcoc-stats-row5-manual-ioc} & \UseMacro{table-lcoc-stats-row5-generated-coc} & \UseMacro{table-lcoc-stats-row5-generated-ioc} & \UseMacro{table-lcoc-stats-row5-pattern-coc} & \UseMacro{table-lcoc-stats-row5-pattern-ioc} & \UseMacro{table-lcoc-stats-row5-reduction-coc} & \UseMacro{table-lcoc-stats-row5-reduction-ioc}\\
\UseMacro{table-lcoc-stats-row6-names-0} & \UseMacro{table-lcoc-stats-row6-manual-coc} & \UseMacro{table-lcoc-stats-row6-manual-ioc} & \UseMacro{table-lcoc-stats-row6-generated-coc} & \UseMacro{table-lcoc-stats-row6-generated-ioc} & \UseMacro{table-lcoc-stats-row6-pattern-coc} & \UseMacro{table-lcoc-stats-row6-pattern-ioc} & \UseMacro{table-lcoc-stats-row6-reduction-coc} & \UseMacro{table-lcoc-stats-row6-reduction-ioc}\\
\UseMacro{table-lcoc-stats-row7-names-0} & \UseMacro{table-lcoc-stats-row7-manual-coc} & \UseMacro{table-lcoc-stats-row7-manual-ioc} & \UseMacro{table-lcoc-stats-row7-generated-coc} & \UseMacro{table-lcoc-stats-row7-generated-ioc} & \UseMacro{table-lcoc-stats-row7-pattern-coc} & \UseMacro{table-lcoc-stats-row7-pattern-ioc} & \UseMacro{table-lcoc-stats-row7-reduction-coc} & \UseMacro{table-lcoc-stats-row7-reduction-ioc}\\
\UseMacro{table-lcoc-stats-row8-names-0} & \UseMacro{table-lcoc-stats-row8-manual-coc} & \UseMacro{table-lcoc-stats-row8-manual-ioc} & \UseMacro{table-lcoc-stats-row8-generated-coc} & \UseMacro{table-lcoc-stats-row8-generated-ioc} & \UseMacro{table-lcoc-stats-row8-pattern-coc} & \UseMacro{table-lcoc-stats-row8-pattern-ioc} & \UseMacro{table-lcoc-stats-row8-reduction-coc} & \UseMacro{table-lcoc-stats-row8-reduction-ioc}\\
\UseMacro{table-lcoc-stats-row9-names-0} & \UseMacro{table-lcoc-stats-row9-manual-coc} & \UseMacro{table-lcoc-stats-row9-manual-ioc} & \UseMacro{table-lcoc-stats-row9-generated-coc} & \UseMacro{table-lcoc-stats-row9-generated-ioc} & \UseMacro{table-lcoc-stats-row9-pattern-coc} & \UseMacro{table-lcoc-stats-row9-pattern-ioc} & \UseMacro{table-lcoc-stats-row9-reduction-coc} & \UseMacro{table-lcoc-stats-row9-reduction-ioc}\\
\UseMacro{table-lcoc-stats-row10-names-0} & \UseMacro{table-lcoc-stats-row10-manual-coc} & \UseMacro{table-lcoc-stats-row10-manual-ioc} & \UseMacro{table-lcoc-stats-row10-generated-coc} & \UseMacro{table-lcoc-stats-row10-generated-ioc} & \UseMacro{table-lcoc-stats-row10-pattern-coc} & \UseMacro{table-lcoc-stats-row10-pattern-ioc} & \UseMacro{table-lcoc-stats-row10-reduction-coc} & \UseMacro{table-lcoc-stats-row10-reduction-ioc}\\
\UseMacro{table-lcoc-stats-row11-names-0} & \UseMacro{table-lcoc-stats-row11-manual-coc} & \UseMacro{table-lcoc-stats-row11-manual-ioc} & \UseMacro{table-lcoc-stats-row11-generated-coc} & \UseMacro{table-lcoc-stats-row11-generated-ioc} & \UseMacro{table-lcoc-stats-row11-pattern-coc} & \UseMacro{table-lcoc-stats-row11-pattern-ioc} & \UseMacro{table-lcoc-stats-row11-reduction-coc} & \UseMacro{table-lcoc-stats-row11-reduction-ioc}\\
\UseMacro{table-lcoc-stats-row12-names-0} & \UseMacro{table-lcoc-stats-row12-manual-coc} & \UseMacro{table-lcoc-stats-row12-manual-ioc} & \UseMacro{table-lcoc-stats-row12-generated-coc} & \UseMacro{table-lcoc-stats-row12-generated-ioc} & \UseMacro{table-lcoc-stats-row12-pattern-coc} & \UseMacro{table-lcoc-stats-row12-pattern-ioc} & \UseMacro{table-lcoc-stats-row12-reduction-coc} & \UseMacro{table-lcoc-stats-row12-reduction-ioc}\\
\cmidrule{1-9}\UseMacro{table-lcoc-stats-row13-names-0} & \multirow{4}{*}{\UseMacro{table-lcoc-stats-row13-manual-coc}} & \multirow{4}{*}{\UseMacro{table-lcoc-stats-row13-manual-ioc}} & \multirow{4}{*}{\UseMacro{table-lcoc-stats-row13-generated-coc}} & \multirow{4}{*}{\UseMacro{table-lcoc-stats-row13-generated-ioc}} & \multirow{4}{*}{\UseMacro{table-lcoc-stats-row13-pattern-coc}} & \multirow{4}{*}{\UseMacro{table-lcoc-stats-row13-pattern-ioc}} & \multirow{4}{*}{\UseMacro{table-lcoc-stats-row13-reduction-coc}} & \multirow{4}{*}{\UseMacro{table-lcoc-stats-row13-reduction-ioc}}\\
\UseMacro{table-lcoc-stats-row13-names-1} &  &  &  &  &  &  &  & \\
\UseMacro{table-lcoc-stats-row13-names-2} &  &  &  &  &  &  &  & \\
\UseMacro{table-lcoc-stats-row13-names-3} &  &  &  &  &  &  &  & \\
\cmidrule{1-9}\UseMacro{table-lcoc-stats-row14-names-0} & \UseMacro{table-lcoc-stats-row14-manual-coc} & \UseMacro{table-lcoc-stats-row14-manual-ioc} & \UseMacro{table-lcoc-stats-row14-generated-coc} & \UseMacro{table-lcoc-stats-row14-generated-ioc} & \UseMacro{table-lcoc-stats-row14-pattern-coc} & \UseMacro{table-lcoc-stats-row14-pattern-ioc} & \UseMacro{table-lcoc-stats-row14-reduction-coc} & \UseMacro{table-lcoc-stats-row14-reduction-ioc}\\
\UseMacro{table-lcoc-stats-row15-names-0} & \UseMacro{table-lcoc-stats-row15-manual-coc} & \UseMacro{table-lcoc-stats-row15-manual-ioc} & \UseMacro{table-lcoc-stats-row15-generated-coc} & \UseMacro{table-lcoc-stats-row15-generated-ioc} & \UseMacro{table-lcoc-stats-row15-pattern-coc} & \UseMacro{table-lcoc-stats-row15-pattern-ioc} & \UseMacro{table-lcoc-stats-row15-reduction-coc} & \UseMacro{table-lcoc-stats-row15-reduction-ioc}\\
\UseMacro{table-lcoc-stats-row16-names-0} & \UseMacro{table-lcoc-stats-row16-manual-coc} & \UseMacro{table-lcoc-stats-row16-manual-ioc} & \UseMacro{table-lcoc-stats-row16-generated-coc} & \UseMacro{table-lcoc-stats-row16-generated-ioc} & \UseMacro{table-lcoc-stats-row16-pattern-coc} & \UseMacro{table-lcoc-stats-row16-pattern-ioc} & \UseMacro{table-lcoc-stats-row16-reduction-coc} & \UseMacro{table-lcoc-stats-row16-reduction-ioc}\\
\cmidrule{1-9}\UseMacro{table-lcoc-stats-row17-names-0} & \multirow{2}{*}{\UseMacro{table-lcoc-stats-row17-manual-coc}} & \multirow{2}{*}{\UseMacro{table-lcoc-stats-row17-manual-ioc}} & \multirow{2}{*}{\UseMacro{table-lcoc-stats-row17-generated-coc}} & \multirow{2}{*}{\UseMacro{table-lcoc-stats-row17-generated-ioc}} & \multirow{2}{*}{\UseMacro{table-lcoc-stats-row17-pattern-coc}} & \multirow{2}{*}{\UseMacro{table-lcoc-stats-row17-pattern-ioc}} & \multirow{2}{*}{\UseMacro{table-lcoc-stats-row17-reduction-coc}} & \multirow{2}{*}{\UseMacro{table-lcoc-stats-row17-reduction-ioc}}\\
\UseMacro{table-lcoc-stats-row17-names-1} &  &  &  &  &  &  &  & \\
\cmidrule{1-9}\UseMacro{table-lcoc-stats-row18-names-0} & \multirow{2}{*}{\UseMacro{table-lcoc-stats-row18-manual-coc}} & \multirow{2}{*}{\UseMacro{table-lcoc-stats-row18-manual-ioc}} & \multirow{2}{*}{\UseMacro{table-lcoc-stats-row18-generated-coc}} & \multirow{2}{*}{\UseMacro{table-lcoc-stats-row18-generated-ioc}} & \multirow{2}{*}{\UseMacro{table-lcoc-stats-row18-pattern-coc}} & \multirow{2}{*}{\UseMacro{table-lcoc-stats-row18-pattern-ioc}} & \multirow{2}{*}{\UseMacro{table-lcoc-stats-row18-reduction-coc}} & \multirow{2}{*}{\UseMacro{table-lcoc-stats-row18-reduction-ioc}}\\
\UseMacro{table-lcoc-stats-row18-names-1} &  &  &  &  &  &  &  & \\
\cmidrule{1-9}\UseMacro{table-lcoc-stats-row19-names-0} & \UseMacro{table-lcoc-stats-row19-manual-coc} & \UseMacro{table-lcoc-stats-row19-manual-ioc} & \UseMacro{table-lcoc-stats-row19-generated-coc} & \UseMacro{table-lcoc-stats-row19-generated-ioc} & \UseMacro{table-lcoc-stats-row19-pattern-coc} & \UseMacro{table-lcoc-stats-row19-pattern-ioc} & \UseMacro{table-lcoc-stats-row19-reduction-coc} & \UseMacro{table-lcoc-stats-row19-reduction-ioc}\\
\UseMacro{table-lcoc-stats-row20-names-0} & \UseMacro{table-lcoc-stats-row20-manual-coc} & \UseMacro{table-lcoc-stats-row20-manual-ioc} & \UseMacro{table-lcoc-stats-row20-generated-coc} & \UseMacro{table-lcoc-stats-row20-generated-ioc} & \UseMacro{table-lcoc-stats-row20-pattern-coc} & \UseMacro{table-lcoc-stats-row20-pattern-ioc} & \UseMacro{table-lcoc-stats-row20-reduction-coc} & \UseMacro{table-lcoc-stats-row20-reduction-ioc}\\
\UseMacro{table-lcoc-stats-row21-names-0} & \UseMacro{table-lcoc-stats-row21-manual-coc} & \UseMacro{table-lcoc-stats-row21-manual-ioc} & \UseMacro{table-lcoc-stats-row21-generated-coc} & \UseMacro{table-lcoc-stats-row21-generated-ioc} & \UseMacro{table-lcoc-stats-row21-pattern-coc} & \UseMacro{table-lcoc-stats-row21-pattern-ioc} & \UseMacro{table-lcoc-stats-row21-reduction-coc} & \UseMacro{table-lcoc-stats-row21-reduction-ioc}\\
\UseMacro{table-lcoc-stats-row22-names-0} & \UseMacro{table-lcoc-stats-row22-manual-coc} & \UseMacro{table-lcoc-stats-row22-manual-ioc} & \UseMacro{table-lcoc-stats-row22-generated-coc} & \UseMacro{table-lcoc-stats-row22-generated-ioc} & \UseMacro{table-lcoc-stats-row22-pattern-coc} & \UseMacro{table-lcoc-stats-row22-pattern-ioc} & \UseMacro{table-lcoc-stats-row22-reduction-coc} & \UseMacro{table-lcoc-stats-row22-reduction-ioc}\\
\UseMacro{table-lcoc-stats-row23-names-0} & \UseMacro{table-lcoc-stats-row23-manual-coc} & \UseMacro{table-lcoc-stats-row23-manual-ioc} & \UseMacro{table-lcoc-stats-row23-generated-coc} & \UseMacro{table-lcoc-stats-row23-generated-ioc} & \UseMacro{table-lcoc-stats-row23-pattern-coc} & \UseMacro{table-lcoc-stats-row23-pattern-ioc} & \UseMacro{table-lcoc-stats-row23-reduction-coc} & \UseMacro{table-lcoc-stats-row23-reduction-ioc}\\
\cmidrule{1-9}\UseMacro{table-lcoc-stats-row24-names-0} & \multirow{4}{*}{\UseMacro{table-lcoc-stats-row24-manual-coc}} & \multirow{4}{*}{\UseMacro{table-lcoc-stats-row24-manual-ioc}} & \multirow{4}{*}{\UseMacro{table-lcoc-stats-row24-generated-coc}} & \multirow{4}{*}{\UseMacro{table-lcoc-stats-row24-generated-ioc}} & \multirow{4}{*}{\UseMacro{table-lcoc-stats-row24-pattern-coc}} & \multirow{4}{*}{\UseMacro{table-lcoc-stats-row24-pattern-ioc}} & \multirow{4}{*}{\UseMacro{table-lcoc-stats-row24-reduction-coc}} & \multirow{4}{*}{\UseMacro{table-lcoc-stats-row24-reduction-ioc}}\\
\UseMacro{table-lcoc-stats-row24-names-1} &  &  &  &  &  &  &  & \\
\UseMacro{table-lcoc-stats-row24-names-2} &  &  &  &  &  &  &  & \\
\UseMacro{table-lcoc-stats-row24-names-3} &  &  &  &  &  &  &  & \\
\cmidrule{1-9}\UseMacro{table-lcoc-stats-row25-names-0} & \multirow{2}{*}{\UseMacro{table-lcoc-stats-row25-manual-coc}} & \multirow{2}{*}{\UseMacro{table-lcoc-stats-row25-manual-ioc}} & \multirow{2}{*}{\UseMacro{table-lcoc-stats-row25-generated-coc}} & \multirow{2}{*}{\UseMacro{table-lcoc-stats-row25-generated-ioc}} & \multirow{2}{*}{\UseMacro{table-lcoc-stats-row25-pattern-coc}} & \multirow{2}{*}{\UseMacro{table-lcoc-stats-row25-pattern-ioc}} & \multirow{2}{*}{\UseMacro{table-lcoc-stats-row25-reduction-coc}} & \multirow{2}{*}{\UseMacro{table-lcoc-stats-row25-reduction-ioc}}\\
\UseMacro{table-lcoc-stats-row25-names-1} &  &  &  &  &  &  &  & \\
\cmidrule{1-9}\UseMacro{table-lcoc-stats-row26-names-0} & \UseMacro{table-lcoc-stats-row26-manual-coc} & \UseMacro{table-lcoc-stats-row26-manual-ioc} & \UseMacro{table-lcoc-stats-row26-generated-coc} & \UseMacro{table-lcoc-stats-row26-generated-ioc} & \UseMacro{table-lcoc-stats-row26-pattern-coc} & \UseMacro{table-lcoc-stats-row26-pattern-ioc} & \UseMacro{table-lcoc-stats-row26-reduction-coc} & \UseMacro{table-lcoc-stats-row26-reduction-ioc}\\
\cmidrule{1-9}\UseMacro{table-lcoc-stats-row27-names-0} & \multirow{2}{*}{\UseMacro{table-lcoc-stats-row27-manual-coc}} & \multirow{2}{*}{\UseMacro{table-lcoc-stats-row27-manual-ioc}} & \multirow{2}{*}{\UseMacro{table-lcoc-stats-row27-generated-coc}} & \multirow{2}{*}{\UseMacro{table-lcoc-stats-row27-generated-ioc}} & \multirow{2}{*}{\UseMacro{table-lcoc-stats-row27-pattern-coc}} & \multirow{2}{*}{\UseMacro{table-lcoc-stats-row27-pattern-ioc}} & \multirow{2}{*}{\UseMacro{table-lcoc-stats-row27-reduction-coc}} & \multirow{2}{*}{\UseMacro{table-lcoc-stats-row27-reduction-ioc}}\\
\UseMacro{table-lcoc-stats-row27-names-1} &  &  &  &  &  &  &  & \\
\cmidrule{1-9}\UseMacro{table-lcoc-stats-row28-names-0} & \multirow{2}{*}{\UseMacro{table-lcoc-stats-row28-manual-coc}} & \multirow{2}{*}{\UseMacro{table-lcoc-stats-row28-manual-ioc}} & \multirow{2}{*}{\UseMacro{table-lcoc-stats-row28-generated-coc}} & \multirow{2}{*}{\UseMacro{table-lcoc-stats-row28-generated-ioc}} & \multirow{2}{*}{\UseMacro{table-lcoc-stats-row28-pattern-coc}} & \multirow{2}{*}{\UseMacro{table-lcoc-stats-row28-pattern-ioc}} & \multirow{2}{*}{\UseMacro{table-lcoc-stats-row28-reduction-coc}} & \multirow{2}{*}{\UseMacro{table-lcoc-stats-row28-reduction-ioc}}\\
\UseMacro{table-lcoc-stats-row28-names-1} &  &  &  &  &  &  &  & \\
\cmidrule{1-9}\UseMacro{table-lcoc-stats-row29-names-0} & \UseMacro{table-lcoc-stats-row29-manual-coc} & \UseMacro{table-lcoc-stats-row29-manual-ioc} & \UseMacro{table-lcoc-stats-row29-generated-coc} & \UseMacro{table-lcoc-stats-row29-generated-ioc} & \UseMacro{table-lcoc-stats-row29-pattern-coc} & \UseMacro{table-lcoc-stats-row29-pattern-ioc} & \UseMacro{table-lcoc-stats-row29-reduction-coc} & \UseMacro{table-lcoc-stats-row29-reduction-ioc}\\
\UseMacro{table-lcoc-stats-row30-names-0} & \UseMacro{table-lcoc-stats-row30-manual-coc} & \UseMacro{table-lcoc-stats-row30-manual-ioc} & \UseMacro{table-lcoc-stats-row30-generated-coc} & \UseMacro{table-lcoc-stats-row30-generated-ioc} & \UseMacro{table-lcoc-stats-row30-pattern-coc} & \UseMacro{table-lcoc-stats-row30-pattern-ioc} & \UseMacro{table-lcoc-stats-row30-reduction-coc} & \UseMacro{table-lcoc-stats-row30-reduction-ioc}\\
\UseMacro{table-lcoc-stats-row31-names-0} & \UseMacro{table-lcoc-stats-row31-manual-coc} & \UseMacro{table-lcoc-stats-row31-manual-ioc} & \UseMacro{table-lcoc-stats-row31-generated-coc} & \UseMacro{table-lcoc-stats-row31-generated-ioc} & \UseMacro{table-lcoc-stats-row31-pattern-coc} & \UseMacro{table-lcoc-stats-row31-pattern-ioc} & \UseMacro{table-lcoc-stats-row31-reduction-coc} & \UseMacro{table-lcoc-stats-row31-reduction-ioc}\\
\cmidrule{1-9}\UseMacro{table-lcoc-stats-row32-names-0} & \multirow{2}{*}{\UseMacro{table-lcoc-stats-row32-manual-coc}} & \multirow{2}{*}{\UseMacro{table-lcoc-stats-row32-manual-ioc}} & \multirow{2}{*}{\UseMacro{table-lcoc-stats-row32-generated-coc}} & \multirow{2}{*}{\UseMacro{table-lcoc-stats-row32-generated-ioc}} & \multirow{2}{*}{\UseMacro{table-lcoc-stats-row32-pattern-coc}} & \multirow{2}{*}{\UseMacro{table-lcoc-stats-row32-pattern-ioc}} & \multirow{2}{*}{\UseMacro{table-lcoc-stats-row32-reduction-coc}} & \multirow{2}{*}{\UseMacro{table-lcoc-stats-row32-reduction-ioc}}\\
\UseMacro{table-lcoc-stats-row32-names-1} &  &  &  &  &  &  &  & \\
\cmidrule{1-9}\UseMacro{table-lcoc-stats-row33-names-0} & \multirow{2}{*}{\UseMacro{table-lcoc-stats-row33-manual-coc}} & \multirow{2}{*}{\UseMacro{table-lcoc-stats-row33-manual-ioc}} & \multirow{2}{*}{\UseMacro{table-lcoc-stats-row33-generated-coc}} & \multirow{2}{*}{\UseMacro{table-lcoc-stats-row33-generated-ioc}} & \multirow{2}{*}{\UseMacro{table-lcoc-stats-row33-pattern-coc}} & \multirow{2}{*}{\UseMacro{table-lcoc-stats-row33-pattern-ioc}} & \multirow{2}{*}{\UseMacro{table-lcoc-stats-row33-reduction-coc}} & \multirow{2}{*}{\UseMacro{table-lcoc-stats-row33-reduction-ioc}}\\
\UseMacro{table-lcoc-stats-row33-names-1} &  &  &  &  &  &  &  & \\
\cmidrule{1-9}\UseMacro{table-lcoc-stats-row34-names-0} & \UseMacro{table-lcoc-stats-row34-manual-coc} & \UseMacro{table-lcoc-stats-row34-manual-ioc} & \UseMacro{table-lcoc-stats-row34-generated-coc} & \UseMacro{table-lcoc-stats-row34-generated-ioc} & \UseMacro{table-lcoc-stats-row34-pattern-coc} & \UseMacro{table-lcoc-stats-row34-pattern-ioc} & \UseMacro{table-lcoc-stats-row34-reduction-coc} & \UseMacro{table-lcoc-stats-row34-reduction-ioc}\\
\UseMacro{table-lcoc-stats-row35-names-0} & \UseMacro{table-lcoc-stats-row35-manual-coc} & \UseMacro{table-lcoc-stats-row35-manual-ioc} & \UseMacro{table-lcoc-stats-row35-generated-coc} & \UseMacro{table-lcoc-stats-row35-generated-ioc} & \UseMacro{table-lcoc-stats-row35-pattern-coc} & \UseMacro{table-lcoc-stats-row35-pattern-ioc} & \UseMacro{table-lcoc-stats-row35-reduction-coc} & \UseMacro{table-lcoc-stats-row35-reduction-ioc}\\
\UseMacro{table-lcoc-stats-row36-names-0} & \UseMacro{table-lcoc-stats-row36-manual-coc} & \UseMacro{table-lcoc-stats-row36-manual-ioc} & \UseMacro{table-lcoc-stats-row36-generated-coc} & \UseMacro{table-lcoc-stats-row36-generated-ioc} & \UseMacro{table-lcoc-stats-row36-pattern-coc} & \UseMacro{table-lcoc-stats-row36-pattern-ioc} & \UseMacro{table-lcoc-stats-row36-reduction-coc} & \UseMacro{table-lcoc-stats-row36-reduction-ioc}\\
\UseMacro{table-lcoc-stats-row37-names-0} & \UseMacro{table-lcoc-stats-row37-manual-coc} & \UseMacro{table-lcoc-stats-row37-manual-ioc} & \UseMacro{table-lcoc-stats-row37-generated-coc} & \UseMacro{table-lcoc-stats-row37-generated-ioc} & \UseMacro{table-lcoc-stats-row37-pattern-coc} & \UseMacro{table-lcoc-stats-row37-pattern-ioc} & \UseMacro{table-lcoc-stats-row37-reduction-coc} & \UseMacro{table-lcoc-stats-row37-reduction-ioc}\\
\UseMacro{table-lcoc-stats-row38-names-0} & \UseMacro{table-lcoc-stats-row38-manual-coc} & \UseMacro{table-lcoc-stats-row38-manual-ioc} & \UseMacro{table-lcoc-stats-row38-generated-coc} & \UseMacro{table-lcoc-stats-row38-generated-ioc} & \UseMacro{table-lcoc-stats-row38-pattern-coc} & \UseMacro{table-lcoc-stats-row38-pattern-ioc} & \UseMacro{table-lcoc-stats-row38-reduction-coc} & \UseMacro{table-lcoc-stats-row38-reduction-ioc}\\
\UseMacro{table-lcoc-stats-row39-names-0} & \UseMacro{table-lcoc-stats-row39-manual-coc} & \UseMacro{table-lcoc-stats-row39-manual-ioc} & \UseMacro{table-lcoc-stats-row39-generated-coc} & \UseMacro{table-lcoc-stats-row39-generated-ioc} & \UseMacro{table-lcoc-stats-row39-pattern-coc} & \UseMacro{table-lcoc-stats-row39-pattern-ioc} & \UseMacro{table-lcoc-stats-row39-reduction-coc} & \UseMacro{table-lcoc-stats-row39-reduction-ioc}\\
\UseMacro{table-lcoc-stats-row40-names-0} & \UseMacro{table-lcoc-stats-row40-manual-coc} & \UseMacro{table-lcoc-stats-row40-manual-ioc} & \UseMacro{table-lcoc-stats-row40-generated-coc} & \UseMacro{table-lcoc-stats-row40-generated-ioc} & \UseMacro{table-lcoc-stats-row40-pattern-coc} & \UseMacro{table-lcoc-stats-row40-pattern-ioc} & \UseMacro{table-lcoc-stats-row40-reduction-coc} & \UseMacro{table-lcoc-stats-row40-reduction-ioc}\\
\midrule
$\sum$ & \UseMacro{table-lcoc-stats-row-sum-manual-coc} & \UseMacro{table-lcoc-stats-row-sum-manual-ioc} & \UseMacro{table-lcoc-stats-row-sum-generated-coc} & \UseMacro{table-lcoc-stats-row-sum-generated-ioc} & \UseMacro{table-lcoc-stats-row-sum-pattern-coc} & \UseMacro{table-lcoc-stats-row-sum-pattern-ioc} & \UseMacro{table-lcoc-stats-row-sum-reduction-coc} & \UseMacro{table-lcoc-stats-row-sum-reduction-ioc}\\
\bottomrule
\end{tabular}
\end{\UseMacro{table-lcoc-stats-fontsize}}
\end{\UseMacro{table-lcoc-stats-float-env}}

Using \Tool to write \ptns instead of directly writing C/C++ code, the
total characters written is decreased from
\UseMacro{table-lcoc-stats-row-sum-manual-coc} to
\UseMacro{table-lcoc-stats-row-sum-pattern-coc}, and the total
identifiers written is decreased from
\UseMacro{table-lcoc-stats-row-sum-manual-ioc} to
\UseMacro{table-lcoc-stats-row-sum-pattern-ioc}.
The characters of hand-written C/C++ code is an underestimate of
actual numbers because in most cases we do not include the additional
lines for declarations of variables due to inconvenience of counting.
Due to the same reason, the identifiers of hand-written C/C++ code
shown in Table~\ref{tab:lcoc-stats} is also an underestimate of actual
numbers. However, using \Tool to write \ptns still shows a significant
\UseMacro{table-lcoc-stats-row-sum-reduction-coc}\% savings in terms of the number of
characters and
\UseMacro{table-lcoc-stats-row-sum-reduction-ioc}\% savings in terms of the number of
identifiers.

There are few groups of \optimizations where we write more characters
and/or identifiers of code to express them in \ptns using \Tool.
For example, \UseMacro{table-lcoc-stats-row32-names-0} is an
\optimization that transforms \CodeIn{(x \char`\>{}\char`\>{} C0)
\char`\<{}\char`\<{} C0} into \CodeIn{x \& -(1 \char`\<{}\char`\<{}
C0)}, and \UseMacro{table-lcoc-stats-row32-names-1} is a very
similar \optimization that transforms \CodeIn{(x \char`\>{}\char`\>{}>
C0) \char`\<{}\char`\<{} C0} into the same result. In \OpenJDK these
two \optimizations are implemented together by including both
\CodeIn{\char`\>{}\char`\>{}} and \CodeIn{\char`\>{}\char`\>{}>}
operators, but we write two separate \ptns using \Tool, in more lines
of code. However, \Tool still saves
\UseMacro{table-lcoc-stats-row32-reduction-coc}\%
characters.
We leave how to express the same simplification as in \OpenJDK using
\Tool as future work.

We also count the number of characters and identifiers of generated
C/C++ code from \Tool (see columns of
``\UseMacro{table-lcoc-stats-col-generated}''). It is unsurprising
that the generated code has much higher numbers of characters and
identifiers than hand-written code, because \Tool's design of C/C++
code generation prefers consistency to flexibility of coding style,
which will benefit future maintenance. For example, \CodeIn{is\_int()}
and \CodeIn{isa\_int()} are used interchangeably in \OpenJDK to check
if a type is of \CodeIn{int}, but \Tool sticks to \CodeIn{isa\_int()},
which is recommended, because it returns \CodeIn{NULL} instead of
throwing an assertion failure when the checked type is not
\CodeIn{int}. After all, as long as generated code keeps the
effectiveness of \optimizations, it is always preferred to increase
maintainability. We will compare performance of generated code and
hand-written code in Section~\ref{sec:eval:performance}.

\subsection{Performance}
\label{sec:eval:performance}

Our objective with RQ2 is to demonstrate that the performance of \JIT
does not substantially change when replacing the hand-written code in
\OpenJDK with code generated from \ptns.

A total of \UseMacro{total-num-patterns-OpenJDK} \optimizations in
\OpenJDK are replaced using code generated from \Tool. To evaluate
their performance, we use the \Renaissance benchmark
suite~\cite{Prokopec19Renaissance} which is a benchmark suite for \JVM
consisting of 27 individual benchmarks. Some of these benchmarks
(\textit{neo4j-analytics}) are incompatible with Java 18 (Java version
used in this paper) and are discarded from the experiment.
Furthermore, some benchmarks exhibit large variance in their execution
times across multiple runs.
Since such large variances can lead to inaccuracies in performance
evaluation, these benchmarks need to be discarded. To identify these
benchmarks, we built a ``vanilla'' version of \OpenJDK termed as
baseline. Each benchmark is executed on the baseline build of \OpenJDK
five times with each execution consisting of
\UseMacro{total-repetitions-per-run} iterations to warm up the JVM and
consequently trigger the JIT optimizations. We then compute the
coefficient of variance (CV)~\cite{Blackburn06DeCapo} for each
benchmark across the five runs by only considering the last
\UseMacro{measured-repetitions-per-run} iterations of each run, when
JVM is fully warmed up. Benchmarks with CV exceeding 10\% are
discarded from the experiment. Following these two filtering stages,
we excluded 17 more benchmarks and the remaining suitable benchmarks
used in the experiment are \textit{log-regression, als, page-rank,
finagle-http, scala-kmeans, fj-kmeans, gauss-mix, par-mnemonics} and
\textit{dec-tree}.

We now describe the approach used to evaluate the performance of an
optimization. The previously identified benchmarks are executed 5
times each on the baseline build with each run once again consisting
of \UseMacro{total-repetitions-per-run} iterations. The baseline
execution time for a benchmark is then computed by averaging the last
\UseMacro{measured-repetitions-per-run} iterations over the 5 runs.
This procedure is then repeated for each optimization, by replacing
the hand-written code in the baseline source with the \Tool generated
code for the corresponding optimization, to yield the execution time
of the benchmarks for that optimization build. A relative difference
measure as shown below is then used to evaluate the performance of an
optimization (\Tool generated) relative to baseline (hand-written):
\[
\left.\frac{\textnormal{time}_\textnormal{hand-written} - \textnormal{time}_{\textnormal{generated}}}{\textnormal{time}_{\textnormal{hand-written}}} \right \rvert_{\textnormal{benchmark}}
\]
where
$\textnormal{time}_{\textnormal{hand-written}}$/$\textnormal{time}_{\textnormal{generated}}$
is the average execution time of a \textit{benchmark} based on
bootstrap re-sampling~\cite{Efron94Bootstrap} from
\UseMacro{total-measured-repetitions} total executions, i.e., last
\UseMacro{measured-repetitions-per-run} iterations over
\UseMacro{num-end-to-end-runs} runs, of hand-written/generated code.
Figure~\ref{fig:performance-speedup-generated-relative-to-manual}
shows the percentage speedup of \Tool generated code relative to
hand-written code for every group of \optimizations in
Table~\ref{tab:lcoc-stats} and the filtered subset of benchmarks. A
positive speedup of generated code relative to hand-written code is
marked with an up arrow and a negative speedup (slowdown) with a down
arrow. Based on the results of significant difference testing using
bootstrap re-sampling, those without statistically significant
difference ($p=0.05$) between \Tool generated and hand-written are
marked with circles. From
Figure~\ref{fig:performance-speedup-generated-relative-to-manual},
most of benchmarks show no significant difference or small differences
within the range of 5\%. Some benchmarks, e.g., \emph{fj-kmeans} and
\emph{gauss-mix}, show more differences. We investigated the
benchmarks and found such differences even existed when comparing
results of baselines between two experiments, which indicates such
benchmarks are more sensitive to noise.
Overall, the execution times of \OpenJDK build for the \Renaissance
suite with \Tool generated code is comparable in performance to that
with hand-written code.

\begin{figure}
\centering
\includegraphics[width=\linewidth]{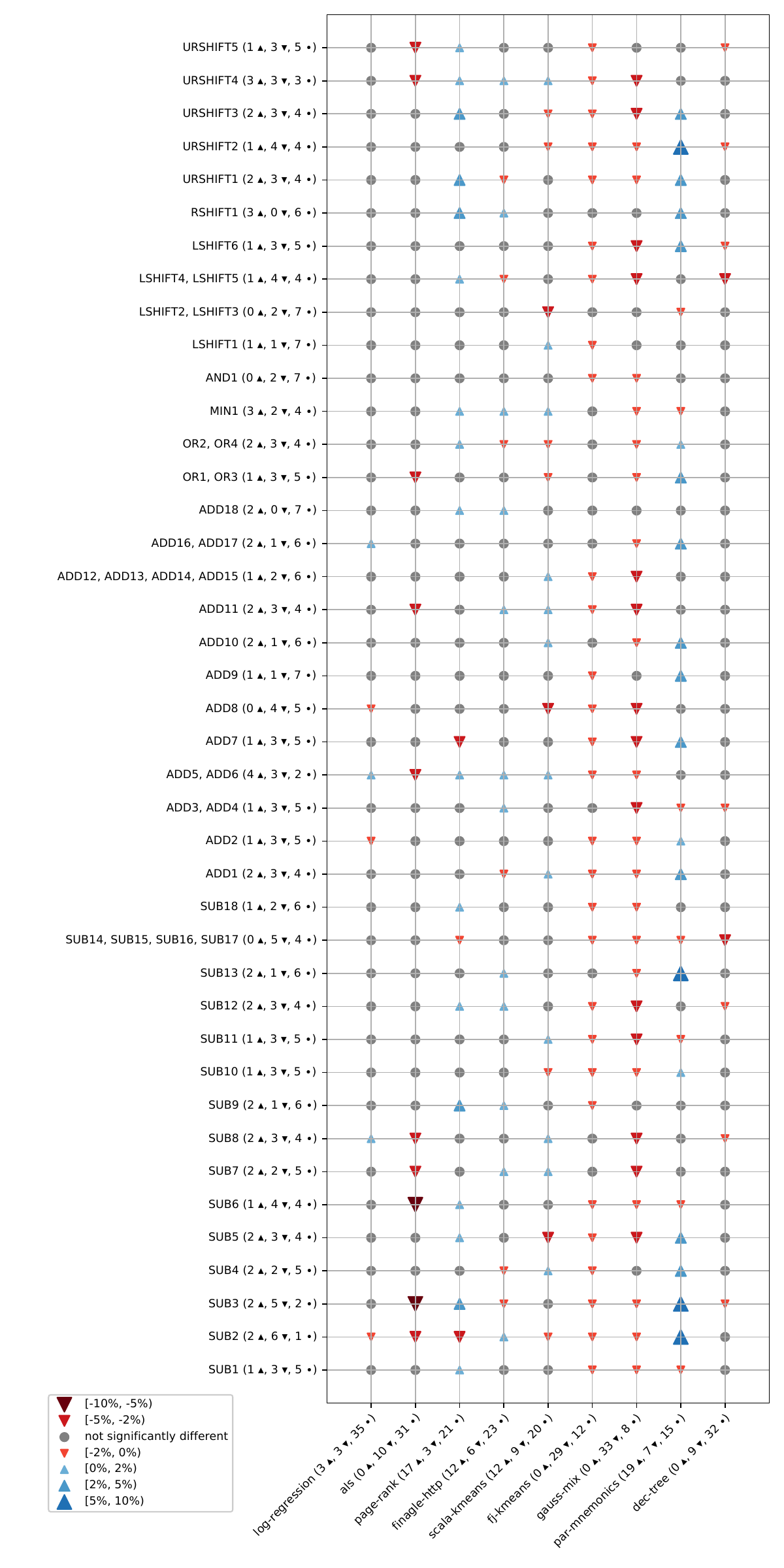}
\vspace{-10pt}
\caption{\UseMacro{figure-performance-speedup-generated-relative-to-manual-caption}}
\end{figure}

\subsection{\Shadow between \Ptns}
\label{sec:eval:shadow}
Recall that \Tool is able to check if one \ptn \shadows another \ptn
(Section~\ref{sec:shadow}).
We check each pair of \ptns (total \UseMacro{total-num-patterns} \ptns) to
evaluate \Tool's effectiveness on detecting \shadows between \ptns.
We set a timeout with \UseMacro{shadow-timeout-per-pair} seconds for
every check of two \ptns and all checks finish within the given time
and return definite results (\textit{YES} or \textit{NO}, without
\textit{UNKNOWN} as defined in Figure~\ref{algo:shadow-deciding}).
Table~\ref{tab:shadow-stats} enumerates all \UseMacro{total-shadows}
pairs of \ptns where one \shadows the other (\Tool returned
\textit{YES}). In every row,
``\UseMacro{table-shadow-stats-col-shadowing}'' \ptn \shadows
``\UseMacro{table-shadow-stats-col-shadowed}'' \ptn. The column of
``\UseMacro{table-shadow-stats-col-before}'' shows the expression in
\CodeIn{before} API, which is the expression to be matched in the
\ptn, and the column of
``\UseMacro{table-shadow-stats-col-precondition}'' shows the
precondition of the \ptn, where $\top$ means no precondition.

\begin{\UseMacro{table-shadow-stats-float-env}}[t]
\centering
\caption{\UseMacro{table-shadow-stats-caption}}
\begin{\UseMacro{table-shadow-stats-fontsize}}
\begin{tabular}{rrrr}
\toprule
\multicolumn{2}{c}{\textbf{\UseMacro{table-shadow-stats-col-shadowing}}} & \multicolumn{2}{c}{\textbf{\UseMacro{table-shadow-stats-col-shadowed}}}\\
\cmidrule(r){1-2}\cmidrule(l){3-4}
\textbf{\UseMacro{table-shadow-stats-col-before}} & \textbf{\UseMacro{table-shadow-stats-col-precondition}} & \textbf{\UseMacro{table-shadow-stats-col-before}} & \textbf{\UseMacro{table-shadow-stats-col-precondition}}\\
\midrule
\UseMacro{table-shadow-stats-row0-shadowing-before} & \UseMacro{table-shadow-stats-row0-shadowing-precondition} & \UseMacro{table-shadow-stats-row0-shadowed-before-0} & \UseMacro{table-shadow-stats-row0-shadowed-precondition-0}\\
\cmidrule{1-4}\multicolumn{1}{r}{\multirow{3}{*}{\UseMacro{table-shadow-stats-row1-shadowing-before}}} & \multicolumn{1}{r}{\multirow{3}{*}{\UseMacro{table-shadow-stats-row1-shadowing-precondition}}} & \UseMacro{table-shadow-stats-row1-shadowed-before-0} & \UseMacro{table-shadow-stats-row1-shadowed-precondition-0}\\
& & \UseMacro{table-shadow-stats-row1-shadowed-before-1} & \UseMacro{table-shadow-stats-row1-shadowed-precondition-1}\\
& & \UseMacro{table-shadow-stats-row1-shadowed-before-2} & \UseMacro{table-shadow-stats-row1-shadowed-precondition-2}\\
\cmidrule{1-4}\UseMacro{table-shadow-stats-row2-shadowing-before} & \UseMacro{table-shadow-stats-row2-shadowing-precondition} & \UseMacro{table-shadow-stats-row2-shadowed-before-0} & \UseMacro{table-shadow-stats-row2-shadowed-precondition-0}\\
\cmidrule{1-4}\UseMacro{table-shadow-stats-row3-shadowing-before} & \UseMacro{table-shadow-stats-row3-shadowing-precondition} & \UseMacro{table-shadow-stats-row3-shadowed-before-0} & \UseMacro{table-shadow-stats-row3-shadowed-precondition-0}\\
\cmidrule{1-4}\multicolumn{1}{r}{\multirow{5}{*}{\UseMacro{table-shadow-stats-row4-shadowing-before}}} & \multicolumn{1}{r}{\multirow{5}{*}{\UseMacro{table-shadow-stats-row4-shadowing-precondition}}} & \UseMacro{table-shadow-stats-row4-shadowed-before-0} & \UseMacro{table-shadow-stats-row4-shadowed-precondition-0}\\
& & \UseMacro{table-shadow-stats-row4-shadowed-before-1} & \UseMacro{table-shadow-stats-row4-shadowed-precondition-1}\\
& & \UseMacro{table-shadow-stats-row4-shadowed-before-2} & \UseMacro{table-shadow-stats-row4-shadowed-precondition-2}\\
& & \UseMacro{table-shadow-stats-row4-shadowed-before-3} & \UseMacro{table-shadow-stats-row4-shadowed-precondition-3}\\
& & \UseMacro{table-shadow-stats-row4-shadowed-before-4} & \UseMacro{table-shadow-stats-row4-shadowed-precondition-4}\\
\cmidrule{1-4}\UseMacro{table-shadow-stats-row5-shadowing-before} & \UseMacro{table-shadow-stats-row5-shadowing-precondition} & \UseMacro{table-shadow-stats-row5-shadowed-before-0} & \UseMacro{table-shadow-stats-row5-shadowed-precondition-0}\\
\cmidrule{1-4}\UseMacro{table-shadow-stats-row6-shadowing-before} & \UseMacro{table-shadow-stats-row6-shadowing-precondition} & \UseMacro{table-shadow-stats-row6-shadowed-before-0} & \UseMacro{table-shadow-stats-row6-shadowed-precondition-0}\\
\cmidrule{1-4}\multicolumn{1}{r}{\multirow{3}{*}{\UseMacro{table-shadow-stats-row7-shadowing-before}}} & \multicolumn{1}{r}{\multirow{3}{*}{\UseMacro{table-shadow-stats-row7-shadowing-precondition}}} & \UseMacro{table-shadow-stats-row7-shadowed-before-0} & \UseMacro{table-shadow-stats-row7-shadowed-precondition-0}\\
& & \UseMacro{table-shadow-stats-row7-shadowed-before-1} & \UseMacro{table-shadow-stats-row7-shadowed-precondition-1}\\
& & \UseMacro{table-shadow-stats-row7-shadowed-before-2} & \UseMacro{table-shadow-stats-row7-shadowed-precondition-2}\\
\cmidrule{1-4}\multicolumn{1}{r}{\multirow{3}{*}{\UseMacro{table-shadow-stats-row8-shadowing-before}}} & \multicolumn{1}{r}{\multirow{3}{*}{\UseMacro{table-shadow-stats-row8-shadowing-precondition}}} & \UseMacro{table-shadow-stats-row8-shadowed-before-0} & \UseMacro{table-shadow-stats-row8-shadowed-precondition-0}\\
& & \UseMacro{table-shadow-stats-row8-shadowed-before-1} & \UseMacro{table-shadow-stats-row8-shadowed-precondition-1}\\
& & \UseMacro{table-shadow-stats-row8-shadowed-before-2} & \UseMacro{table-shadow-stats-row8-shadowed-precondition-2}\\
\bottomrule
\end{tabular}
\end{\UseMacro{table-shadow-stats-fontsize}}
\end{\UseMacro{table-shadow-stats-float-env}}

In order to see if the reported shadow causes any real world issue, we
then manually inspected if any \shadowed \ptn is placed after \shadowing
\ptn (in execution order) in \OpenJDK if both \ptns are implemented in
\OpenJDK, because in this case the \shadowed \ptn would be entirely
\shadowed and thus never be reached in the \optimization pass (see
Section~\ref{sec:shadow}). We found that \ptn
\UseMacro{table-shadow-stats-row1-shadowing-before} \shadows both \ptn
\UseMacro{table-shadow-stats-row1-shadowed-before-1} and \ptn
\UseMacro{table-shadow-stats-row1-shadowed-before-2} while both
\shadowed \ptns are put after the \shadowing \ptns in \OpenJDK. We
reported this to \OpenJDK developers and they confirmed this issue and
accepted our pull request to reorder the \ptns. We discuss more
details for the pull requests in the next section.

\subsection{Test Generation \& Pull Requests}
\label{sec:eval:pr}

We use \Tool to generate IR tests for all
\UseMacro{total-num-patterns-OpenJDK} \ptns adapted from existing
\optimizations in \OpenJDK (so that we can run those tests with
\OpenJDK). Excluding the \ptns with preconditions for which \Tool does
not support generating tests, we successfully generate
\UseMacro{test-num-methods} tests. We also generate
\UseMacro{test-num-classes} test classes each of which wraps all the
tests for patterns with the same operator, e.g., class
\CodeIn{TestSubNode} includes \CodeIn{test\UseMacro{ptn-pSub1}},
\CodeIn{test\UseMacro{ptn-pSub2}}, etc. Next we put the test classes
in \CodeIn{test/hotspot/jtreg/compiler/c2/irTests/}, and
build and run the tests with \OpenJDK. All the
\UseMacro{test-num-methods} tests pass. During our testing, we found
\UseMacro{num-tests-in-test-PR} tests were missing in \OpenJDK; in
other words the corresponding \optimizations were not tested in
\OpenJDK. Thus we opened \UseMacro{total-num-test-PRs} pull
request~\cite{PRAddTests} to add those generated
\UseMacro{num-tests-in-test-PR} tests to existing test suites of
\OpenJDK. This pull request has been integrated into the master branch
of \OpenJDK.

We opened \UseMacro{total-num-non-test-PRs} more pull requests for
\OpenJDK, so far.
The first category of \UseMacro{total-num-new-optimization-PRs} pull
requests was introducing new \JIT \optimizations.
Figure~\ref{fig:pr-patterns:pNewSubAddSub1574}--\ref{fig:pr-patterns:pNewSub_XOrY_Minus_XXorY_}
shows the \Tool \ptns for new \optimizations that we contributed as
PRs; we contributed the C/C++ code generated by \Tool (and not the
\ptns). Also, we contributed the IR tests generated by \Tool for the
\ptns.
Note that one pull request could contain more than one
\ptn/\optimization.
We adapted \ptn \UseMacro{ptn-pNewSubAddSub1574}
(Figure~\ref{fig:pr-patterns:pNewSubAddSub1574}) from \LLVM. We then
added generated C/C++ code from the \ptn to \CodeIn{SubNode::Ideal}
method of \OpenJDK, and opened a pull
request~\cite{PRNewSubAddSub1574} for the changes. Similarly, we
opened another three pull requests~\cite{PRNewAddAddSub1156,
PRNewAddAddSub1202, PRNewXPlus_ConMinusY_} that added generated
C/C++ code from \ptn \UseMacro{ptn-pNewAddAddSub1156}
(Figure~\ref{fig:pr-patterns:pNewAddAddSub1156}), \ptn
\UseMacro{ptn-pNewAddAddSub1202}
(Figure~\ref{fig:pr-patterns:pNewAddAddSub1202}), \ptn
\UseMacro{ptn-pNewXPlus_ConMinusY_}
(Figure~\ref{fig:pr-patterns:pNewXPlus_ConMinusY_}) and \ptn
\UseMacro{ptn-pNewXPlus_ConMinusY_Sym}
(Figure~\ref{fig:pr-patterns:pNewXPlus_ConMinusY_Sym}) to
\CodeIn{AddNode::Ideal} method. Note \ptn
\UseMacro{ptn-pNewXPlus_ConMinusY_} and \ptn
\UseMacro{ptn-pNewXPlus_ConMinusY_Sym} are included in a single pull
request. We also opened one pull request~\cite{PRNewSubAddSub1564} for
\ptn \UseMacro{ptn-pNewSubAddSub1564}
(Figure~\ref{fig:pr-patterns:pNewSubAddSub1564}). All the
\UseMacro{num-integrated-new-optimization-PRs} pull requests have been
integrated into the master branch of \OpenJDK. We opened one pull
request~\cite{PRNewSub_XOrY_Minus_XXorY_} for \ptn
\UseMacro{ptn-pNewSub_XOrY_Minus_XXorY_}
(Figure~\ref{fig:pr-patterns:pNewSub_XOrY_Minus_XXorY_}) and this pull
request is under review.

The second category of pull requests we opened was reordering existing
\JIT \optimizations.
Figure~\ref{fig:pr-patterns:pAdd2}--\ref{fig:pr-patterns:pAdd6} shows
the associated \optimizations as \ptns.
As we mentioned in Section~\ref{sec:eval:shadow}, \ptn
\UseMacro{ptn-pAdd2} \shadows both \ptn \UseMacro{ptn-pAdd5} and \ptn
\UseMacro{ptn-pAdd6}, which means any expression matched by either of
\ptn \UseMacro{ptn-pAdd5} or \ptn \UseMacro{ptn-pAdd6} must be matched
by \ptn \UseMacro{ptn-pAdd2}. Meanwhile in \CodeIn{AddNode::Ideal}
method the C/C++ code implementing \ptn \UseMacro{ptn-pAdd5} and
\UseMacro{ptn-pAdd6} are located after the code for \ptn
\UseMacro{ptn-pAdd2} (in execution order). Therefore \ptn
\UseMacro{ptn-pAdd5} or \UseMacro{ptn-pAdd6} would not be reached
unless they were moved before \ptn \UseMacro{ptn-pAdd2}. We
reported this issue to \OpenJDK developers and initially proposed
reordering in our pull request~\cite{PRRemoveAddNodeP5P6}. \OpenJDK
developers confirmed the issue and then they realized the effects of
the two \ptns have been done by applying two other \optimizations
sequentially and thus it is no longer necessary to have these two
\ptns. \Ptn \UseMacro{ptn-pAdd5} and \ptn \UseMacro{ptn-pAdd6} were
removed from \OpenJDK when the pull request was integrated.
As future work, we plan to extend our \shadow determining algorithm to
detect duplicate \optimization sequences~\cite{CodeSemanticCTOS}.

\begin{figure*}[t]
\begin{subfigure}[b]{0.50\textwidth}
\begin{lstlisting}[language=jog]
@Pattern
public void (*@\UseMacro{ptn-pNewSubAddSub1574}@*)(int x, @Constant int c0, @Constant int c1) {
  before(c0 - (x + c1));
  if (Lib.okToConvert(x + c1, c0)) {
    after((c0 - c1) - x); } }
\end{lstlisting}
\caption{\UseMacro{figure-pr-patterns-pNewSubAddSub1574}}
\end{subfigure}
~
\begin{subfigure}[b]{0.25\textwidth}
\begin{lstlisting}[language=jog]
@Pattern
public void (*@\UseMacro{ptn-pNewAddAddSub1156}@*)(int x) {
  before(x + x);
  after(x << 1); }
\end{lstlisting}
\caption{\UseMacro{figure-pr-patterns-pNewAddAddSub1156}}
\end{subfigure}
~
\begin{subfigure}[b]{0.24\textwidth}
\begin{lstlisting}[language=jog]
@Pattern
public void (*@\UseMacro{ptn-pNewAddAddSub1202}@*)(int x,
         @Constant int c) {
  before((x ^ -1) + c);
  after((c - 1) - x); }
\end{lstlisting}
\caption{\UseMacro{figure-pr-patterns-pNewAddAddSub1202}}
\end{subfigure}
\Hrule
\begin{subfigure}[b]{0.25\textwidth}
\begin{lstlisting}[language=jog]
@Pattern
public void (*@\UseMacro{ptn-pNewXPlus_ConMinusY_}@*)(int x, int y,
         @Constant int con) {
  before(x + (con - y));
  after((x - y) + con); }
\end{lstlisting}
\caption{\UseMacro{figure-pr-patterns-pNewXPlus_ConMinusY_}}
\end{subfigure}
~
\begin{subfigure}[b]{0.25\textwidth}
\begin{lstlisting}[language=jog]
@Pattern
public void (*@\UseMacro{ptn-pNewXPlus_ConMinusY_Sym}@*)(int x, int y,
         @Constant int con) {
  before((con - y) + x);
  after((x - y) + con); }
\end{lstlisting}
\caption{\UseMacro{figure-pr-patterns-pNewXPlus_ConMinusY_Sym}}
\end{subfigure}
~
\begin{subfigure}[b]{0.25\textwidth}
\begin{lstlisting}[language=jog]
@Pattern
public void (*@\UseMacro{ptn-pNewSubAddSub1564}@*)(int x,
         @Constant int c) {
  before(c - (x ^ -1));
  after(x + (c + 1)); }
\end{lstlisting}
\caption{\UseMacro{figure-pr-patterns-pNewSubAddSub1564}}
\end{subfigure}
~
\begin{subfigure}[b]{0.24\textwidth}
\begin{lstlisting}[language=jog]
@Pattern
public void (*@\UseMacro{ptn-pNewSub_XOrY_Minus_XXorY_}@*)(int x,
                    int y) {
  before((x | y) - (x ^ y));
  after(x & y); }
\end{lstlisting}
\caption{\UseMacro{figure-pr-patterns-pNewSub_XOrY_Minus_XXorY_}}
\end{subfigure}
\Hrule
\begin{subfigure}[b]{0.35\textwidth}
\begin{lstlisting}[language=jog]
@Pattern
public void (*@\UseMacro{ptn-pAdd2}@*)(int a, int b, int c, int d) {
  before((a - b) + (c - d));
  after((a + c) - (b + d)); }
\end{lstlisting}
\caption{\UseMacro{figure-pr-patterns-pAdd2}}
\end{subfigure}
~
\begin{subfigure}[b]{0.31\textwidth}
\begin{lstlisting}[language=jog]
@Pattern
public void (*@\UseMacro{ptn-pAdd5}@*)(int a, int b, int c) {
  before((a - b) + (b - c));
  after(a - c); }
\end{lstlisting}
\caption{\UseMacro{figure-pr-patterns-pAdd5}}
\end{subfigure}
~
\begin{subfigure}[b]{0.31\textwidth}
\begin{lstlisting}[language=jog]
@Pattern
public void (*@\UseMacro{ptn-pAdd6}@*)(int a, int b, int c) {
  before((a - b) + (c - a));
  after(c - b); }
\end{lstlisting}
\caption{\UseMacro{figure-pr-patterns-pAdd6}}
\end{subfigure}
\caption{\UseMacro{figure-pr-patterns}}
\end{figure*}

\section{Limitations}
\label{sec:threats}

\MyPara{Internal validity}
Our experiments on comparing performance of generated and hand-written
\optimizations may suffer the threat from noise. To mitigate the
threat, we did \numberstringnum{\UseMacro{num-end-to-end-runs}}
end-to-end runs and selected only last
\UseMacro{measured-repetitions-per-run} fully warmed-up repetitions
for measurement. We filtered stable benchmarks by coefficient of
variance and we did significant difference test using bootstrap
re-sampling for time from all \UseMacro{total-measured-repetitions}
repetitions measured. Although we tried to find the numbers of
repetitions and end-to-end runs which are large enough to minimize
noise but remains practical for our experiments, choosing different
numbers could impact the experimental results.

\MyPara{Construct validity}
We used the number of characters and number of identifiers as a metric
to quantify code complexity of \ptns written using \Tool compared to
hand-written implementation of \optimizations. This metric may
or may not reflect complexity for every developer and thus may impact
our conclusion on code complexity of \optimizations written using
\Tool.

\MyPara{External validity}
We used only \Renaissance benchmark suites to evaluate performance of
\optimizations. Although \Renaissance is the state-of-the-art
benchmarks for JVM, to the best of our knowledge, it may still not
reflect all use cases of \optimizations in real world. Also, the \ptns
we wrote for evaluation cannot cover all the peephole \optimizations.
Despite our efforts to increase variety of \ptns, such as different
compilers, different operations, etc., we cannot ensure the results
can be generalized to all peephole \optimizations. Last, \Tool is
designed and developed for Java \JIT peephole \optimizations in
OpenJDK (HotSpot). Thus, \Tool will require major changes to be
directly generalized to other implementations of Java \JIT, e.g.,
OpenJ9, or other compilers, e.g., \LLVM.
However, the proposed algorithm of detecting shadow relations between
\optimizations can be easily implemented for other compilers.

\MyPara{Implementation}
Our current implementation of \Tool does not generate tests for the
\ptns with preconditions that specify invariants between variables.
For example, one of the \ptns written from \OpenJDK that transforms
\CodeIn{(x \char`\>{}\char`\>{}> C0) + C1} to \CodeIn{(x + (C1
\char`\<{}\char`\<{} C0)) \char`\>{}\char`\>{}> C0)} requires a
precondition \CodeIn{C0 < 5 \&\& -5 < C1 \&\& C1 < 0 \&\& x >= -(y
\char`\<{}\char`\<{} C0)}~\cite{OpenJDKOptimizationWithComplicatedPrecondition}.
We plan to leverage constraint solvers to obtain valid test inputs for
such tests in our future work.

\section{Related Work}
\label{sec:related}

We describe related work on (1)~\DSLs for \optimizations,
(2)~relation between \optimizations, (3)~finding new \optimizations
and (4)~benchmarking Java \JIT.

\MyPara{\DSLs for \optimizations}
A notable area of research addressing the ease of implementing
compiler optimizations is in the application of domain specific
languages (DSL) for specifying peephole optimizations.
One of the first projects~\cite{GosDsl} introduced a DSL called
\textit{Gospel} for specifying compiler optimizations.
Cobalt~\cite{Lerner03Cobalt} and Rhodium~\cite{Lerner05Rhodium} are
frameworks to specify peephole \optimizations and dataflow analyses,
and PEC~\cite{Kundu09PEC} extends to support loop \optimizations.
More recently, GCC's \textit{Match and
Simplify}~\cite{GCCMatchAndSimplify} introduces a DSL to write
expression simplifications from which code targeting GIMPLE and
GENERIC is auto-generated. 
Similarly, Alive~\cite{Lopes15Alive} is a DSL that can be used for
specifying peephole optimizations targeting LLVM. Alive can also be
used to generate C/C++ code that can be directly included into LLVM's
optimization passes.
CompCert~\cite{Leroy09CompCert} is a formalized and verified C
compiler in Coq. There is research on verifying SSA-based middle-end
optimizer~\cite{Barthe14Formal}, peephole
\optimizations~\cite{Mullen16Peek}, polyhedral model-based
\optimizations~\cite{Courant21Polyhedral}. Both
Alive-FP~\cite{Menendez16AliveFP} and
LifeJacket~\cite{Notzli16LifeJacket} prove correctness of
floating-point \optimizations.
The aforementioned tools are designed to verify correctness of
\optimizations over intermediate representation and introduced
DSLs work on the intermediate representation level, while \Tool focuses on
developers productivity and allows developers to write \optimizations
in a high-level language (Java), using the existing approach that
tests for \optimizations are written. Although \Tool does not verify
\optimizations as aforementioned tools, \Tool presents an approach to detecting
shadow relations between \optimizations and \Tool can generate IR tests
from \optimizations specified in \ptns.

There has been research that does not focus on verifying \optimizations.
CAnDL~\cite{Ginsbach18CAnDL} is a \DSL for \LLVM analysis and it
supports use cases beyond peephole \optimizations, such as control
flows. However, CAnDL requires developers to write constraints to
specify \optimizations, which is more complicated than high-level
expressions \Tool takes; also the generated compiler pass is
independent to existing code structure in \LLVM and thus more
difficult to be integrated.
COpt~\cite{Venkat18COpt} is a high-level \DSL that allows compiler
developers to specify a set of ten high-level \optimizations. In
contrast to \Tool that applies to peephole \optimizations, COpt
applies on high-level \optimizations such as global value numbering,
common subexpression elimination, function inlining, etc.

\MyPara{Relation between \optimizations}
There is research~\cite{Nishida18Loop, Menendez16AliveLoops} on
detecting non-termination bugs due to a suite of peephole
\optimizations applied repeatedly. Termination checking involves
determining whether two \optimizations can be composited, which is a
quantifier-free problem, while the shadow determining problem \Tool
solves involves with universal quantifiers.
Lopes et al.~\cite{Lopes18FutureDirectionsforOptimizingCompilers}
advocated implementation of solver-based tools for finding groups of
\optimizations that subsume each other to improve existing peephole
\optimizations. \Tool addresses that problem and we plan to explore more
relations between \optimizations in future work, e.g., duplicate
\optimizations check.

\MyPara{Finding new \optimizations}
Optgen~\cite{Buchwald15Optgen} exhaustively generates all local
\optimization rules up to a given cost limit.
Barany~\cite{Barany18Finding} and Theodoridis et
al.'s~\cite{Theodoridis22Finding} work compare different compilers'
output to find missed \optimizations. Unlike them, \Tool does not
automatically find new \optimizations but provides developers a way to
easily develop \optimizations.

\MyPara{Benchmarking Java \JIT}
\Renaissance~\cite{Prokopec19Renaissance} is recent benchmark suites
for \JVM, which shows more significant performance differences on
evaluating impacts of \JIT compiler \optimizations than older
benchmarks such as DaCapo~\cite{Blackburn06DeCapo} and
SPECjvm2008~\cite{SPECjvm2008}; therefore we used \Renaissance to
evaluate performance of generated \optimization passes from \Tool.

\section{Conclusion}
\label{sec:conclusion}

Writing peephole \optimizations requires substantial effort.
The current approach of hand-written implementation in Java \JIT is
not scalable and it is prone to bugs.
We presented \Tool, a framework that facilitates developing Java \JIT
peephole \optimizations. Compiler developers can write \ptns in the
same language as the compiler (i.e., Java), using the existing
approach for writing tests for peephole \optimizations.
\Tool translates every \ptn into C/C++ code that can be integrated as
a \JIT \optimization pass, and generates tests from the \ptn as
needed. We wrote \UseMacro{total-num-patterns} \ptns for
\optimizations found in \OpenJDK, \LLVM, as well as some that we
designed. Our evaluation shows that \Tool reduces the code size and
code complexity when compared to hand-written implementation
of \optimizations while maintaining the effectiveness of
\optimizations. \Tool can also automatically detect possible \shadow
relations between pairs of \optimizations. We utilized this to find a
bug in Java \JIT as two \optimizations could never be triggered as a
consequence of being shadowed by another. We opened
\UseMacro{total-num-PRs} pull requests for \OpenJDK, including
\UseMacro{total-num-new-optimization-PRs} on new \optimizations,
\UseMacro{total-num-shadow-PRs} on removing of \shadowed
\optimizations, and \UseMacro{total-num-test-PRs} on new tests of
existing untested \optimizations, of which
\UseMacro{total-num-integrated-PRs} PRs have been integrated into the
master branch of \OpenJDK, so far.
We believe that \Tool will have significant impact in further
developments of Java \JIT compilers.

\begin{acks}
We thank Nader Al Awar, Yu Liu, Pengyu Nie, August Shi, Jiyang Zhang,
and the anonymous reviewers for their comments and feedback.
This work is partially supported by a Google Faculty Research Award, a
grant from the Army Research Office, and the US National Science
Foundation under Grant Nos.~CCF-1652517, CCF-2107291, CCF-2217696.
\end{acks}

\balance
\bibliography{bib}

\end{document}